\documentclass[12pt,english,longbibliography,aps,prb,preprint,nofootinbib,superscriptaddress,showpacs,showkeys,12pt,sort&compress]{revtex4}
\usepackage{ae,aecompl}
\usepackage[T1]{fontenc}
\usepackage[latin9]{inputenc}
\usepackage{amsmath}
\usepackage{graphicx}

\makeatletter

\providecommand{\tabularnewline}{\\}

\@ifundefined{textcolor}{}
{%
 \definecolor{BLACK}{gray}{0}
 \definecolor{WHITE}{gray}{1}
 \definecolor{RED}{rgb}{1,0,0}
 \definecolor{GREEN}{rgb}{0,1,0}
 \definecolor{BLUE}{rgb}{0,0,1}
 \definecolor{CYAN}{cmyk}{1,0,0,0}
 \definecolor{MAGENTA}{cmyk}{0,1,0,0}
 \definecolor{YELLOW}{cmyk}{0,0,1,0}
}

\usepackage{amsfonts}
\usepackage{aecompl}

\date{\today}

\makeatother

\usepackage{babel}
\begin{document}

\title{An optimized interatomic potential for silicon and its application
to thermal stability of silicene}

\author{G. P. Purja Pun}

\address{Department of Physics and Astronomy, MSN 3F3, George Mason University,
Fairfax, Virginia 22030, USA}

\author{Y. Mishin}

\address{Department of Physics and Astronomy, MSN 3F3, George Mason University,
Fairfax, Virginia 22030, USA}
\begin{abstract}
An optimized interatomic potential has been constructed for silicon
using a modified Tersoff model. The potential reproduces a wide range
of properties of Si and improves over existing potentials with respect
to point defect structures and energies, surface energies and reconstructions,
thermal expansion, melting temperature and other properties. The proposed
potential is compared with three other potentials from the literature.
The potentials demonstrate reasonable agreement with first-principles
binding energies of small Si clusters as well as single-layer and
bilayer silicenes. The four potentials are used to evaluate the thermal
stability of free-standing silicenes in the form of nano-ribbons,
nano-flakes and nano-tubes. While single-layer silicene is mechanically
stable at zero Kelvin, it is predicted to become unstable and collapse
at  room temperature. By contrast, the bilayer silicene demonstrates
a larger bending rigidity and remains stable at and even above room
temperature. The results suggest that bilayer silicene might exist
in a free-standing form at ambient conditions. 
\end{abstract}

\maketitle

\section{Introduction\label{sec:Introduction}}

Silicon is one of the most important functional materials widely used
in electronic, optical, energy conversion and many other applications.
Not surprisingly, Si has been the subject of many classical molecular
dynamics (MD) and other large-scale atomistic computer studies for
almost three decades. Although classical atomistic simulations cannot
access electronic or magnetic properties, they are indispensable for
gaining a better understanding of the atomic structures, thermal and
mechanical properties of the crystalline, liquid and amorphous Si
and various nano-scale objects such as nano-wires and nano-dots. Atomistic
simulations rely on semi-empirical interatomic potentials. The accuracy
of the results delivered by atomistic simulations depends critically
on the reliability of interatomic potentials.

Several dozen semi-empirical potentials have been developed for Si.
Although none of them reproduces all properties accurately, there
is a trend towards a gradual improvement in their reliability as more
sophisticated potential generation methods are developed and larger
experimental and first-principles datasets become available for the
optimization and testing. The most popular Si potentials were proposed
by Stillinger and Weber (SW)\citep{Stillinger85} and Tersoff.\citep{Tersoff88,Tersoff:1988dn,Tersoff:1989wj}
The original Tersoff potentials were modified by several authors by
slightly changing the analytical functions and improving the optimization.\citep{Dodson1987,Ramana-Murty:1995fk,Kumagai:2007ly,Yu:2007on,Monteverde:2012wu,Martinez:2013aa}
Other Si potential formats include the environment-dependent interatomic
potential,\citep{Justo:1998fu} the modified embedded atom method
(MEAM) potentials,\citep{Baskes89,Lenosky:2000rt,Ryu:2009dn,Timonova:2010aa,Timonova:2011aa,Jelinek:2012ij,Liu:2015jw}
and bond-order potentials.\citep{Gillespie:2007vl,Oloriegbe_PhD:2008aa}

One of the most significant drawbacks of the existing Si potentials
is the overestimation of the melting temperature $T_{m}$, in many
cases by hundreds of degrees. Other typical problems include underestimated
vacancy and surface energies and positive Cauchy pressure $(c_{12}-c_{44})$,
which in reality is negative ($c_{ij}$ being elastic constants).
Kumagai et al.\citep{Kumagai:2007ly} constructed a significantly
improved Tersoff potential that predicts $T_{m}=1681$ K in close
agreement with the experimental value of 1687 K, gives the correct
Cauchy pressure, and is accurate with respect to many other properties.
This potential, usually referred to as MOD,\citep{Kumagai:2007ly}
is probably the most advanced Tersoff-type potential for Si available
today. However, it still suffers from a low vacancy formation energy,
low surface energies, and overestimated thermal expansion at high
temperatures and the volume effect of melting.

The goal of this work was twofold. The first goal was to further improve
on the MOD potential\citep{Kumagai:2007ly} by addressing its shortcomings
with a minimal impact on other properties. This was achieved by slightly
modifying the potential format and performing a deeper optimization.
When testing the new potential, we compare it not only with MOD but
also with the popular SW potential.\citep{Stillinger85} We further
include the MEAM potential developed by Ryu et al.\citep{Ryu:2009dn}
to represent a different potential format. To our knowledge, this
is the only MEAM potential whose melting point is close to experimental.

The second goal was to test the four potentials for their ability
to predict the energies of low-dimensional structures, such as small
Si clusters and single- and double-layer forms of silicene (2D allotrope
of Si). Si potentials are traditionally considered to be incapable
of reproducing low-dimensional structures. This view is largely based
on testing the SW potential. The MOD and MEAM potentials have not
been tested for the properties of clusters or silicenes in any systematic
manner. Such tests were conducted in this work using all four potentials.
The results suggest that the present potential, MOD and MEAM do capture
the main trends and in many cases agree with first-principles density
functional theory (DFT) calculations. As such, they can be suitable
for exploratory studies of thermal and mechanical stability of Si
clusters and 2D structural forms of Si. In this work we apply them
to evaluate the stability of free-standing single-layer and bilayer
silicenes at room temperature.

\section{Potential generation procedures}

The total energy of a collection of atoms is represented in the form
\[
E=\dfrac{1}{2}\sum_{i\neq j}\phi_{ij}(r_{ij}),
\]
where $r_{ij}$ is distance between atoms $i$ and $j$ and the bond
energy $\phi_{ij}$ is taken as 
\begin{equation}
\phi_{ij}=f_{c}(r_{ij})\left[A\exp(-\lambda_{1}r_{ij})-b_{ij}B\exp(-\lambda_{2}r_{ij})+c_{0}\right].
\end{equation}
Here, the bond order $b_{ij}$ is given by 
\[
b_{ij}=\left(1+\xi_{ij}^{\eta}\right)^{-\delta},
\]
where 
\[
\xi_{ij}=\sum_{k\ne i,j}f_{c}(r_{ij})g(\theta_{ijk})\exp\left[\alpha(r_{ij}-r_{ik})^{\beta}\right].
\]
The term $(1+\xi_{ij})$ represent an effective coordination number
of atom $i$ and $f_{c}(r_{ij})$ is a cutoff function. The latter
has the form 
\[
f_{c}(r)=\begin{cases}
1, & r\le R_{1}\\
\dfrac{1}{2}+\dfrac{9}{16}\cos\left(\pi\dfrac{r-R_{1}}{R_{2}-R_{1}}\right)-\dfrac{1}{16}\cos\left(3\pi\dfrac{r-R_{1}}{R_{2}-R_{1}}\right), & R_{1}<r<R_{2}\\
0, & r\ge R_{2},
\end{cases}
\]
where $R_{1}$ and $R_{2}$ are cutoff radii. The outer cutoff $R_{2}$
is chosen between the first and second coordination shells of the
diamond cubic  structure. The angular function $g(\theta_{ijk})$
has the generalized form 
\[
g(\theta)=c_{1}+\dfrac{c_{2}(h-\cos\theta)^{2}}{c_{3}+(h-\cos\theta)^{2}}\left\{ 1+c_{4}\exp\left[-c_{5}(h-\cos\theta)^{2}\right]\right\} ,
\]
where $\theta_{ijk}$ is the angle between the bonds $ij$ and $ik$.
These functional forms are the same as for the MOD potential,\citep{Kumagai:2007ly}
except for the new coefficient $c_{0}$ that was added to better control
the attractive part of the potential.

The adjustable parameters of the potential are $A$, $B$, $\alpha$,
$h$, $\eta$, $\lambda_{1}$, $\lambda_{2}$, $R_{1}$, $R_{2}$,
$\delta$, $c_{0}$, $c_{1}$, $c_{2}$, $c_{3}$, $c_{4}$ and $c_{5}$.
The power $\beta$ is a fixed odd integer. In the original Tersoff
potential\citep{Tersoff88,Tersoff:1988dn,Tersoff:1989wj} $\beta=3$,
whereas Kumagai et al.\citep{Kumagai:2007ly} chose $\beta=1$. We
tried both numbers and found that $\beta=3$ gives a better potential.

The free parameters of the potential were trained to reproduce basic
physical properties of the diamond cubic (A4) structure and the energies
of several alternate structures. Specifically, the fitting database
included the experimental lattice parameter $a$, cohesive energy
$E_{c}$, elastic constants $c_{ij}$, and the vacancy formation energy
$E_{v}^{f}$. The alternate structures were: simple cubic (SC), $\beta$-Sn
(A5), face-centered cubic (FCC), hexagonal closed pack (HCP), body-centered
cubic (BCC), simple hexagonal (HEX), wurtzite (B4), BC8, ST12, and
clathrate (cP46). Their energies obtained by DFT calculations are
available from open-access databases such as Materials Project,\citep{Materials_Project}
OQMD\citep{Wolverton2013} and AFLOW.\citep{Curtarolo:2012kq,Curtarolo2012}
Some of these structures were found experimentally as Si polymorphs
under high pressure, others were only generated in the computer for
testing purposes. The parameter optimization process utilized a simulated
annealing algorithm. The objective function was the sum of weighted
squares of deviations of properties from their target values. Numerous
optimization runs were conducted using the weights as a tool to achieve
the most meaningful distribution of the errors over different properties.
Several versions of the potential were generated and the version deemed
to be most reasonable was selected as final.

The optimized potential parameters are listed in Table \ref{tab:parameters}.
The potential has been incorporated in the molecular dynamics package
LAMMPS (Large-scale Atomic/Molecular Massively Parallel Simulator)\citep{Plimpton95}
as the pair style \texttt{tersoff/mod/c}.

The transferability of the new potential was evaluated by computing
a number of physical properties that were not included in the training
database and comparing the results with experimental data and/or DFT
calculations available in the literature. The same comparison was
made for the MOD, MEAM and SW potentials to demonstrate their strengths
and weaknesses relative to the new potential. We utilized the MOD
and SW potential files from the LAMMPS potential library. The MEAM
potential file was obtained from the developers.\citep{Ryu:2009dn}
The potential testing results are reported in the next Section.

\section{Properties of solid Si}

Table \ref{tab:fundamental_prop} summarizes some of the properties
of crystalline Si predicted by the four potentials. All properties
have been computed in this work unless otherwise is indicted by citations.
The defect energies are reported after full atomic relaxation.

\subsection{Lattice properties}

\begin{figure}
\noindent \begin{centering}
\textbf{(a)}\enskip{}\includegraphics[width=0.69\textwidth]{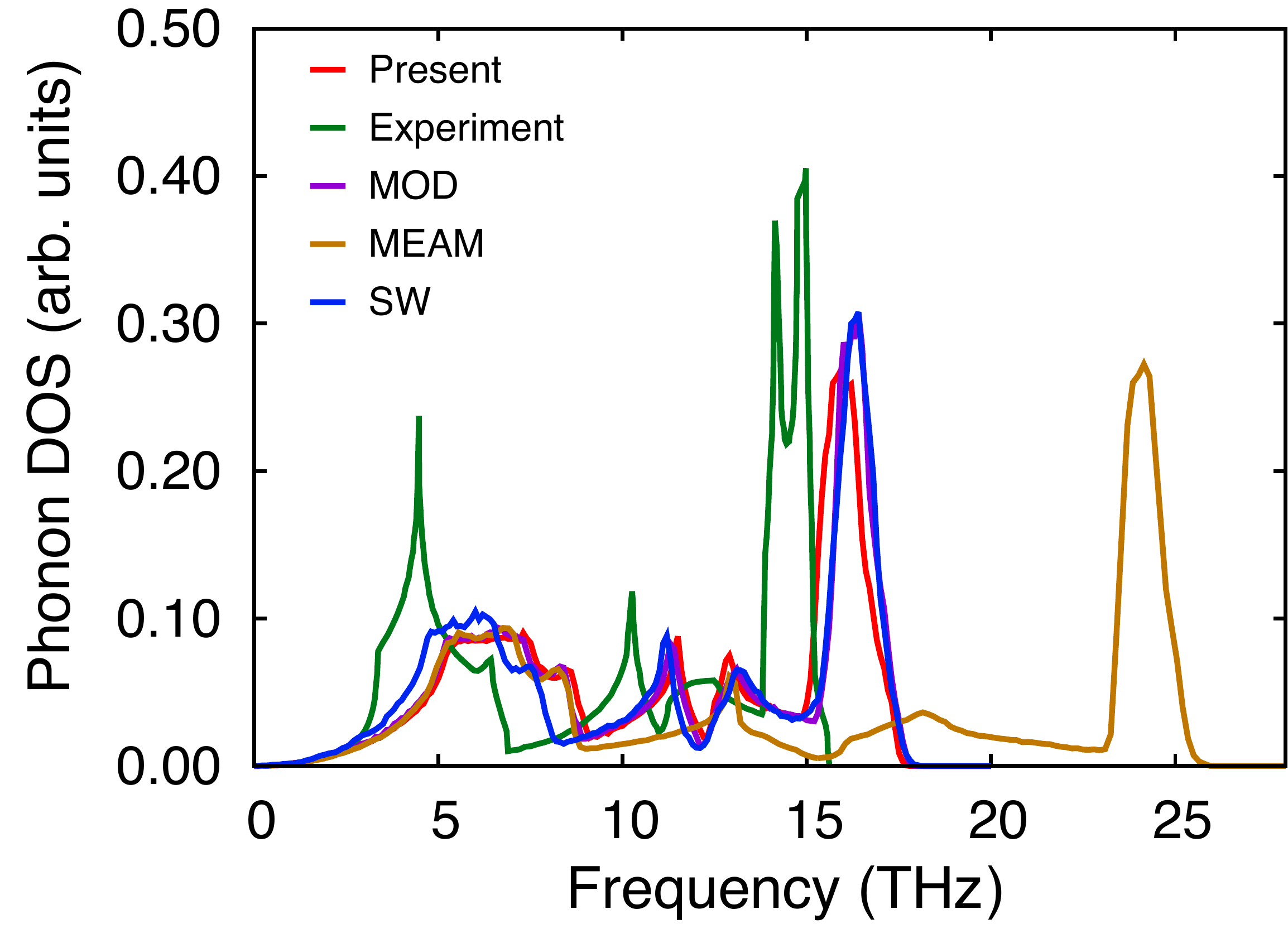} 
\par\end{centering}
\bigskip{}
 \bigskip{}
 \bigskip{}

\noindent \begin{centering}
\textbf{(b)}\enskip{}\includegraphics[width=0.7\textwidth]{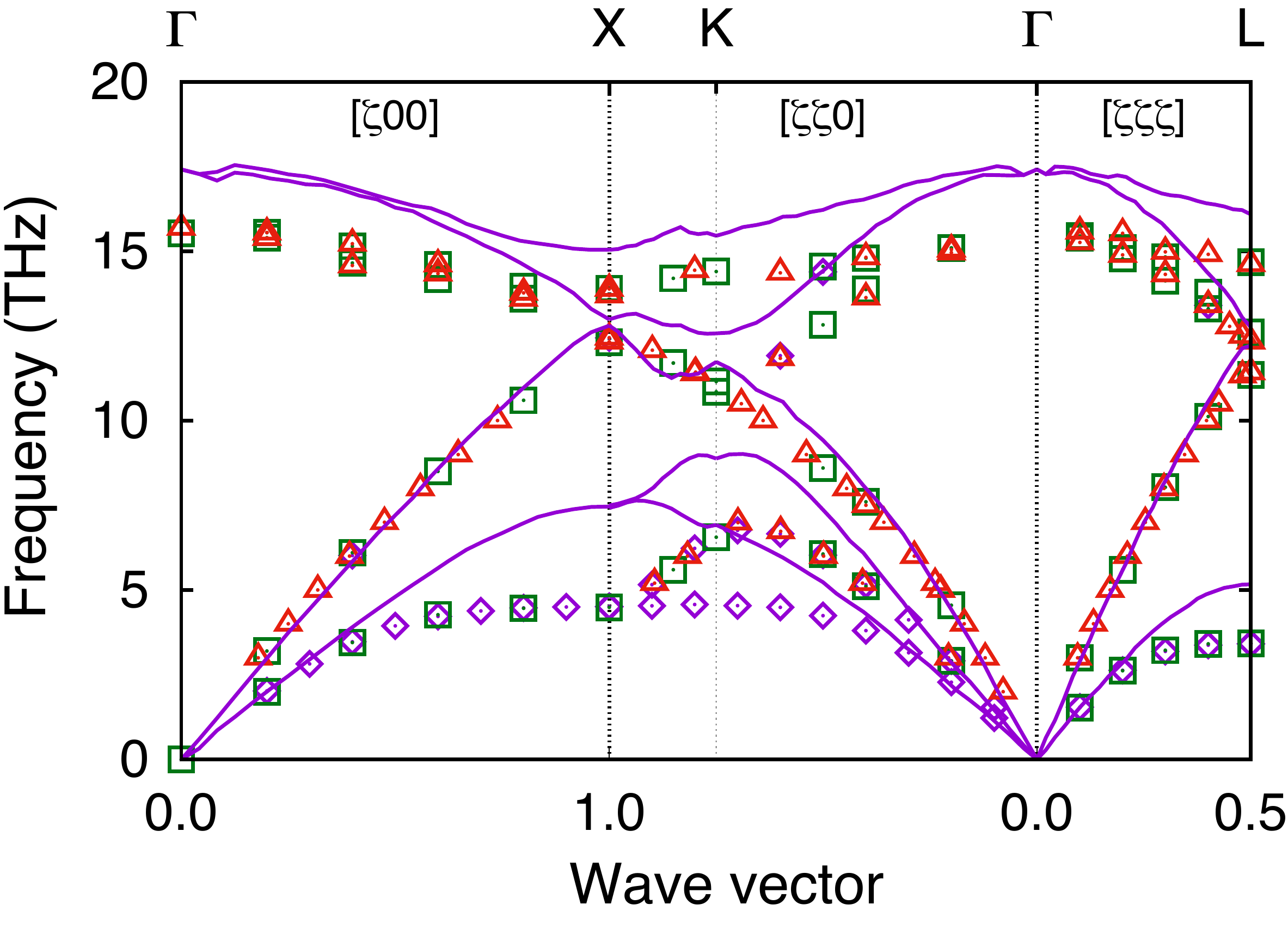} 
\par\end{centering}
\caption{Phonons properties of diamond cubic Si. (a) Density of states calculated
with different interatomic potentials in comparison with experimental
data.\citep{Dolling:1963} (b) Dispersion relations at room temperature
computed with the present potential in comparison with experiment.\citep{Dolling:1963,Nilsson:1972ve,Zdetsis:1979cs,Kulda:1994rc}
\label{fig:phonons} }
\end{figure}

The present potential, MOD and MEAM accurately reproduce the elastic
constants. The SW potential gives less accurate elastic constants
and a positive Cauchy pressure contrary to experiment.\citep{Smithells-metals2004,Skimin1951}
The phonon density of states (DOS) and phonon dispersion relations
were computed by the method developed by Kong\citep{Kong2011} and
implemented in LAMMPS. The MD simulation was performed at 300 K utilizing
a primitive $16\times16\times16$ supercell with 8192 atoms. The DOS
plots are shown in Fig.~\ref{fig:phonons}(a) and the respective
zone-center optical frequencies $\nu_{\textrm{max}}$ are indicated
in Table \ref{tab:fundamental_prop}. The present potential, MOD and
SW predict surprisingly similar $\nu_{\textrm{max}}$ values that
underestimate the experimental frequency by about 2 THz. The MEAM
potential overshoots $\nu_{\textrm{max}}$ by about 10 THz and the
entire DOS is stretched by a factor of 1.63. Note that none of the
four potentials reproduces the sharp peak at about 5 THz arising from
the acoustic zone-boundary phonons.

Fig.~\ref{fig:phonons}(b) displays the phonon dispersion curves
predicted by the present potential. While general agreement with experiment\citep{Dolling:1963,Nilsson:1972ve,Zdetsis:1979cs,Kulda:1994rc}
is evident and the longitudinal acoustic branches are reproduced accurately,
the potential overestimates the transverse acoustic zone-boundary
frequencies and the optical frequencies.

\begin{figure}
\noindent \centering{}\includegraphics[width=0.7\textwidth]{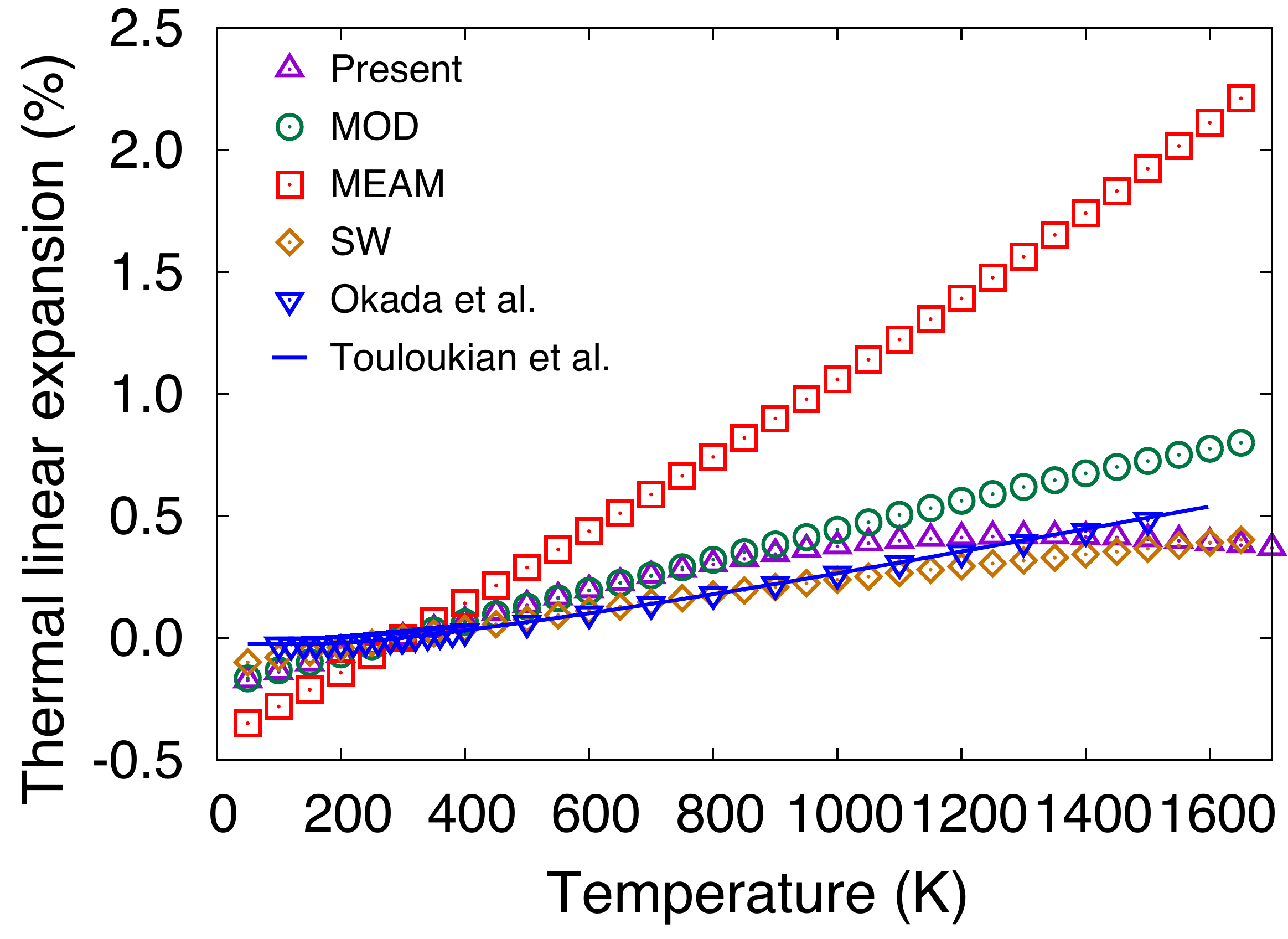}
\caption{Linear thermal expansion of Si lattice, $(a-a_{0})/a_{0}$, relative
to room temperature ($a_{0}$ at 295 K) predicted by four interatomic
potentials in comparison with experimental measurements.\citep{Expansion,Okada:1984aa}
\label{fig:linearfactor} }
\end{figure}

The cubic lattice parameter $a$ was computed as a function of temperature
by zero-pressure MD simulations. The linear thermal expansion coefficient
$(a-a_{0})/a_{0}$ relative to room temperature ($a_{0}$ at 295 K)
is compared with experimental data in Fig.~\ref{fig:linearfactor}.
The SW potential demonstrates exceptionally good agreement with experiment.
The present potential slightly overestimates the experiment at temperatures
below 1300 K and underestimates at higher temperatures. The negative
slope at high temperatures is unphysical, but the overall agreement
with experiment is reasonable. The MOD potential gives a similar thermal
expansion at low temperatures but over-predicts it at high temperatures.
The MEAM potential grossly overestimates the thermal expansion. Given
also the poor agreement for phonons, care should be exercised when
using this potential for thermodynamic calculations of crystalline
Si. Note that neither phonon properties nor thermal expansion were
not included in the fitting databases of the potentials.

\subsection{Lattice defects}

According to DFT calculations,\citep{Puska:1998kx,Goedecker:2002yf,Centoni:2005sd,Wright:2006sf,Dabrowski:2015ss}
a Si vacancy can exist in several metastable structures. In the lowest-energy
structure, the four neighbor atoms slightly move towards the vacant
site preserving the tetrahedral ($T_{d}$) symmetry and leaving four
dangling bonds. A slightly less favorable structure is obtained when
one of the four atoms moves towards the other three and forms six
identical bonds. This configuration has a hexagonal ($D_{3d}$) symmetry
and is referred to as the ``dimerized'' or ``split'' vacancy.
This vacancy reconstruction eliminates the dangling bonds but increases
the elastic strain in the surrounding lattice. The present potential
and MEAM correctly predict the split vacancy to be less stable than
the $T_{d}$ vacancy. The latter has the formation energy within the
range of DFT calculations and consistent with the experimental value
of 3.6 eV.\citep{Dannefaer:1986qd} (It should be noted, though, that
the experiments are performed at high temperatures at which the vacancy
structure is unknown.) The MOD and SW potentials significantly under-predict
the formation energy of the $T_{d}$ vacancy. In addition, with the
MOD potential the split vacancy spontaneously transforms to a $D_{2d}$
structure with the energy of 3.41 eV (the DFT value is 3.46 eV)\citep{Sholihun:2015aa},
whereas the SW potential predicts the split vacancy to be mechanically
unstable and spontaneously transform to the $T_{d}$ structure.

Self-interstitials can exist in four distinct configurations: hexagonal
(hex), tetrahedral ($T_{d}$), bond center (B) and $\langle110\rangle$
split (Table\,\ref{tab:fundamental_prop}). Given the large scatter
of the DFT formation energies, all four potentials perform almost
equally well. There is one exception: the MEAM and SW potentials predict
the hexagonal interstitial to be mechanically unstable and spontaneously
transform to the tetrahedral configuration. Both potentials overestimate
the B-interstitial energy.

Surface energies were computed for the low-index orientations \{100\},
\{110\} and \{111\}. Experiments have shown that these surfaces can
undergo reconstructions to several different structures.\citep{Tong:1988dk,Tromp:1985ya,Qian:1987sp}
Reconstructions of the \{110\} and \{111\} surfaces are accompanied
by a modest energy reduction of about 0.3-0.4 J/m$^{2}$. In this
paper, these surfaces were tested in unreconstructed states. By contrast,
the dimer reconstruction of the \{100\} surface to the more stable
$2\times1$ structure reduces the surface energy by almost 1 J/m$^{2}$.
In this case, both reconstructed and unreconstructed structures were
compared with DFT calculations. Table\,\ref{tab:fundamental_prop}
shows that the SW potential does an excellent job reproducing the
DFT surface energies. The MOD potential is the least accurate: it
systematically underestimates the surfaces energies for all orientations.
The present potential demonstrates a substantial improvement over
MOD: all energies are higher and closer to the DFT data. The MEAM
potential is equally good for all surfaces except for the unreconstructed
\{100\} structure. The latter is mechanically unstable with this potential
and reconstructs to the $2\times1$ structure spontaneously during
static relaxation at 0 K. This instability was not observed in the
DFT calculations.\citep{Stekolnikov:2002zr} The surface energy of
1.74 J/m$^{2}$ shown in the table was obtained by constrained relaxation
of this surface, in which the atoms were only allowed to move in the
direction normal to the surface to prevent the dimerization. With
the potential proposed in this work, the unreconstructed \{100\} surface
is stable at 0 K and forms symmetrical rows of dimers corresponding
to the $2\times1$ reconstruction upon heating to 1000 K and slowly
cooling down to 0 K.

As another test of the potentials, unstable stacking fault energies
$\gamma_{us}$ were calculated for the \{111\} and \{100\} crystal
planes. Such faults are important for the description of dislocation
core structures. In silicon, dislocations glide predominantly on \{111\}
planes. The spacing between \{111\} planes alternates between wide
and narrow. In the former case the chemical bonds are normal to the
planes while in the latter they are at 19.47$^{\circ}$ angles. A
generalized stacking fault is obtained by translation of one half-crystal
relative to the other in a chosen direction parallel to a \{111\}
plane. Depending on whether the cutting plane passes between widely
spaced or narrowly spaced atomic layers, the stacking fault is called
shuffle type or glide type, respectively. After each increment of
crystal translation, the atoms are allowed to minimize the total energy
by local displacements normal (but no parallel) to the fault plane.
The excess energy per unit surface area plotted as a function of the
translation vector is called the gamma-surface. If the dislocation
Burgers vector is parallel to a crystallographic direction $\left\langle hkl\right\rangle $,
then its core structure is dictated by the $\left\{ 111\right\} \left\langle hkl\right\rangle $
cross-sections of the gamma-surface. The unstable stacking fault energy
$\gamma_{us}$ is the maximum energy in this cross-section.

\begin{figure}
\textbf{(a)}\enskip{}\includegraphics[angle=-90,width=0.435\textwidth]{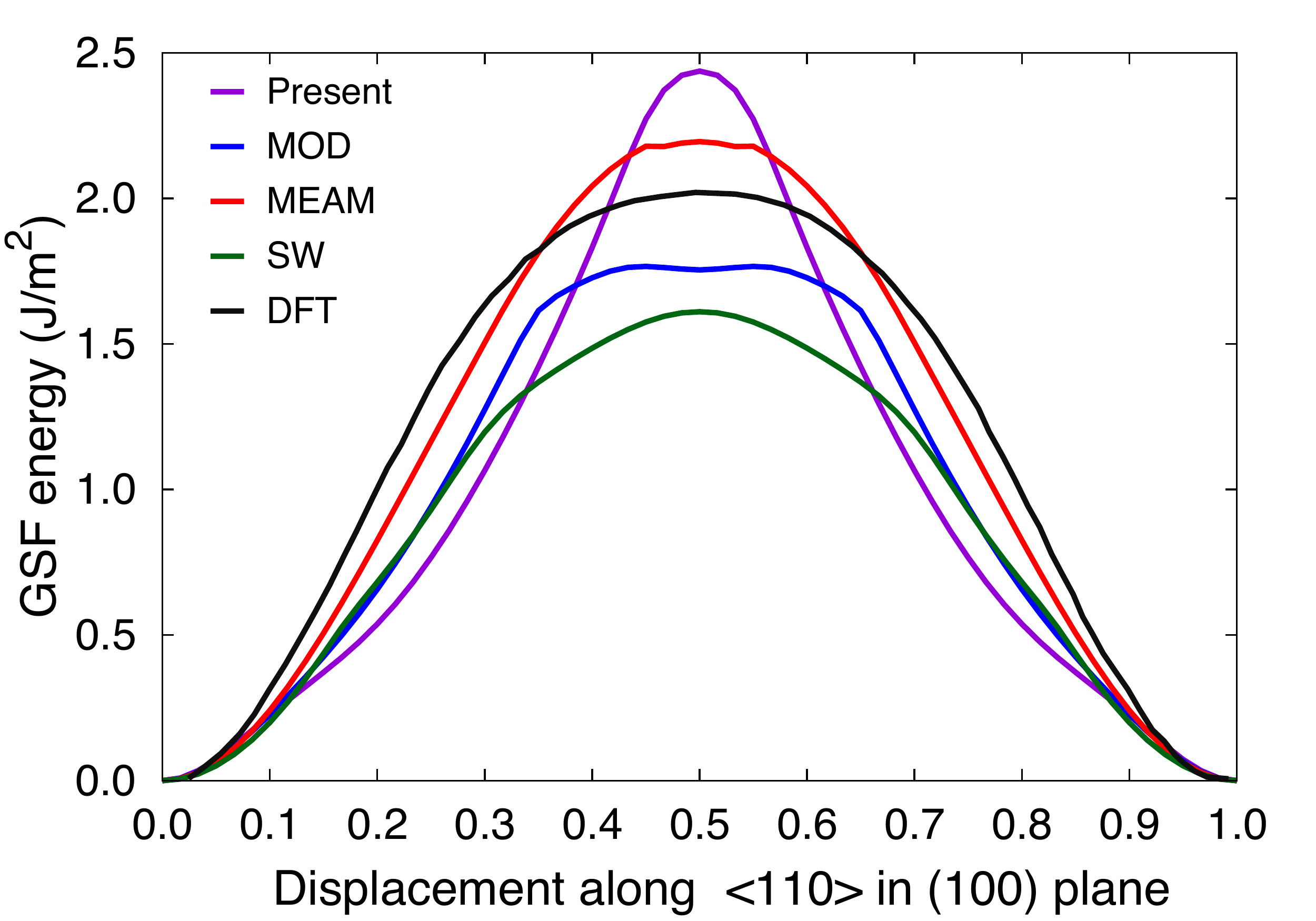}\enskip{}\enskip{}\enskip{}\textbf{(b)}\enskip{}\includegraphics[angle=-90,width=0.435\textwidth]{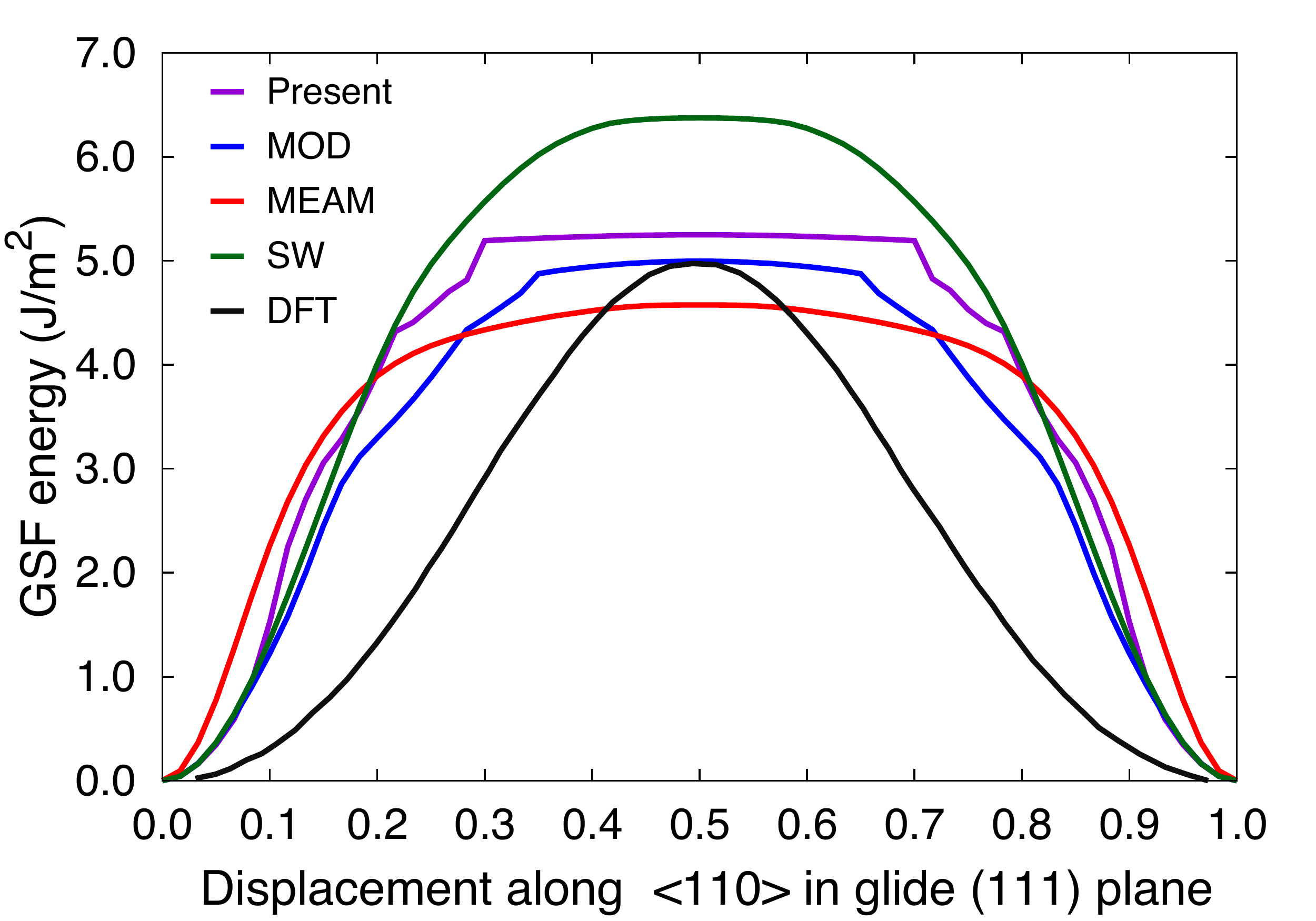}\bigskip{}
 \bigskip{}

\textbf{(c)}\enskip{}\includegraphics[angle=-90,width=0.435\textwidth]{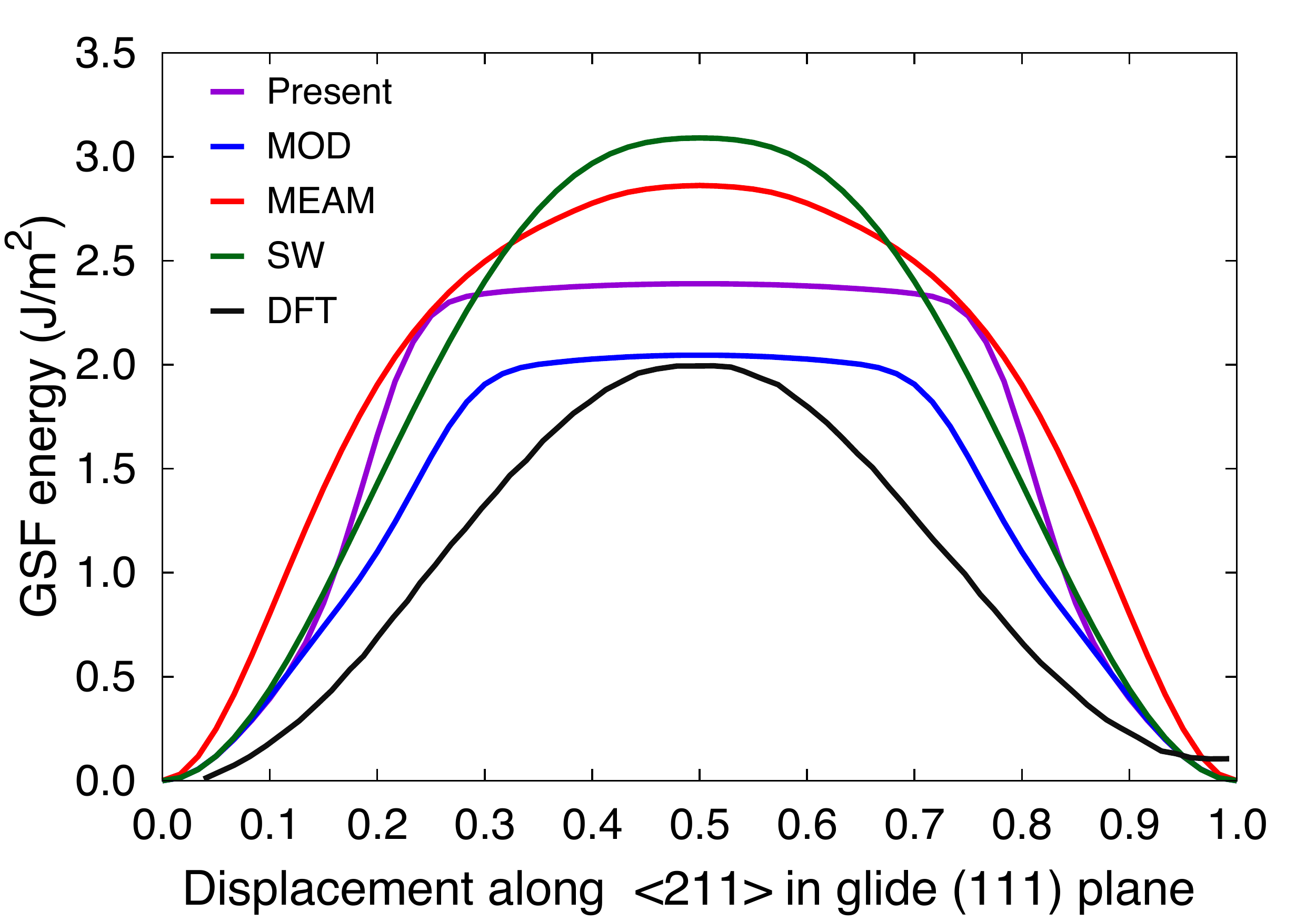}\enskip{}\enskip{}\enskip{}\textbf{(d)}\enskip{}\includegraphics[angle=-90,width=0.435\textwidth]{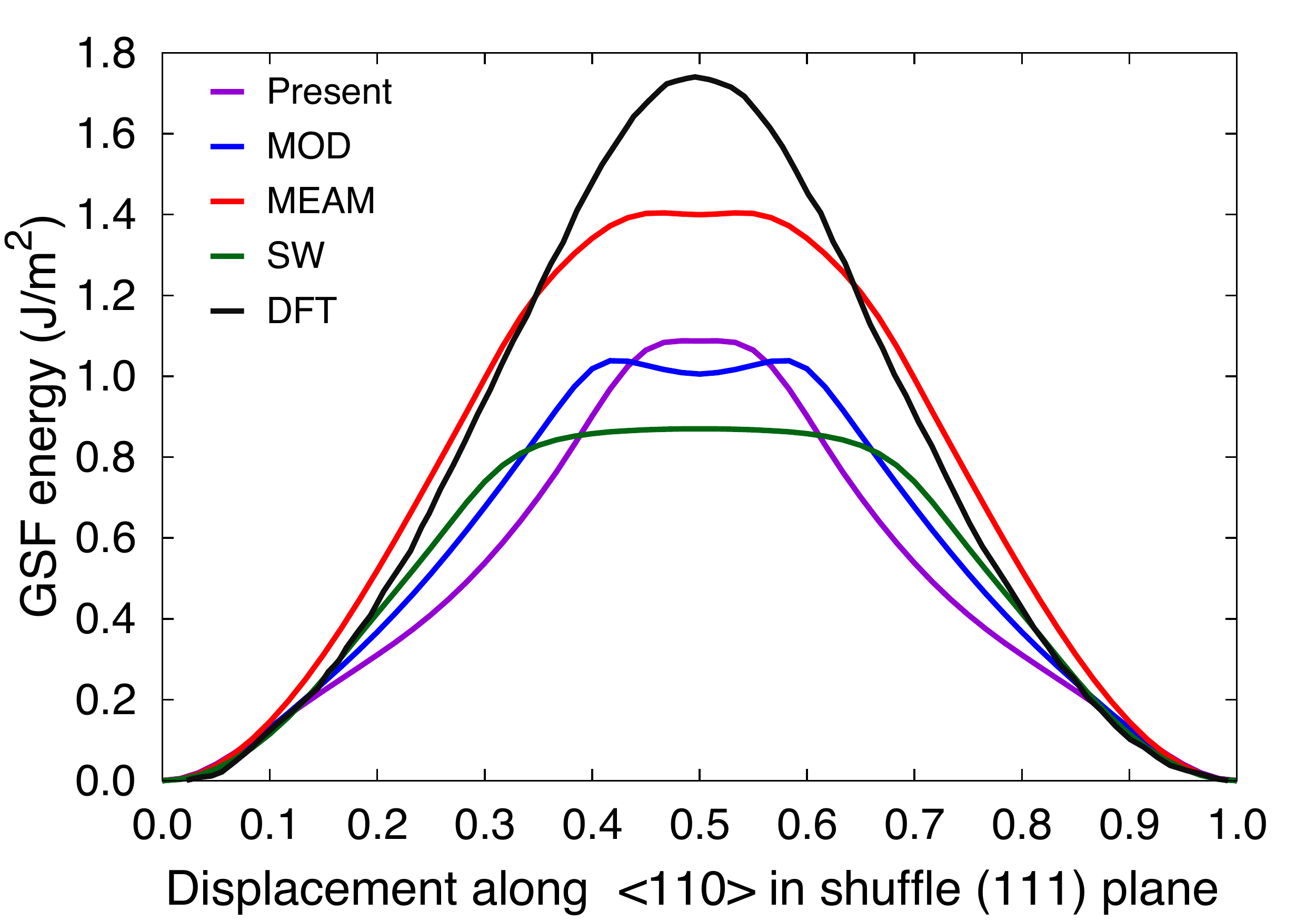}

\caption{Selected cross-sections of the \{111\} and \{100\} gamma surfaces
predicted by the present potential in comparison with other potentials
and DFT calculations.\citep{Kaxiras:1993fk,Juan:1996uo} \label{fig:GSF}}
\end{figure}

Figure \ref{fig:GSF} displays three cross-sections of the \{111\}
gamma surface computed with the four potentials in comparison with
DFT calculations. The figure additionally includes the $\left\{ 100\right\} \left\langle 110\right\rangle $
cross-section for which DFT data is available. The respective $\gamma_{us}$
values are summarized in Table \ref{tab:GSF}. While none of the potentials
reproduces the DFT curves well, the SW potential tends to be the least
accurate. For some of the cross-sections, the Tersoff-type potentials
``chop off'' the tip of the curve due to the short range of atomic
interactions and a relatively sharp cutoff. It should also be noted
that the potentials do not reproduce the stable stacking fault predicted
by DFT calculations {[}Fig.~\ref{fig:GSF}(c){]}. This fault arises
due to long-range interactions and is not captured by these potentials.

\section{Melting temperature and liquid properties of Si}

\begin{figure}
\noindent \begin{centering}
\includegraphics[width=0.7\textwidth]{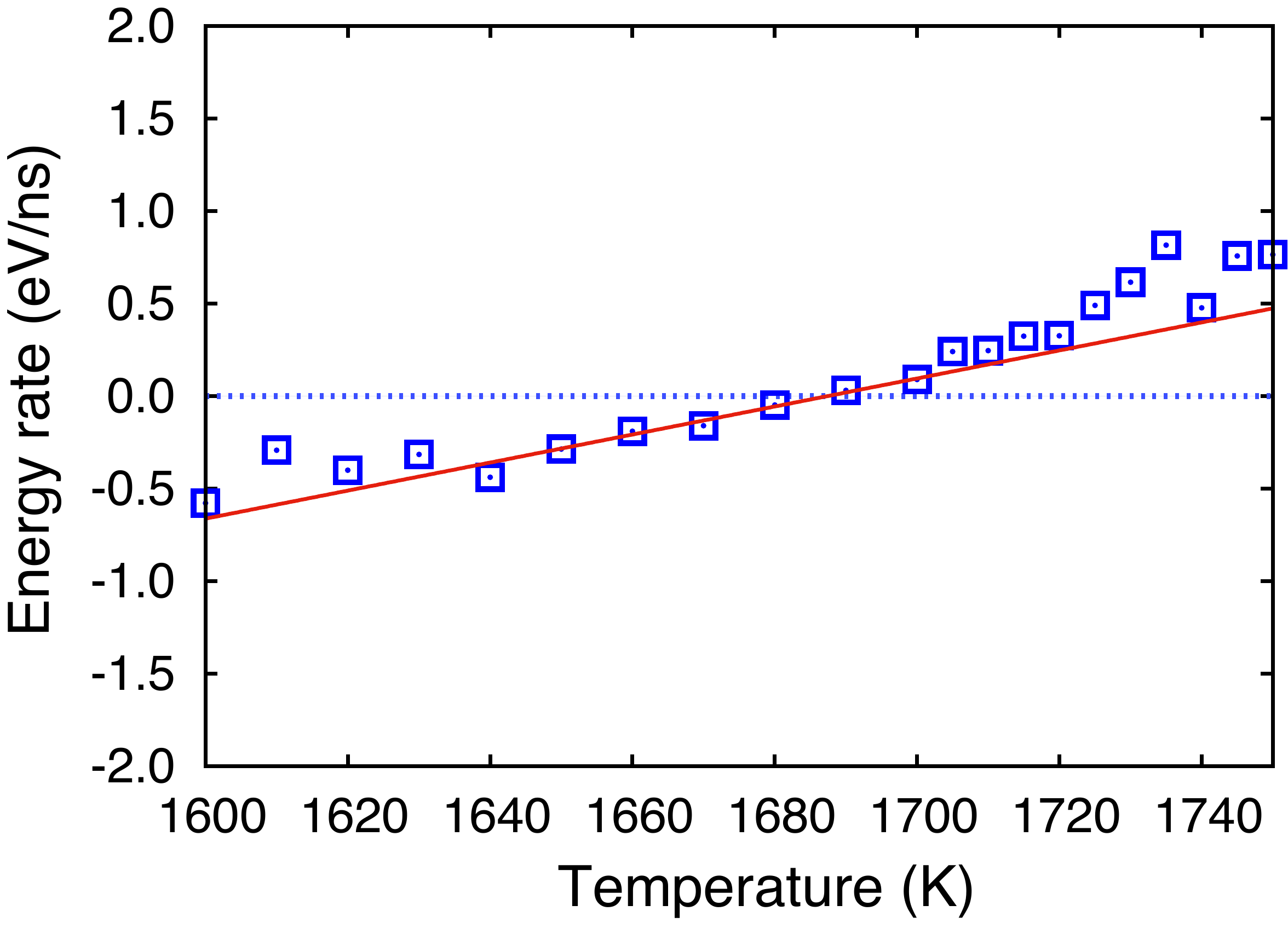} 
\par\end{centering}
\caption{Rate of energy charge as a function of temperature during melting
and crystallization of Si modeled with the present potential. The
line is the linear fit to determine the melting temperature.\label{fig:melting}}
\end{figure}

The melting temperature was computed by the interface velocity method.
A periodic simulation block containing a (111) solid-liquid interface
was subject to a series of isothermal MD simulations in the NPT ensemble
(zero pressures in all directions) at several different temperatures.
The interface migrated towards one phase or the other, depending on
whether the temperature was above or below the melting point. The
total energy of the system was monitored in this process and was found
to be a nearly linear function of time. The slope of this function
gives the rate of the energy change due to the phase transformation.
A plot of this energy rate as a function of temperature was used to
find the melting point by linear interpolation to the zero rate (Fig.~\ref{fig:melting}).
For the present potential, the melting temperature obtained was found
to be $T_{m}=1687\pm4$\,K (the error bar is the standard deviation
of the linear fit). This temperature is in excellent agreement with
the experimental melting point of 1687\,K, even though it was not
included in the fitting procedure. 

To verify our methodology, similar calculations were performed for
the MOD potential. The result was $T_{m}=1682\pm4$\,K, which matches
1681 K reported by the potential developers.\citep{Kumagai:2007ly}
For the SW potential, the same method gives $T_{m}=1677\pm4$\,K.
This number is consistent (within the error bars) with $T_{m}=1691\pm20$\,K
obtained by thermodynamic calculations.\citep{Broughton:1987ez} The
energy rate versus temperature plots for the MOD and SW potentials
can be found in the Supplemental Material to this paper.\citep{Supplementary-Si-Tersoff}

Table \ref{tab:fundamental_prop} summarizes the predictions of the
four potentials for the latent heat of melting $L$ and the volume
effect of melting $\Delta V_{m}$ relative to the volume of the solid
$V_{\textrm{solid}}$. None of the potentials reproduces these properties
well. The present potential gives the most accurate volume effect
$\Delta V_{m}/V_{\textrm{solid}}$ but the least accurate latent heat
$L$. The MOD potential predicts a better value of $L$ but overestimated
the volume effect a factor of two.

Prediction of structural properties of liquid Si presents a significant
challenge to interatomic potentials. The nature of atomic bonding
in Si changes from covalent to metallic upon melting,\citep{Muller:1978fk}
causing an increase in density. In this work, the structure of liquid
Si was characterized by the pair correlation function $g(r)$ and
the bond-angle distribution function $g(\theta,r)$. These functions
were averaged over 300 uncorrelated snapshots from NPT MD simulations
under zero pressure at 1750\,K using a simulation block containing
6912 atoms. The angular distribution $g(\theta,r)$ was computed for
bonds within the radius $r_{m}$ of the first minimum of $g(r)$ and
normalized by unit area under the curve. 

\begin{figure}[htbp]
\textbf{(a)}\enskip{}\includegraphics[width=0.65\textwidth]{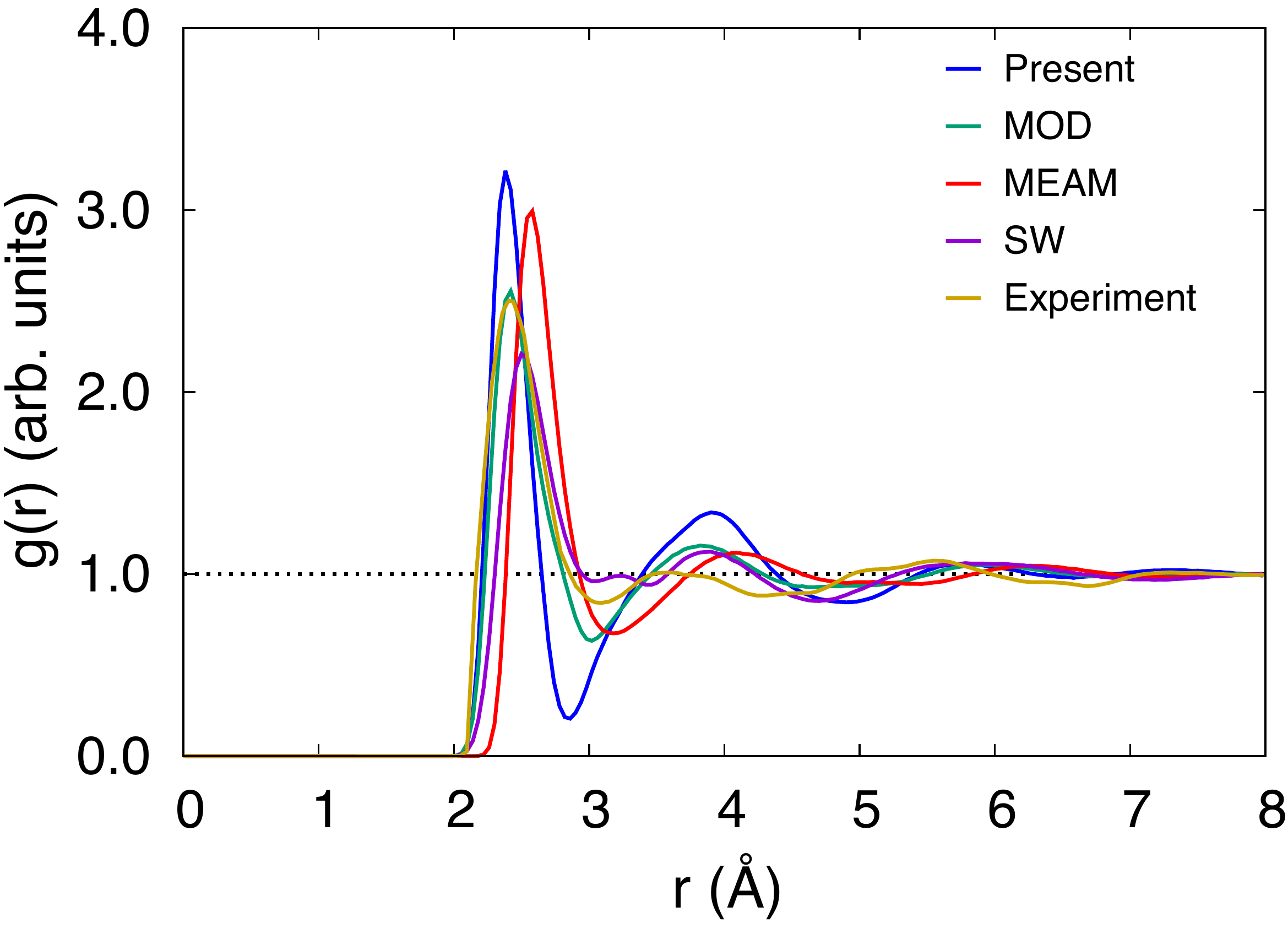}

\bigskip{}

\textbf{(b)}\enskip{}\includegraphics[width=0.65\textwidth]{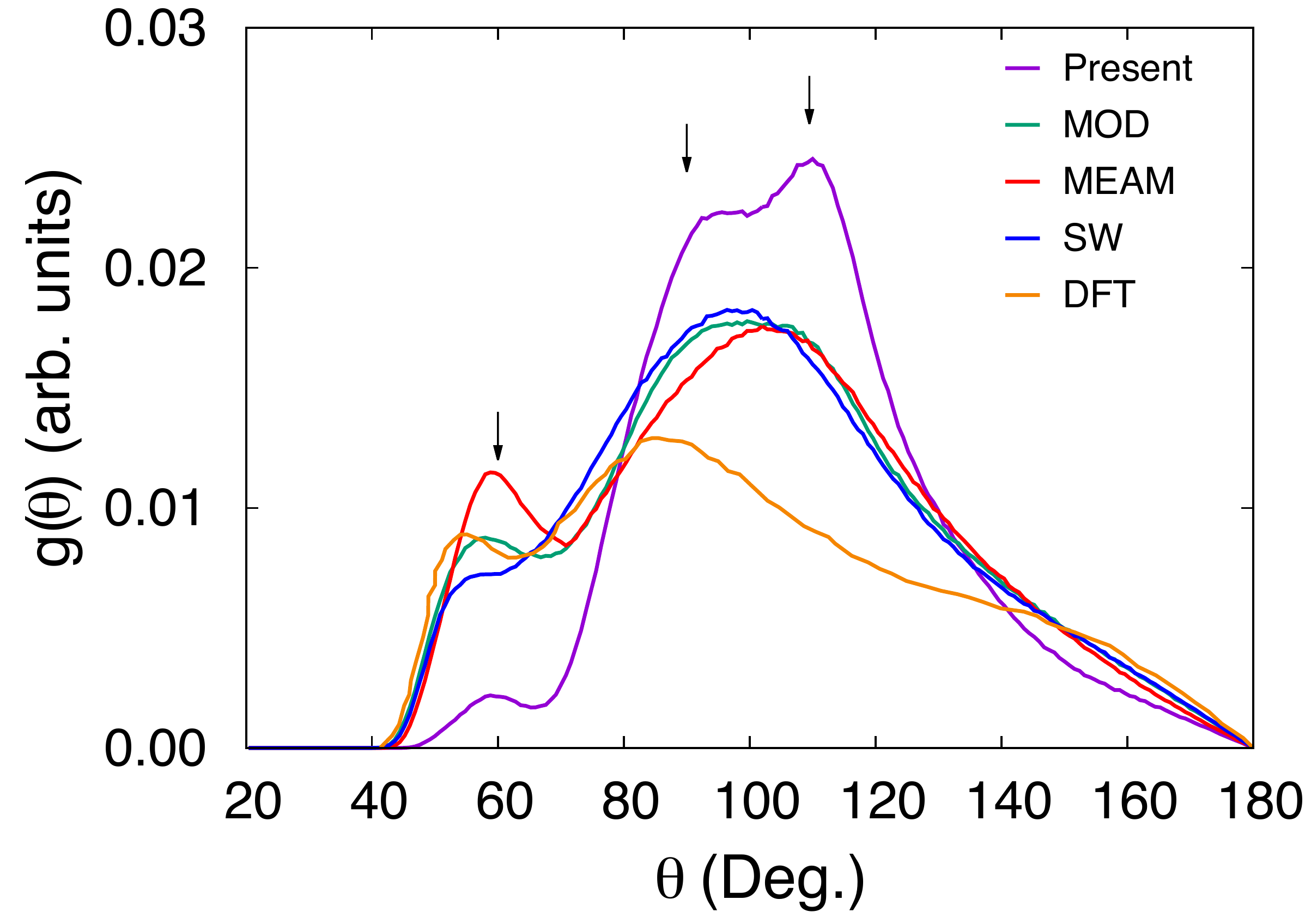}

\caption{Structure of liquid Si: (a) pair correlation function $g(r)$ and
(b) bond-angle distribution $g(\theta,r)$ computed with the present
interatomic potential at the temperature of 1750 K in comparison with
first-principles calculation at 1767 K,\citep{Jank:1990qp} experimental
data at 1733 K,\citep{Waseda:1995ur} and the MOD, MEAM and SW potentials
at 1767 K. The arrows indicate the angles of 60$^{\circ}$, 90$^{\circ}$
and 109.47$^{\circ}$. \label{fig:liquid_prop} }
\end{figure}

The results are shown in Fig.\,\ref{fig:liquid_prop}. The present
potential turns out to be the least accurate for the liquid properties.
The first maximum of $g(r)$ is too high and the first minimum too
deep in comparison with experiment.\citep{Waseda:1995ur} The other
potentials perform better but still show significant departures from
the experiment. For the bond-angle distribution, the results computed
with the four potentials are very different and none agrees with the
DFT simulations. The DFT simulations (\emph{ab initio} MD)\citep{Stich:1989rg,Jank:1990qp}
yield a broader distribution with two peaks of comparable height centered
at $60^{\circ}$ and $90^{\circ}$. The present potential strongly
underestimates the $60^{\circ}$ peak, overestimates the peak at 90$^{\circ}$,
and creates another peak at the tetrahedral angle of 109.47$^{\circ}$.
Using the other potentials, the position of the large peak varies
between 90$^{\circ}$and 109.47$^{\circ}$. Overall, our potential
overestimates the degree of structural order in the liquid phase.
This seems somewhat surprising given that this potential predicts
the most accurate volume effect of melting.

\section{Alternate crystal structures of Si}

Tables \ref{tab:alternate_strucs_iso_vol} and \ref{tab:vol_strucs}
show the equilibrium energies of several crystal structures of Si
relative to the diamond cubic structure and the respective equilibrium
atomic volumes. All these structures were included in the potential
fitting procedure except for two. The h-Si$_{6}$ structure was recently
found by DFT calculations as a new mechanically stable polymorph of
Si attractive for optoelectric applications due to its direct band
gap of 0.61 eV and interesting transport and optical properties.\citep{Guo:2015aa}
The h-Si$_{6}$ structure is composed of Si triangles forming a hexagonal
unit cell with the P6$_{3}$/mmc space group. Si$_{24}$ is another
mechanically stable polymorph that has recently been synthesized by
removing Na from the Na$_{4}$Si$_{24}$ precursor.\citep{Kim:2015aa}
The orthorhombic Cmcm structure of Si$_{24}$ contains open channels
composed of 6 and 8-member rings. This polymorph has a quasi-direct
1.3 eV band gap and demonstrates unique electronic and optical properties
making it a promising candidate for photovoltaic and other applications.
The h-Si$_{6}$ and Si$_{24}$ structures were used for testing purposes
to evaluate the transferability of the potentials. All structures
were equilibrated by isotropic volume relaxation without local displacements
of atoms. For the HCP and Wurtzite structures, the $c/a$ ratios were
fixed at the ideal values. For the simple hexagonal, $\beta$-Sn and
h-Si$_{6}$ structures, $c/a$ was fixed at the DFT values of 0.94,
0.552 and 0.562, respectively. It is worth mentioning that the present
potential and MOD predict the wurtzite phase to be mechanically unstable
at 0 K, which appears to be a generic feature of Tersoff-type potentials.

\begin{figure}
\textbf{(a)}\enskip{}\includegraphics[angle=-90,width=0.435\textwidth]{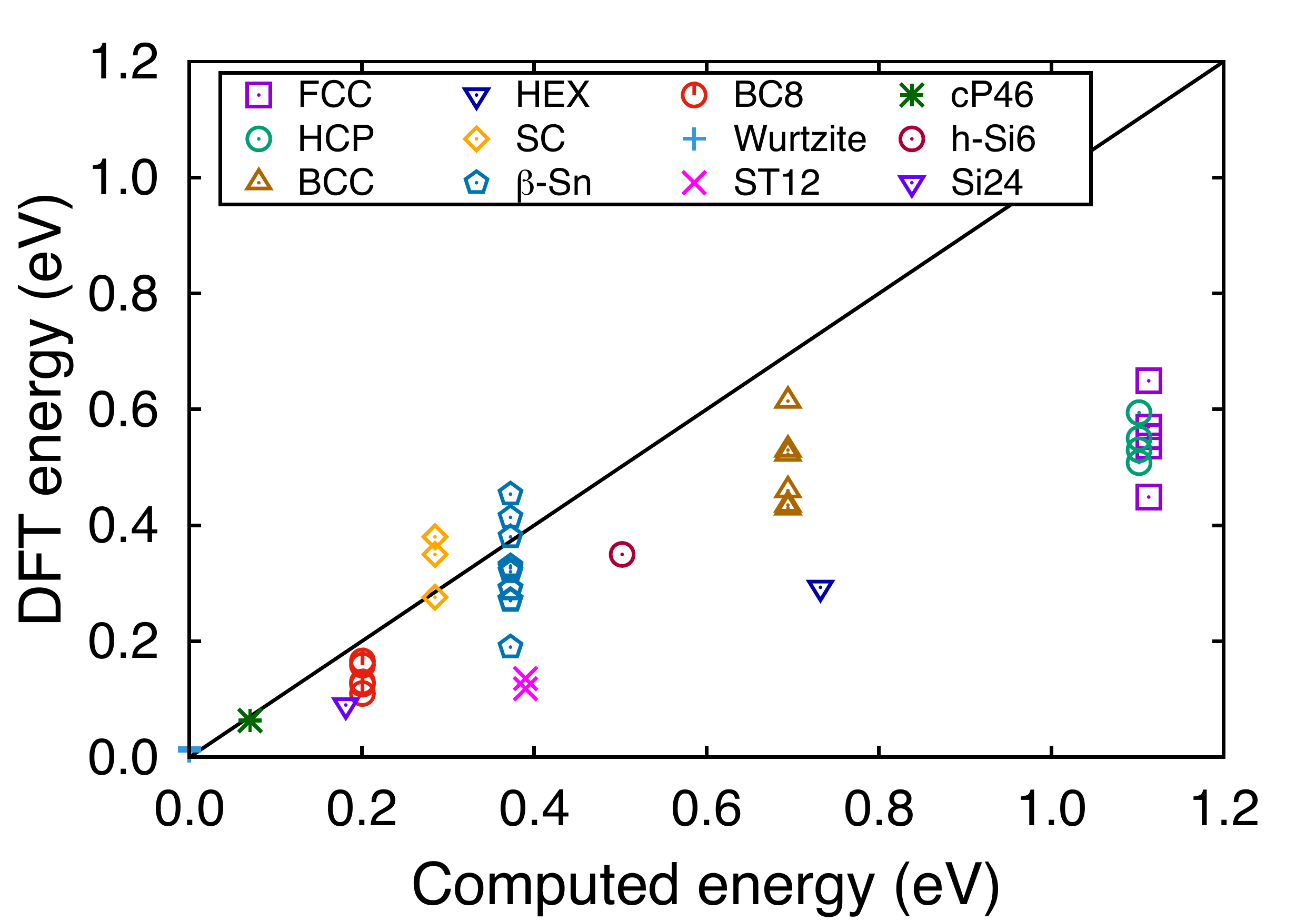}\enskip{}\enskip{}\enskip{}\textbf{(b)}\enskip{}\includegraphics[angle=-90,width=0.435\textwidth]{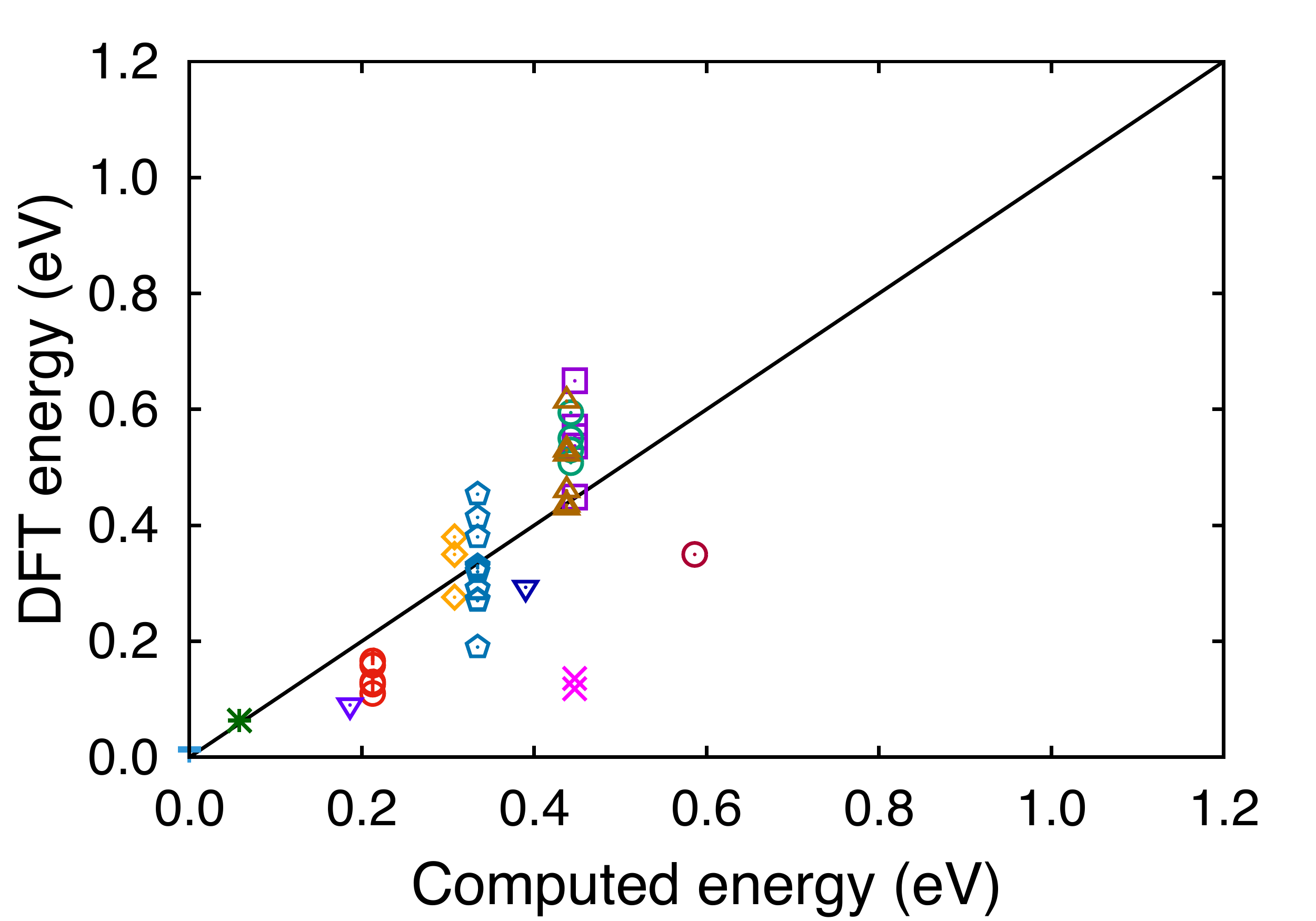}\bigskip{}
 \bigskip{}

\textbf{(c)}\enskip{}\includegraphics[angle=-90,width=0.435\textwidth]{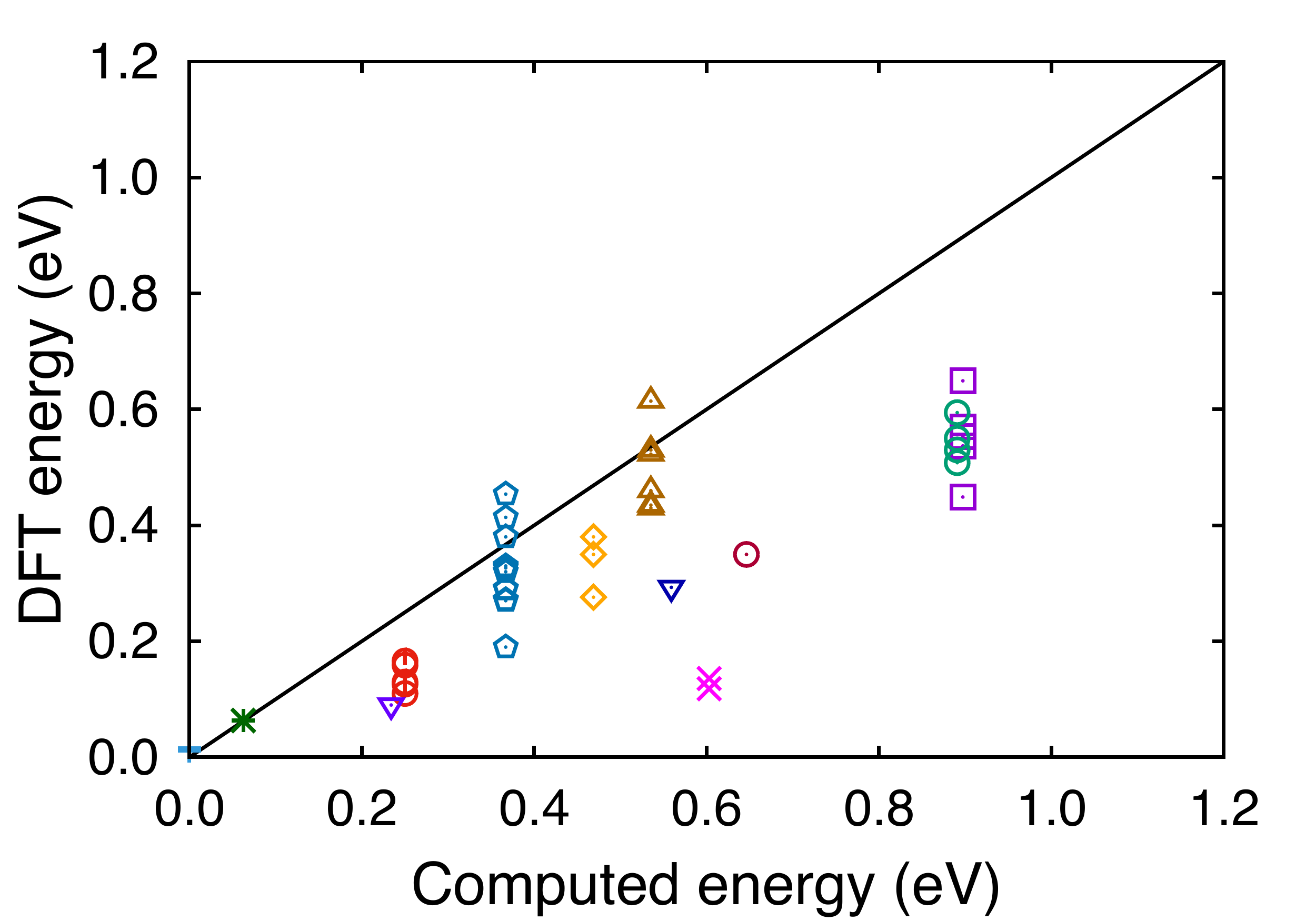}\enskip{}\enskip{}\enskip{}\textbf{(d)}\enskip{}\includegraphics[angle=-90,width=0.435\textwidth]{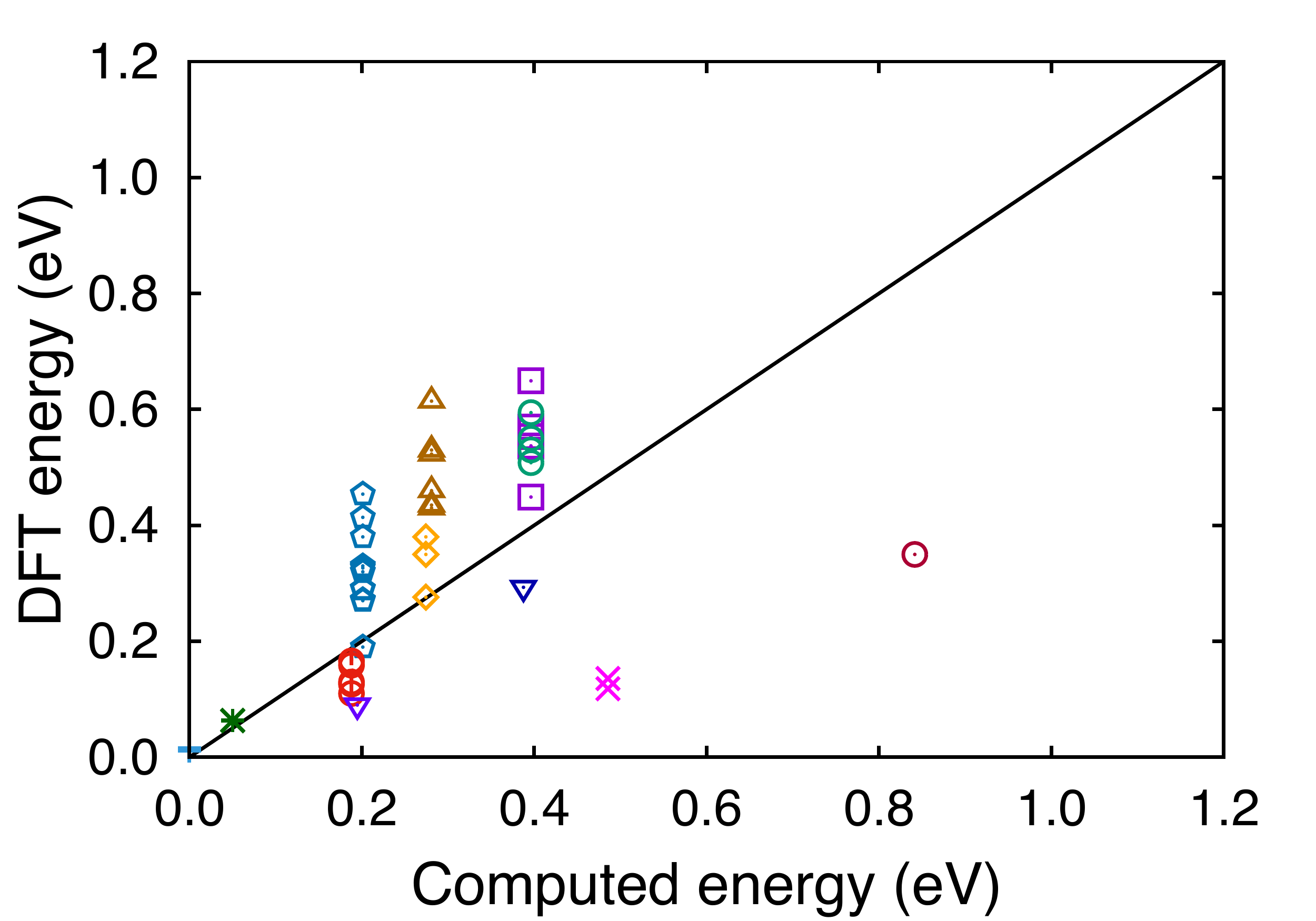}

\caption{DFT energies of crystal structures of Si versus the energies predicted
by interatomic potentials: (a) present potential, (b) MOD potential,\citep{Kumagai:2007ly}
(c) MEAM potential,\citep{Ryu:2009dn}, and (d) SW potential.\citep{Stillinger85}
The energies are counted per atom relative to the diamond cubic structure.
The line of perfect correlation is indicated.\label{fig:DFT-energies}}
\end{figure}

\begin{figure}
\textbf{(a)}\enskip{}\includegraphics[angle=-90,width=0.435\textwidth]{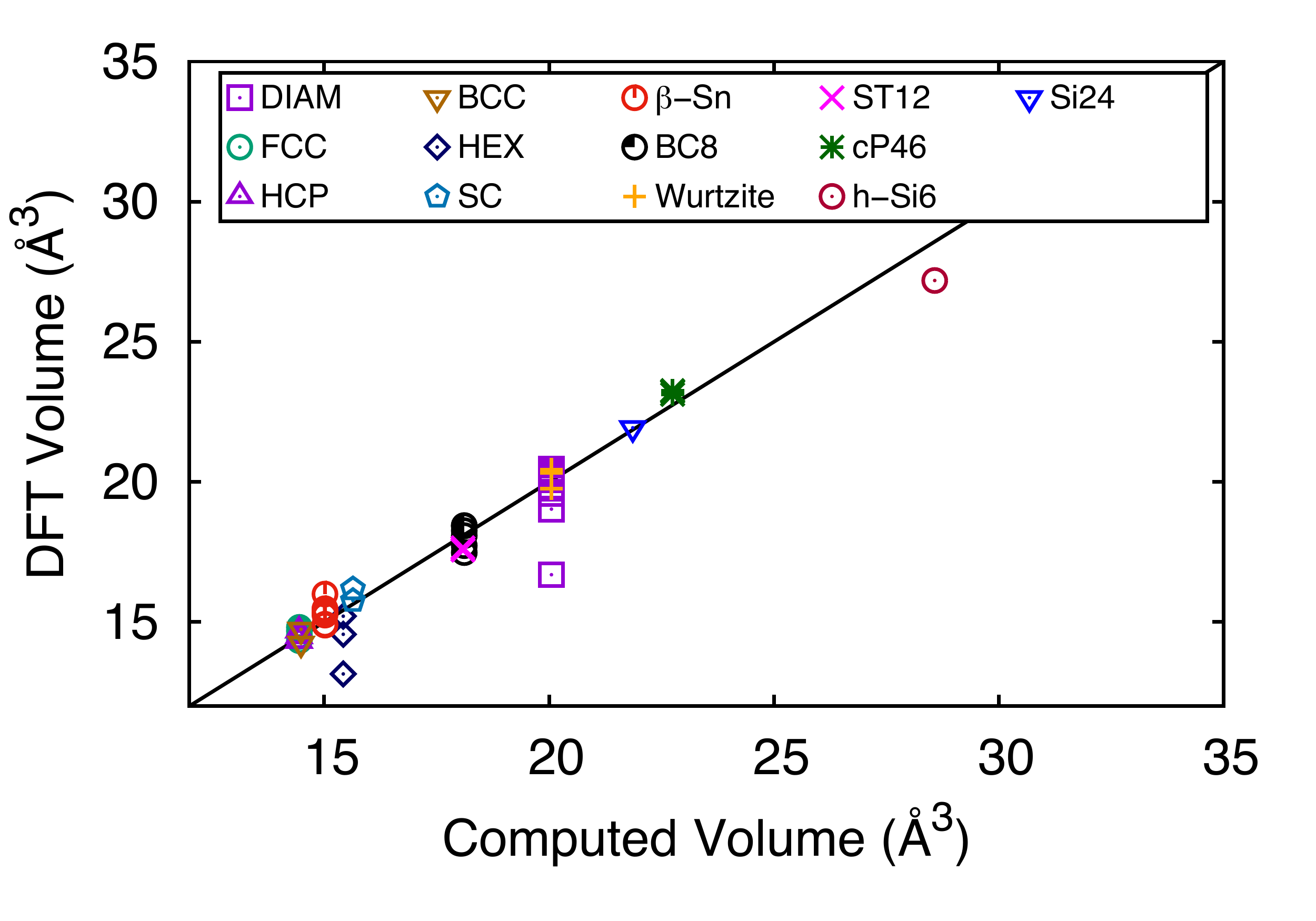}\enskip{}\enskip{}\enskip{}\textbf{(b)}\enskip{}\includegraphics[angle=-90,width=0.435\textwidth]{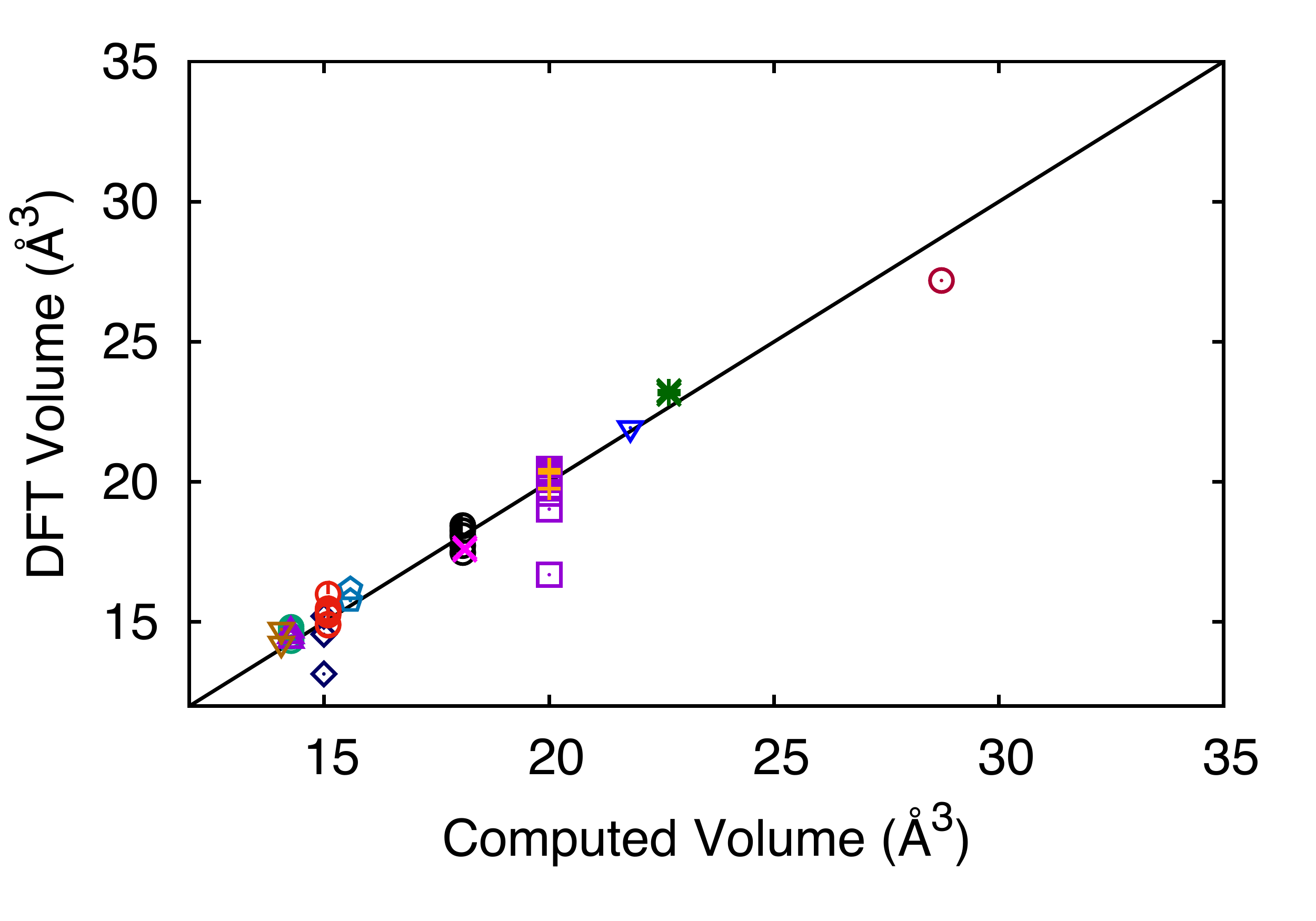}\bigskip{}
 \bigskip{}

\textbf{(c)}\enskip{}\includegraphics[angle=-90,width=0.435\textwidth]{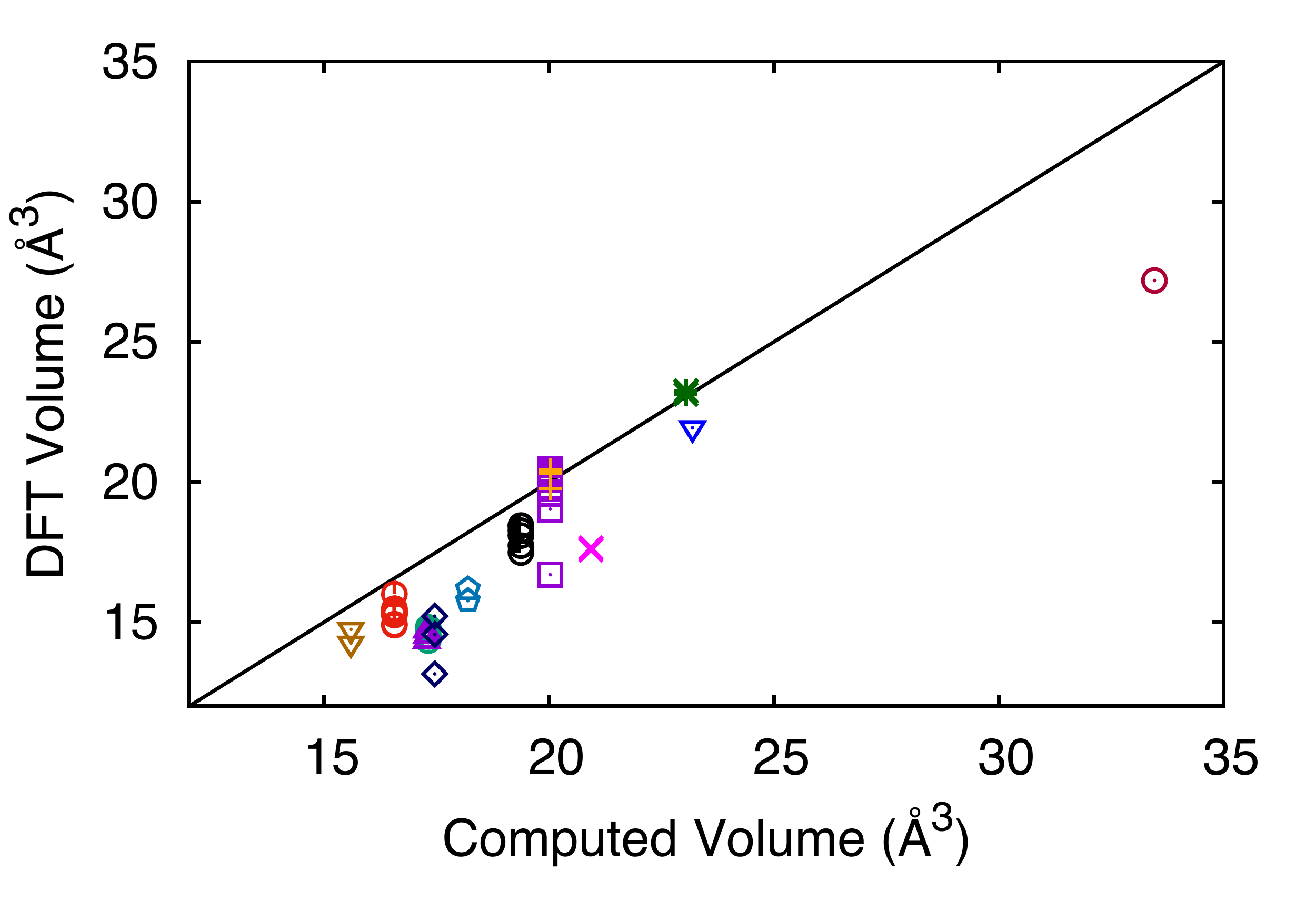}\enskip{}\enskip{}\enskip{}\textbf{(d)}\enskip{}\includegraphics[angle=-90,width=0.435\textwidth]{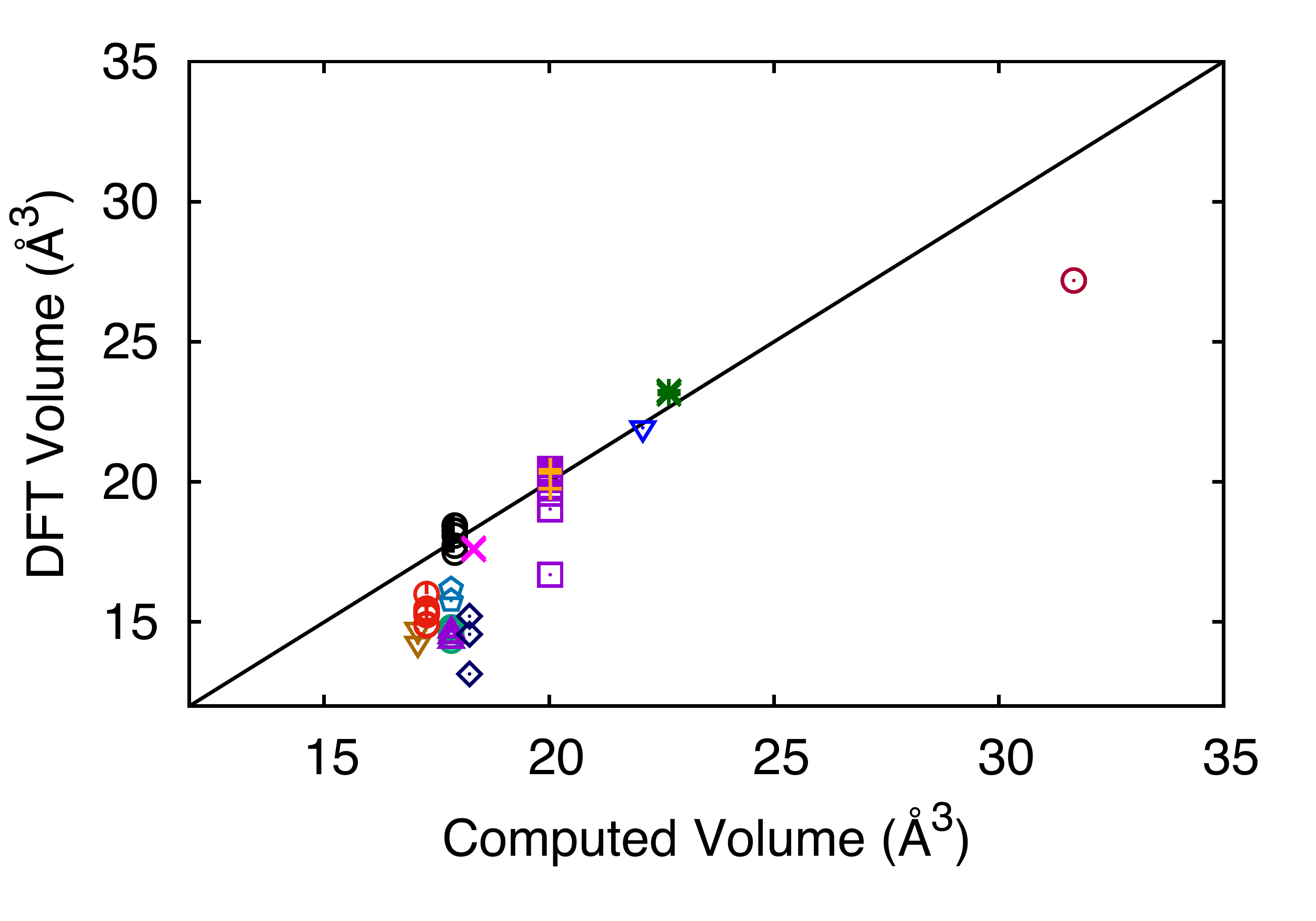}

\caption{DFT atomic volumes of crystal structures of Si versus the atomic volumes
predicted by interatomic potentials: (a) present potential, (b) MOD
potential,\citep{Kumagai:2007ly} (c) MEAM potential,\citep{Ryu:2009dn},
and (d) SW potential.\citep{Stillinger85} The line of perfect correlation
is indicated.\label{fig:DFT-volumes}}
\end{figure}

In Tables \ref{tab:alternate_strucs_iso_vol} and \ref{tab:vol_strucs},
we compare the predictions of the four potentials with DFT calculations
available in the literature. Since the tables are overloaded with
numerical data, we found it instructive to recast this information
in a graphical format. In Figs.~\ref{fig:DFT-energies} and \ref{fig:DFT-volumes}
we plot the energies (volumes) predicted by each potential against
the respective DFT energies (volumes) computed by different authors.
The bisecting line is the line of perfect correlation. The first thing
to notice is the large scatter of the DFT data reported by different
sources, which makes a comparison with potentials somewhat ambiguous.
For each potential, the agreement was quantified by the root-mean-square
(RMS) deviation of the data points from the bisecting line. The RMS
deviations obtained are shown in the last row of Tables \ref{tab:alternate_strucs_iso_vol}
and \ref{tab:vol_strucs}. It should emphasized that these RMS deviations
reflect not only the differences between the potentials and the DFT
calculations but also the scatter of the DFT points themselves. Thus,
only comparison of relative values of the RMS deviations makes sense.
It should also be noted that the energy deviations are strongly dominated
by high-energy structures, such as the close-packed FCC and HCP phases.
With this in mind, it is evident that the present potential is the
least successful in reproducing the structural energies, whereas the
MOD potential is the most successful. For the atomic volumes, however,
the present potential and MOD are equally accurate, while the SW and
MEAM potentials show significantly larger deviations. 

It is interesting to note that the present potential gives the most
accurate predictions for the energy and volume of the novel h-Si$_{6}$
and Si$_{24}$ structures that were not included in the fitting database.
The MOD potential comes close second, whereas the MEAM and SW potentials
are significantly less accurate. The energy-volume plots for several
selected structures can be found in the Supplemental Material to this
article.\citep{Supplementary-Si-Tersoff}

\section{Silicon clusters}

Structure and properties of small Si clusters offer a stringent test
of interatomic potentials. Potentials are usually optimized for bulk
properties, whereas the clusters display very different and much more
open environments in which the coordination number and the type of
bounding may change very significantly from one structure to another.
Si potentials are traditionally considered to be incapable of reproducing
cluster properties, unless such properties are specifically included
in the fitting process as in the case of the Boulding and Andersen
potential.\citep{Boulding:1990aa} It was thus interesting to compare
the predictions of the four potentials with first-principles calculations.

\begin{figure}
\noindent \begin{centering}
\includegraphics[width=0.55\textwidth]{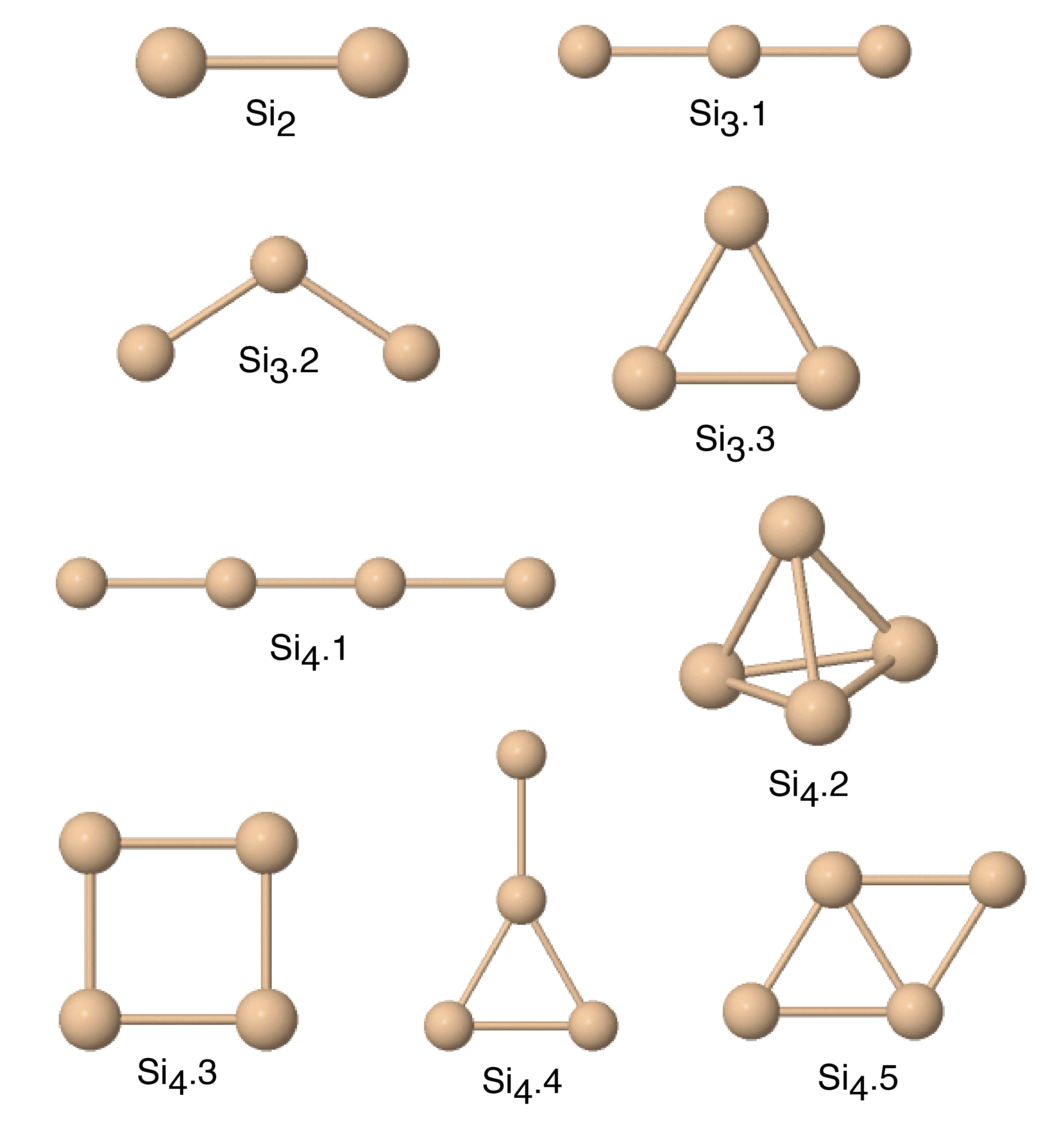} 
\par\end{centering}
\noindent \begin{centering}
\includegraphics[width=0.55\textwidth]{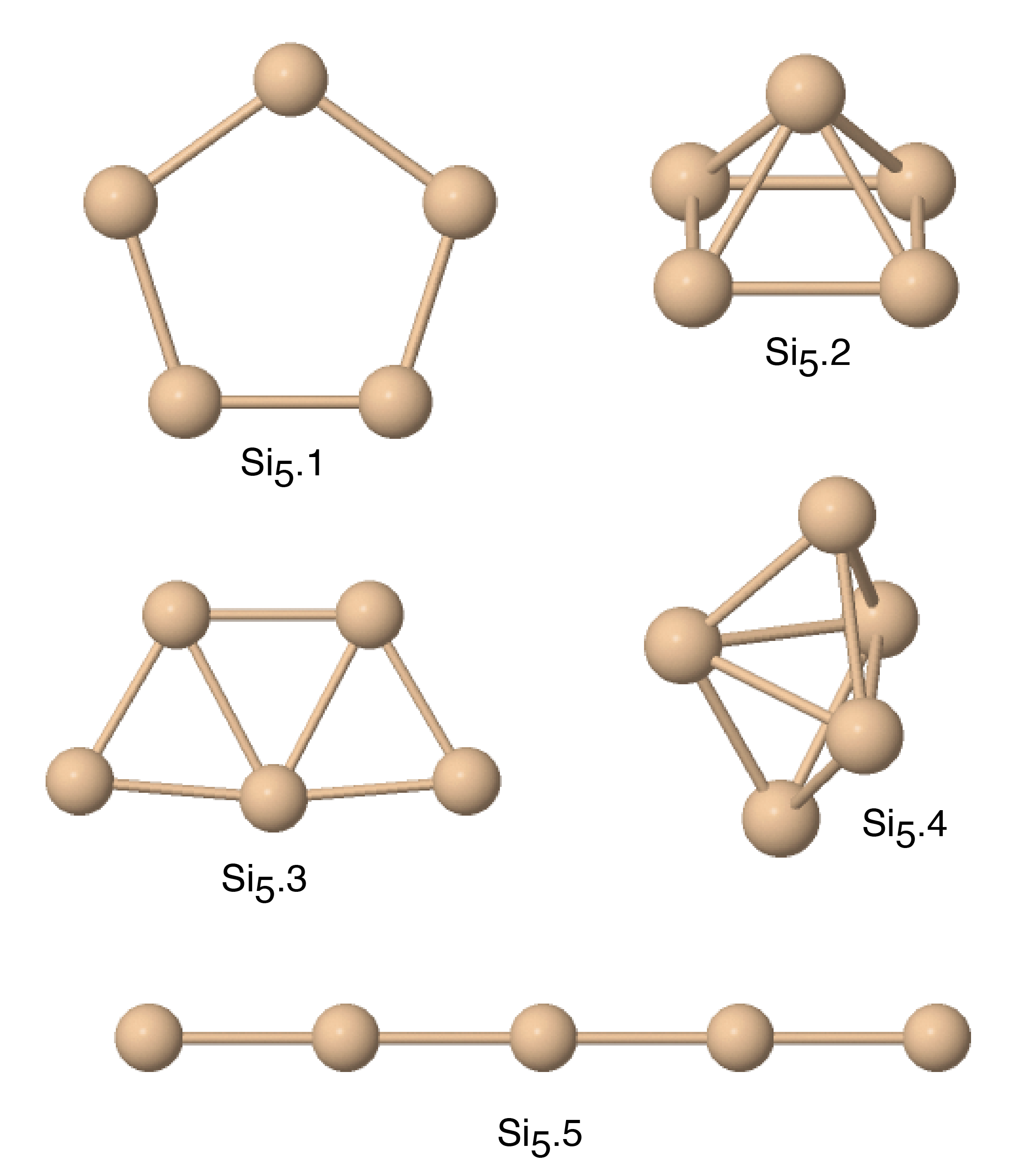} 
\par\end{centering}
\caption{Structures of dimer, trimer, tetramer and pentamer Si clusters tested
in this work. The labels indicate the cluster notations.\label{fig:Si_clusters_1}}
\end{figure}

\begin{figure}
\noindent \begin{centering}
\includegraphics[width=0.55\textwidth]{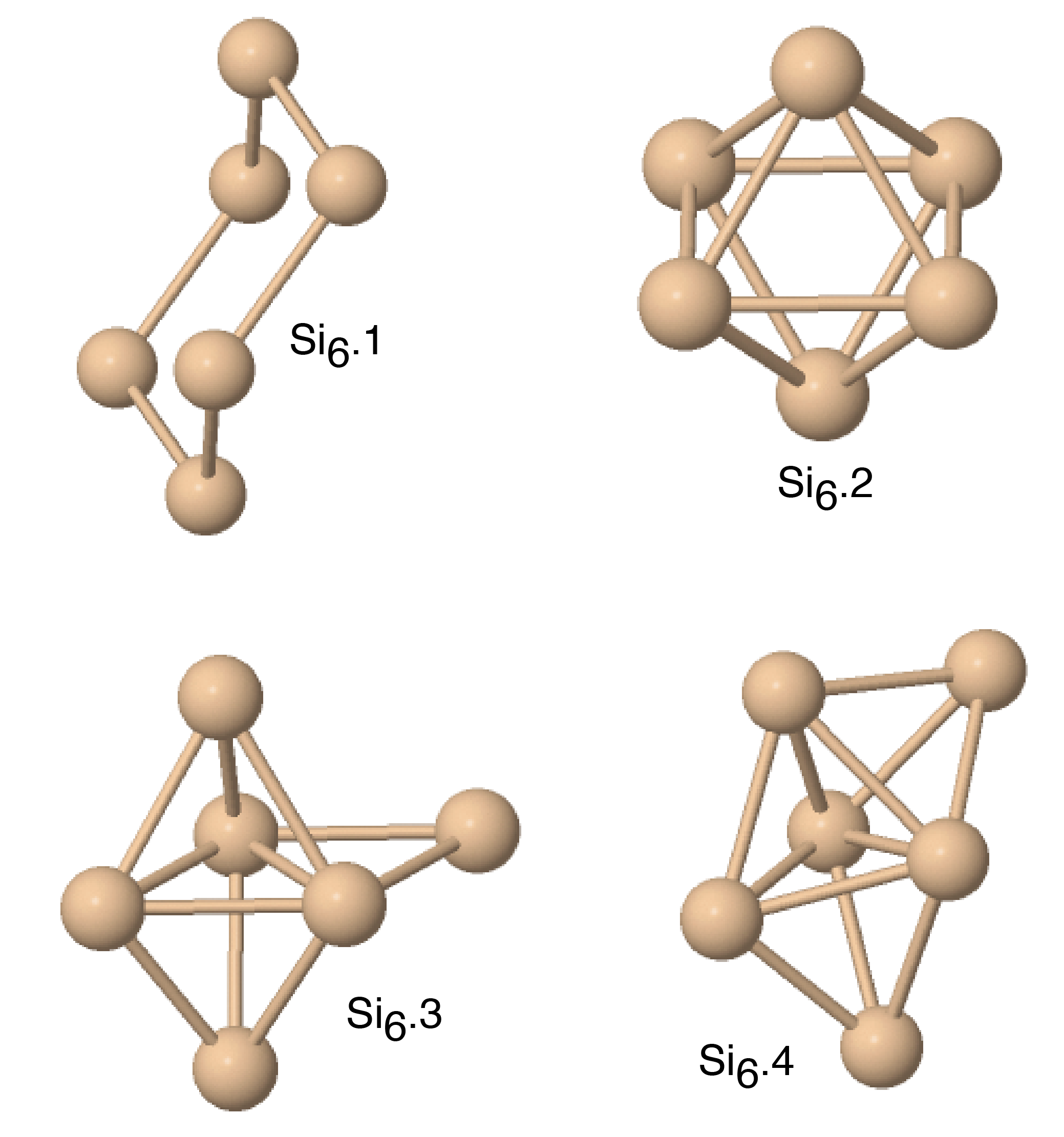} 
\par\end{centering}
\noindent \begin{centering}
\includegraphics[width=0.55\textwidth]{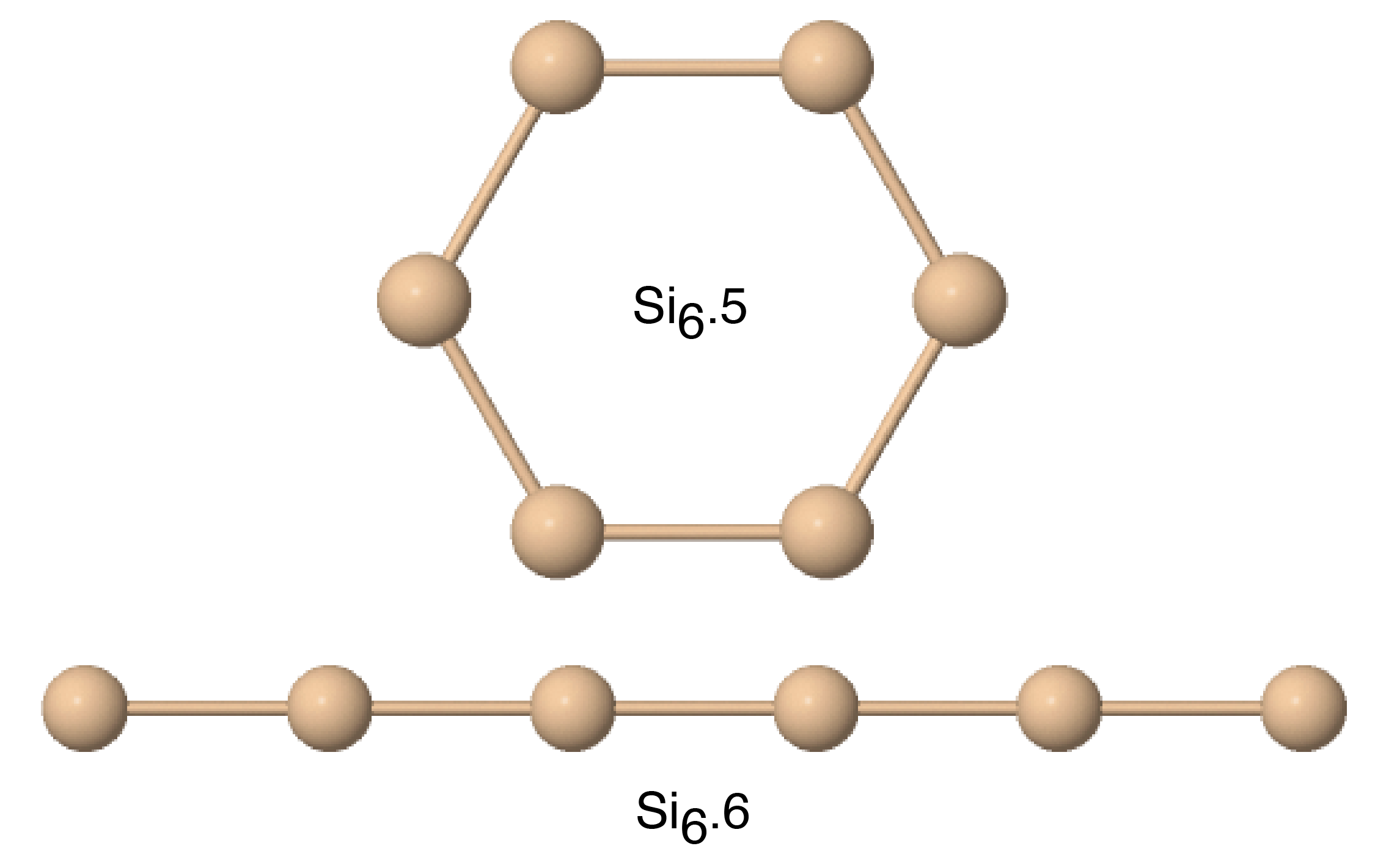} 
\par\end{centering}
\noindent \begin{centering}
\includegraphics[width=0.55\textwidth]{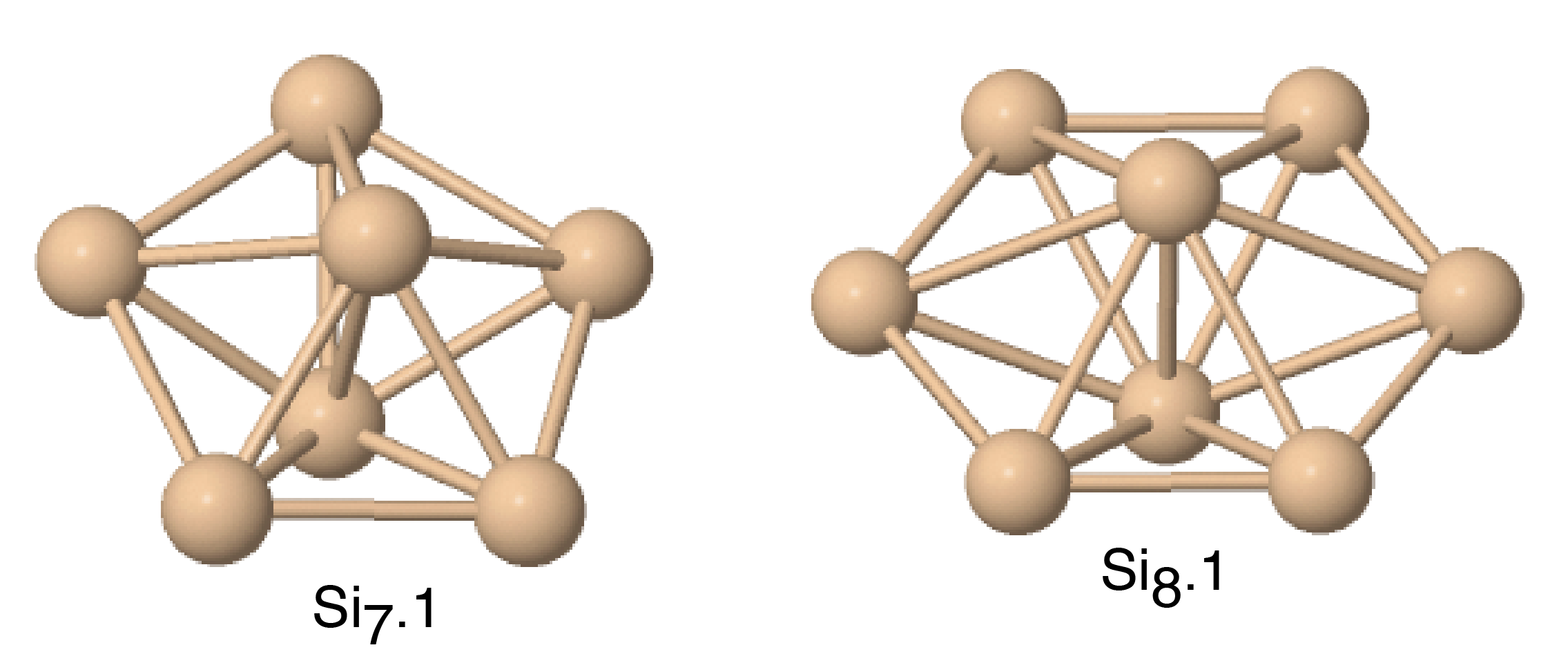} 
\par\end{centering}
\caption{Structures of hexamer, heptamer and octamer Si clusters tested in
this work. The labels indicate the cluster notations.\label{fig:Si_clusters_2}}
\end{figure}

Figs.~\ref{fig:Si_clusters_1} and \ref{fig:Si_clusters_2} show
the structures of the Si$_{n}$ ($n=2-8$) clusters tested in this
work. Several different structures are included for each cluster size
$n$ whenever first-principles data is available. Such structures
are labeled by index $m$ in the Si$_{n}.m$ format in the order of
increasing cohesive (binding) energy according to the DFT calculations.\citep{Fournier:1992tw}
Thus, the structure labeled Si$_{n}.1$ represents the DFT-predicted
ground state for each cluster size $n$ (except for the dimer Si$_{2}$
that has a single structure). In addition to the DFT calculations,\citep{Fournier:1992tw}
we included the results of quantum-chemical (QC) calculations on the
Hartree-Fock level.\citep{Raghavachari:1986yg} Such calculations
are more accurate but the energy scale is not fully compatible with
that of the DFT calculations. To enable comparison, we followed the
proposal\citep{Raghavachari:1985rm,Raghavachari:1988sh} that the
QC energies be scaled by a factor of 1.2 to ensure agreement with
experiment for the dimer energy.

\begin{figure}
\textbf{(a)}\enskip{}\includegraphics[angle=-90,width=0.435\textwidth]{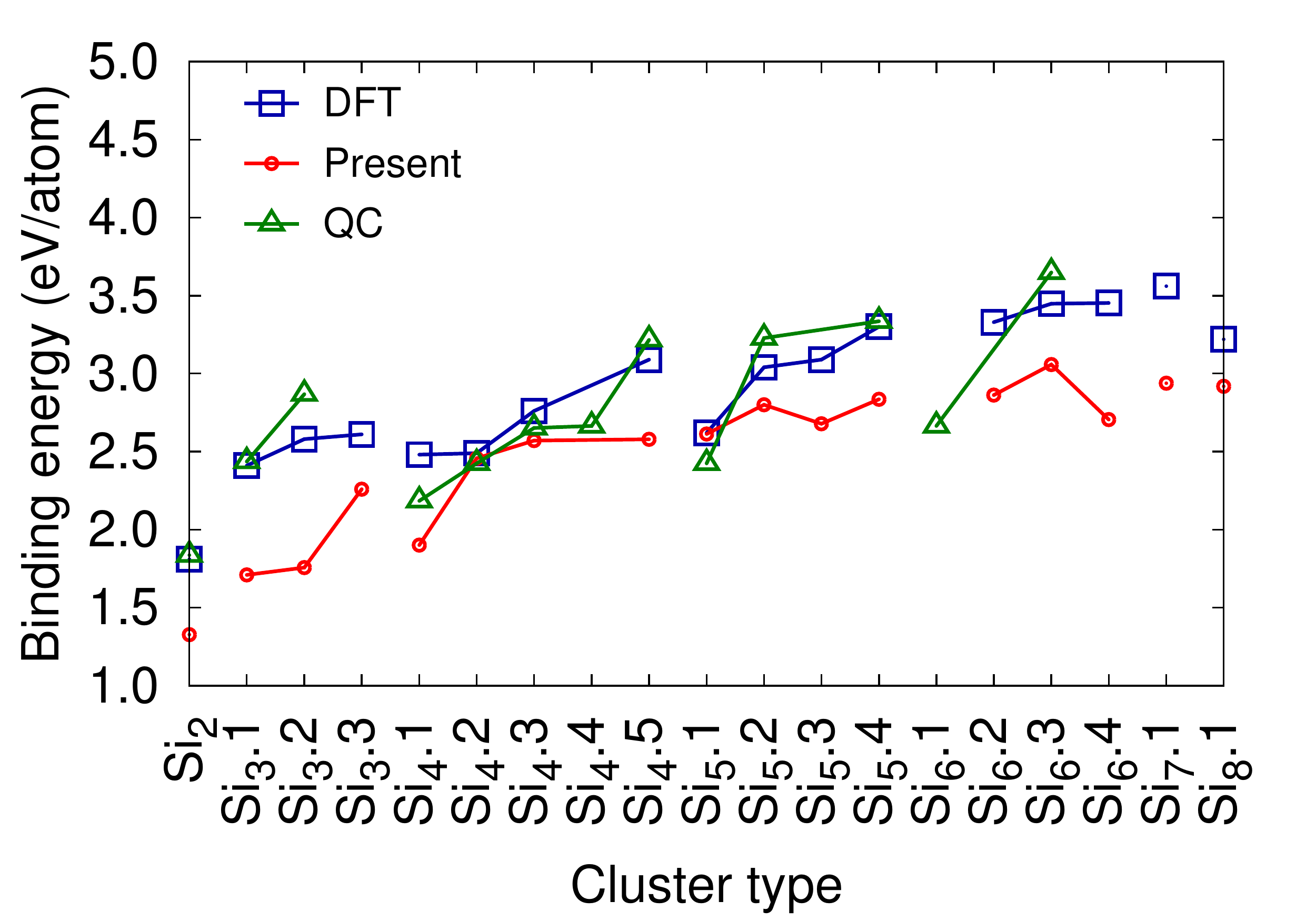}\enskip{}\enskip{}\enskip{}\textbf{(b)}\enskip{}\includegraphics[angle=-90,width=0.435\textwidth]{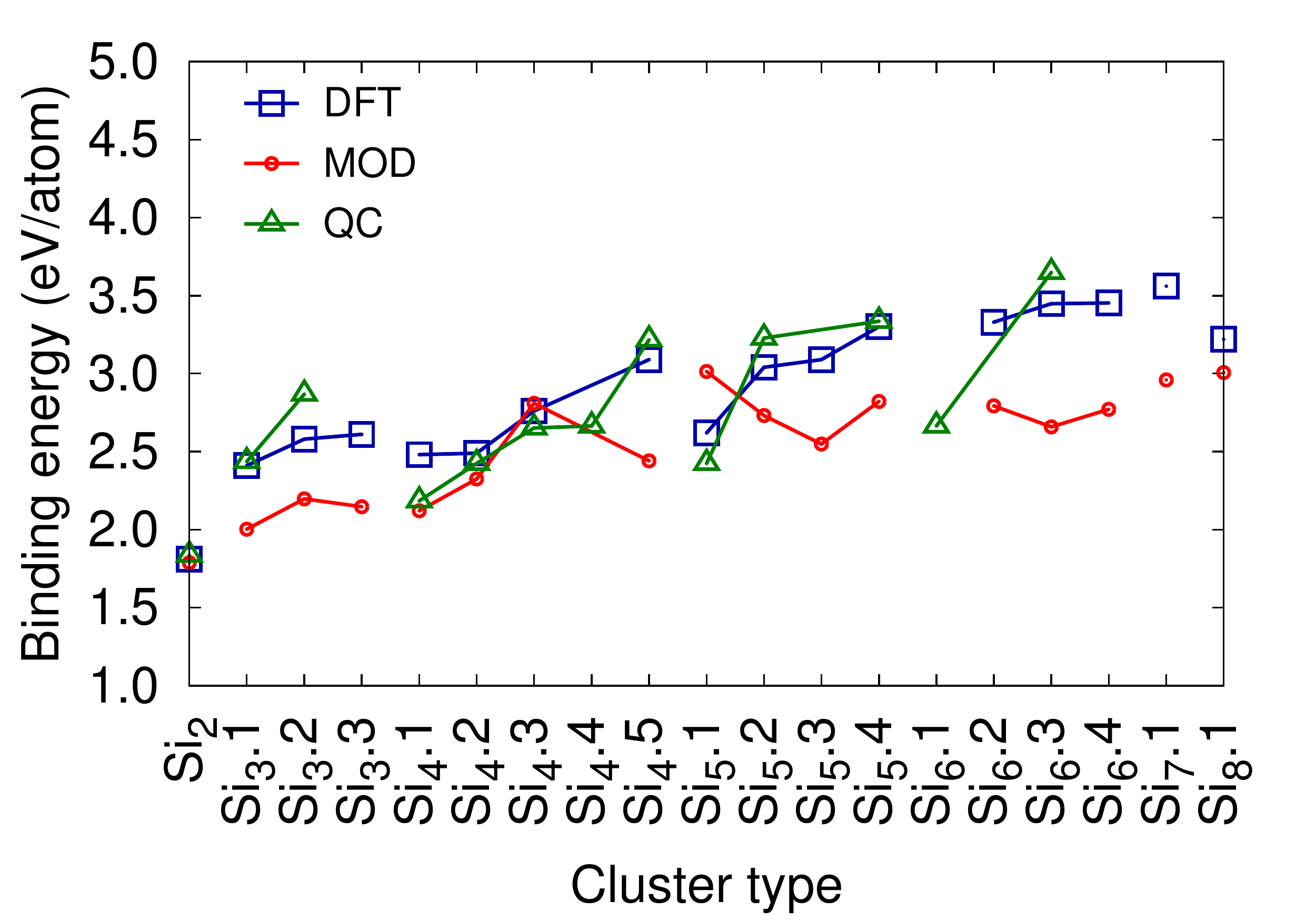}\bigskip{}
 \bigskip{}

\textbf{(c)}\enskip{}\includegraphics[angle=-90,width=0.435\textwidth]{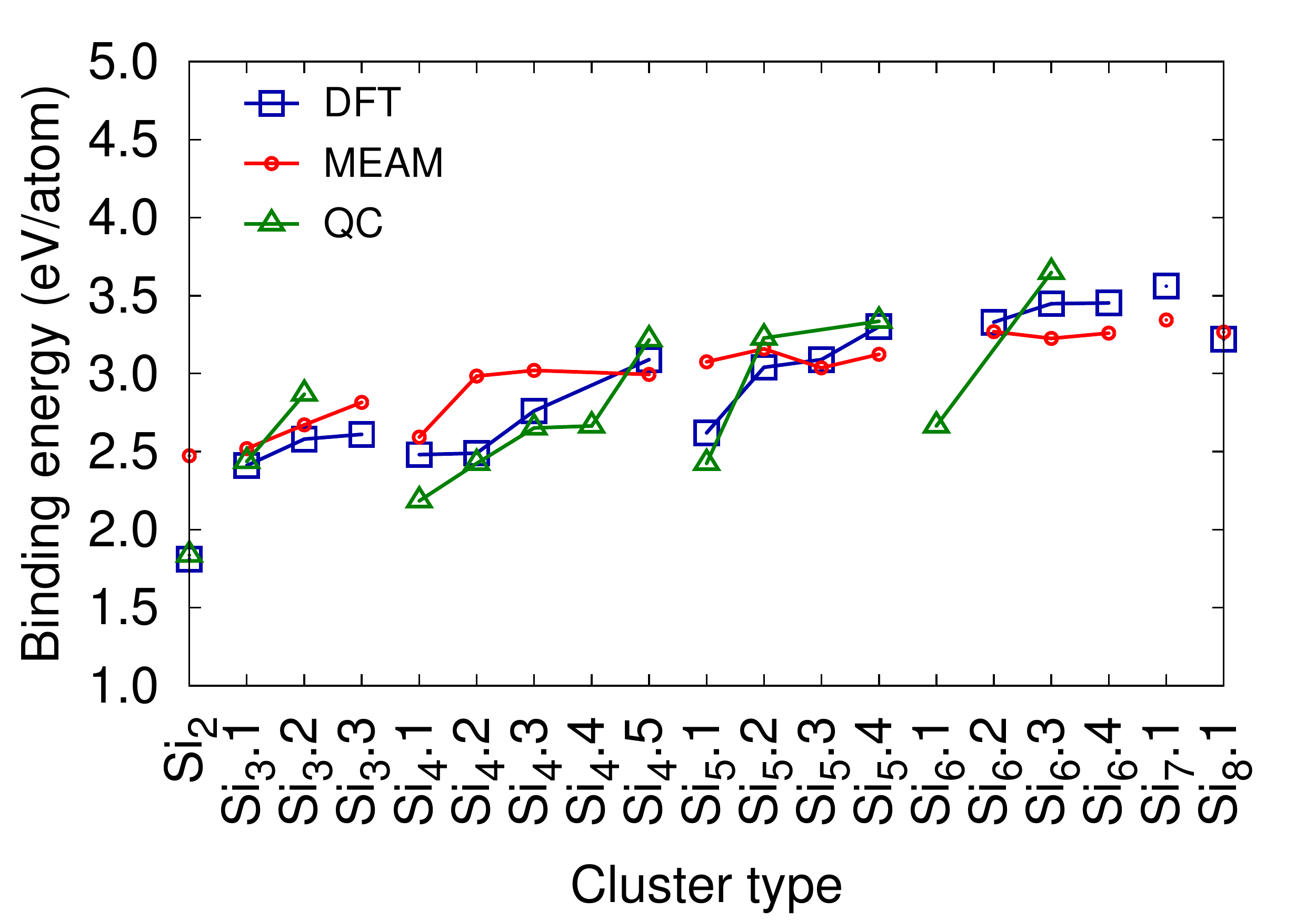}\enskip{}\enskip{}\enskip{}\textbf{(d)}\enskip{}\includegraphics[angle=-90,width=0.435\textwidth]{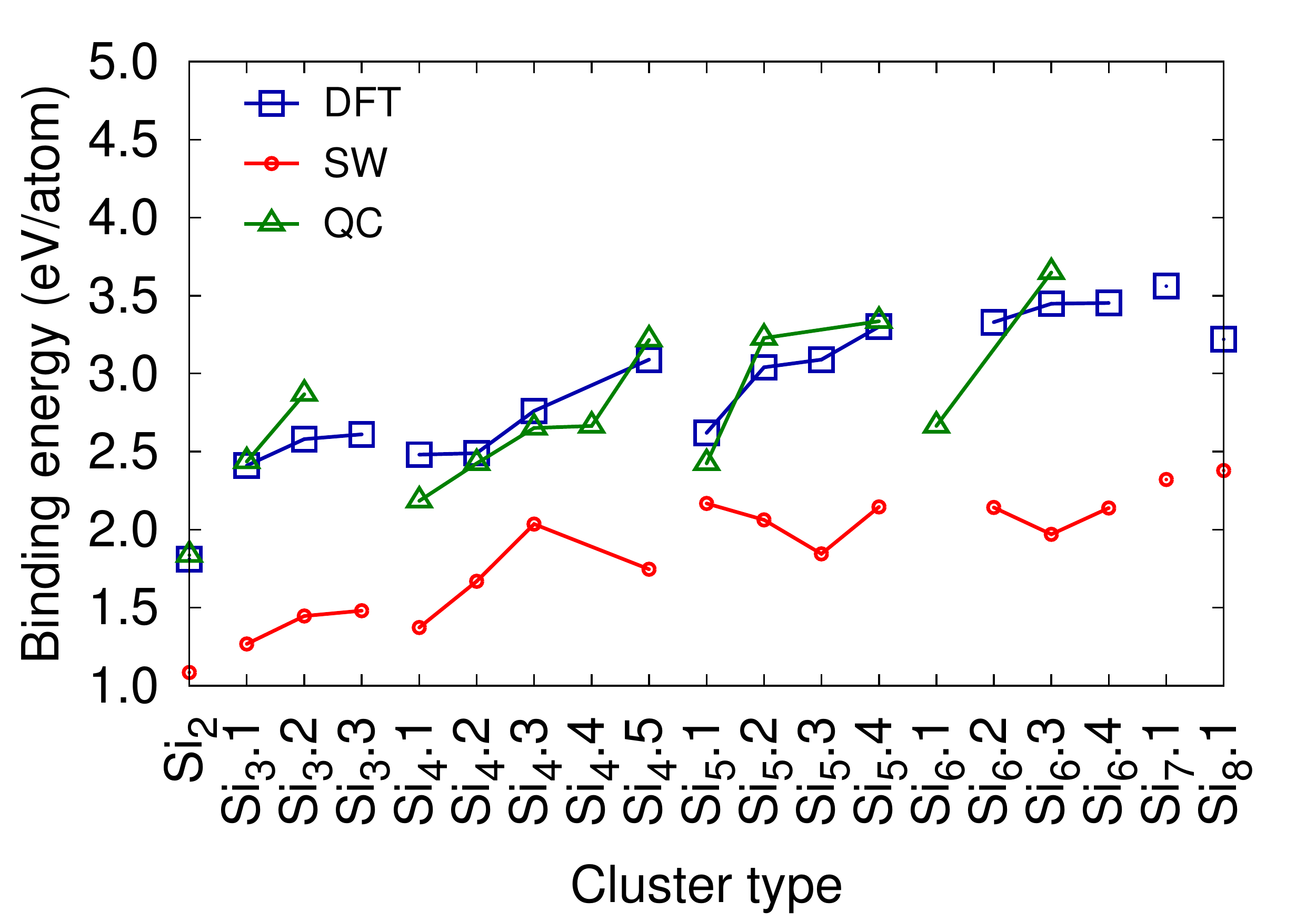}

\caption{Binding energies of Si clusters predicted by interatomic potentials:
(a) present potential, (b) MOD potential,\citep{Kumagai:2007ly} (c)
MEAM potential,\citep{Ryu:2009dn}, and (d) SW potential.\citep{Stillinger85}
First-principles energies computed by DFT and QC methods are shown
for comparison. The clusters are divided into groups corresponding
to the same number of atoms and are ordered with increasing binding
energy. The cluster structures are shown in Figs.~\ref{fig:Si_clusters_1}
and \ref{fig:Si_clusters_2}. \label{fig:cluster_energies}}
\end{figure}

Table \ref{tab:cluster_coh_eng} summarizes the predictions of the
four potentials in comparison with DFT calculations\citep{Fournier:1992tw}
and unscaled QC energies.\citep{Raghavachari:1986yg} In addition
to the clusters, we included an infinitely long linear chain for the
sake of comparison. To aid visual comparison, Fig.~\ref{fig:cluster_energies}
shows the cluster energies grouped by the cluster size (same-size
clusters are connected by straight lines). The QC energies are plotted
in the scaled format. Note that the scaling does indeed bring the
QC and DFT energies to general agreement with each other. Despite
the significant scatter of the individual energies on the level of
0.2-0.4 eV/atom, both calculation methods predict the same ground
state for trimers, tetramers and pentamers. None of the potentials
predicts the correct ordering for all DFT/QC energies. The present
potential and MOD show about the same level of accuracy, but the present
potential makes less mistakes in the ordering. Both potentials tend
to slightly under-bind the clusters. The MEAM potential is the most
successful in reproducing the cluster energies, except for the dimer
energy for which it is least accurate. There are mistakes in the ordering,
but overall the deviations from the first-principles calculations
are about the same as the difference between the two first-principles
methods. The SW potential performs poorly: for some of the clusters,
the binding energy is underestimated by more than 1 eV/atom. For the
infinite atomic chain, the present potential and MOD are in closest
agreement with the DFT/QC energies (Table \ref{tab:cluster_coh_eng}).

This comparison leads to the conclusion that, at least for the cluster
structures tested here, the present potential, MOD and MEAM are quite
capable of predicting the general trends of the cluster energies with
a reasonable accuracy without fitting.

\section{2D silicon structures}

\subsection{Single-layer silicenes}

Silicenes are 2D allotropes of Si that have recently attracted much
attention due to their interesting physical properties and potential
device applications.\citep{Kara:2012wu,Roome:2014aa,Grazianetti:2016aa,Kaloni:2016rc,Lew-Yan-Voon:2016ai}
By contrast to carbon, the sp$^{3}$ hybridized Si would seem to be
an unlikely candidate for a 2D material. Nevertheless, epitaxial honeycomb
Si layers have been found on metallic substrates such as (111)Ag.\citep{Lalmi:2010rw,Kara:2012wu,Vogt:2012xe,Lin:2012kl,Feng:2012bs,Gao:2012rp,Acun:2013kn,Arafune:2013cj,Sone:2014la,Roome:2014aa,Grazianetti:2016aa}.
Unlike in graphene, some of the 2D forms of Si can have a band gap
and could be incorporated in Si-based microelectronics. In particular,
electric field applied to the buckled honeycomb structure of silicene,
which is normally semi-metallic, can open a band gap whose magnitude
increases with the field. It was predicted,\citep{Ni:2012aa} and
recently demonstrated,\citep{Tao:2015aa} that single-layer silicene
can work as a field-effect transistor.\citep{Ni:2012aa} Experimentally,
it has not been possible so far to isolate free-standing silicenes.
They are presently considered hypothetic 2D materials and have only
been studied by DFT calculations. Such calculations predict that single-layer
silicene can possess remarkable electric, optical and magnetic properties,\citep{Ezawa:2012aa,Liu:2013aa,Liu:2011aa,Xu:2012aa}
in addition to ultra-low thermal conductivity.\citep{Peng:2016aa}

\begin{figure}
\begin{centering}
\textbf{(a)}\enskip{}\enskip{}\enskip{}\enskip{}\enskip{}\enskip{}\includegraphics[width=0.16\textwidth]{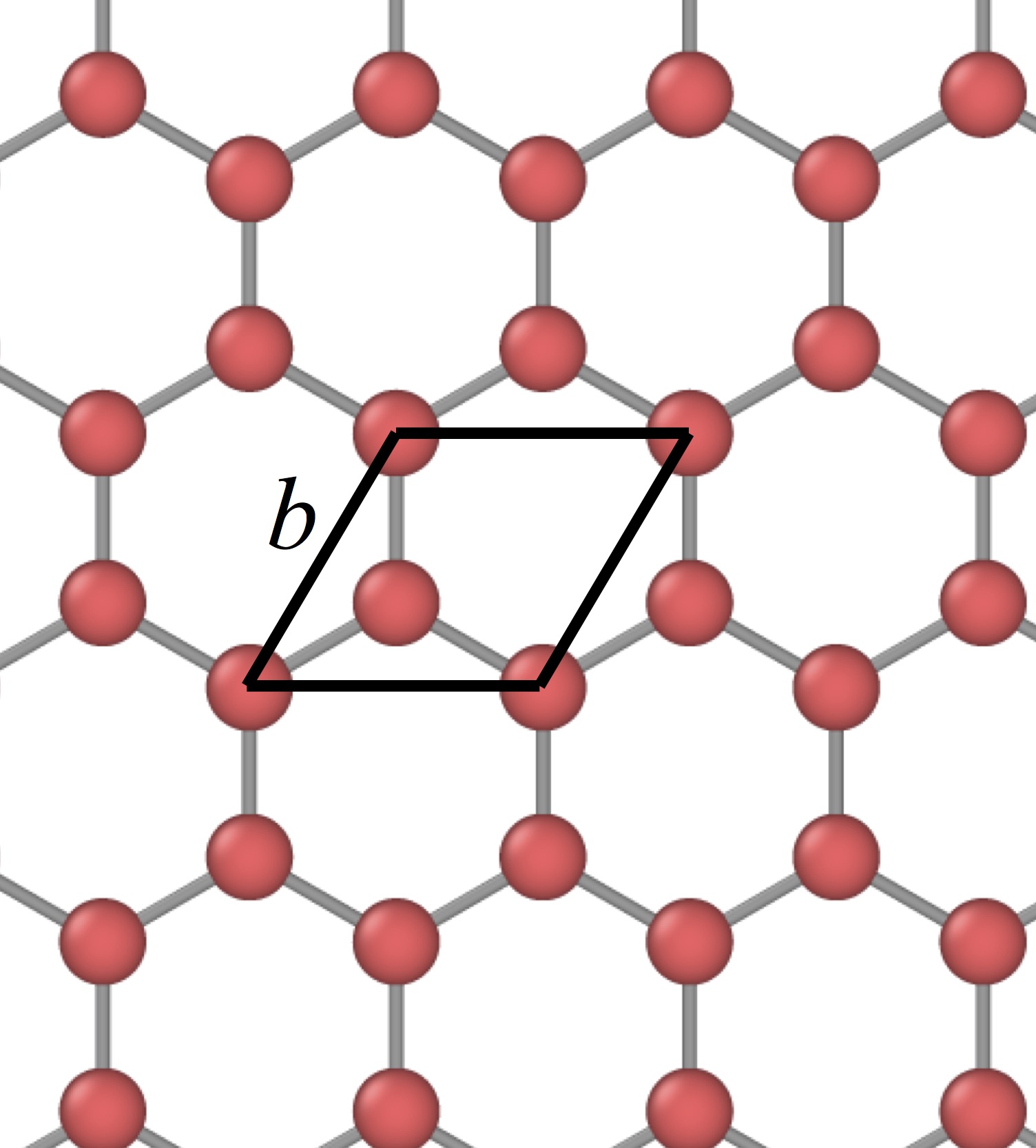}\enskip{}\enskip{}\enskip{}\enskip{} 
\par\end{centering}
\begin{centering}
\bigskip{}
 
\par\end{centering}
\begin{centering}
\textbf{(b)}\enskip{}\includegraphics[width=0.29\textwidth]{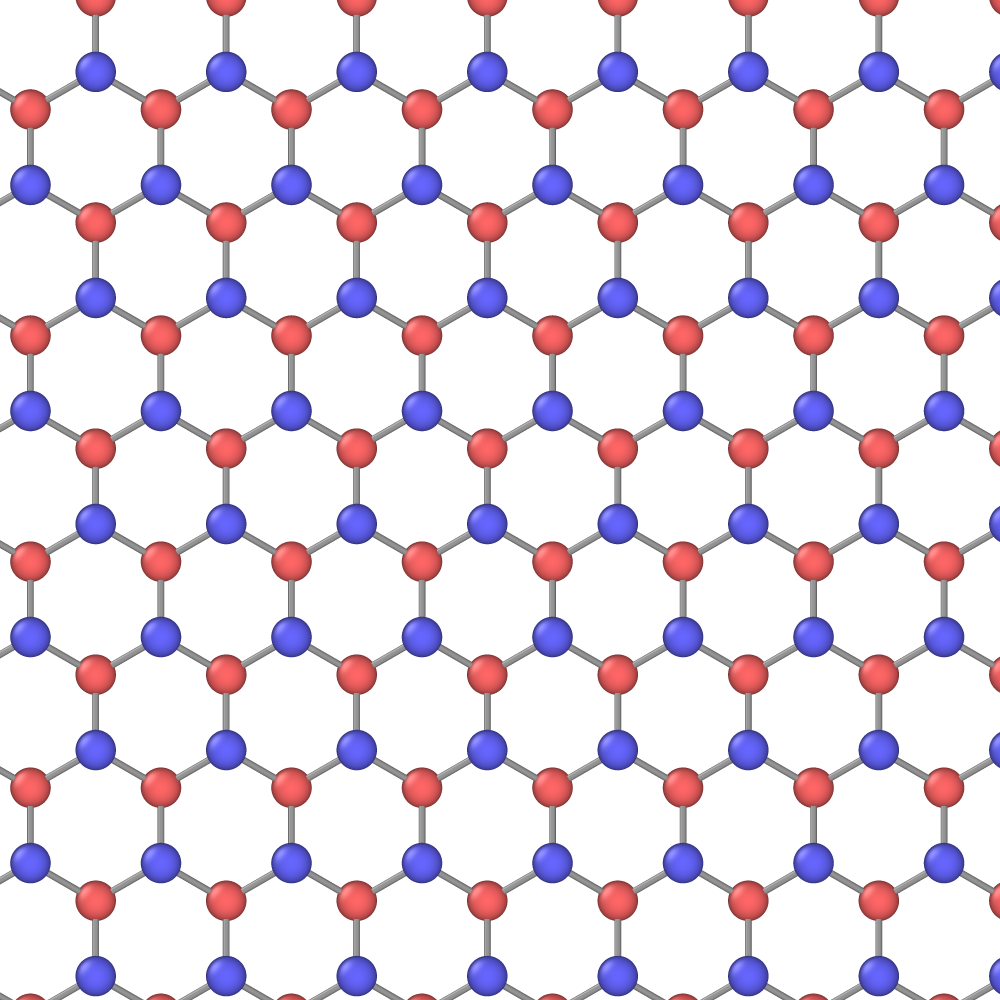} 
\par\end{centering}
\begin{centering}
\bigskip{}
 
\par\end{centering}
\begin{centering}
\textbf{(c)}\enskip{}\includegraphics[width=0.29\textwidth]{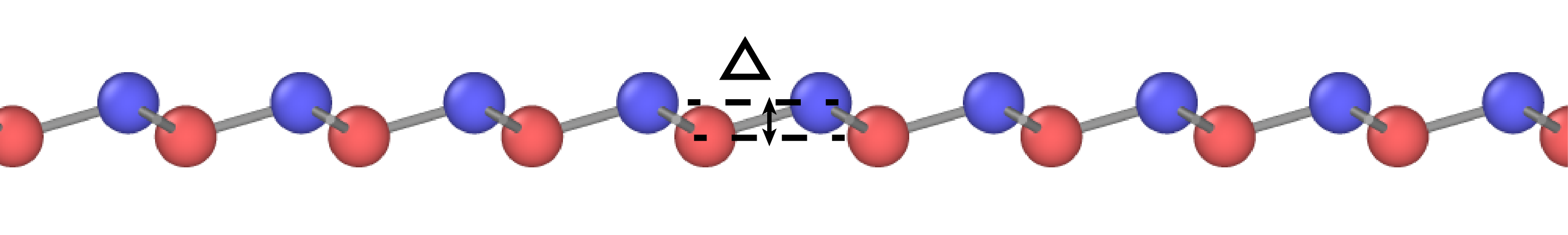} 
\par\end{centering}
\begin{centering}
\bigskip{}
 
\par\end{centering}
\noindent \begin{centering}
\textbf{(d)}\enskip{}\includegraphics[width=0.29\textwidth]{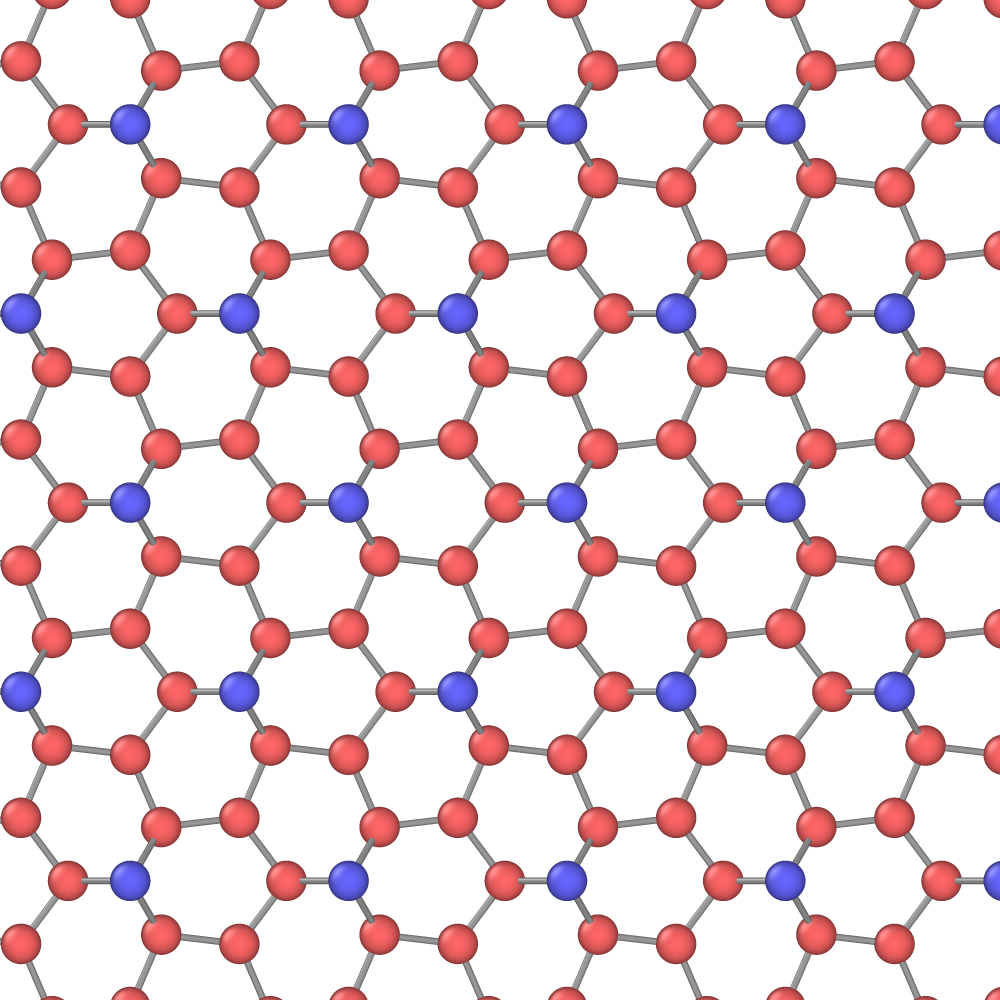} 
\par\end{centering}
\begin{centering}
\bigskip{}
 
\par\end{centering}
\noindent \begin{centering}
\textbf{(e)}\enskip{}\includegraphics[width=0.29\textwidth]{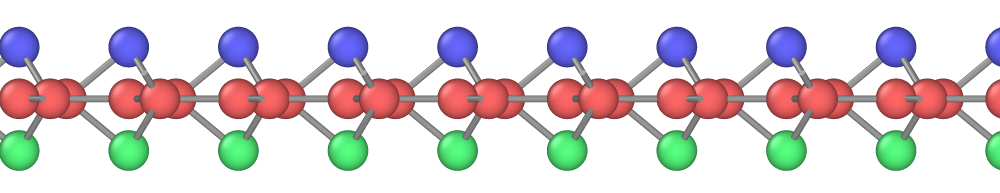} 
\par\end{centering}
\begin{centering}
\bigskip{}
 
\par\end{centering}
\noindent \begin{centering}
\textbf{(f)}\enskip{}\includegraphics[width=0.31\textwidth]{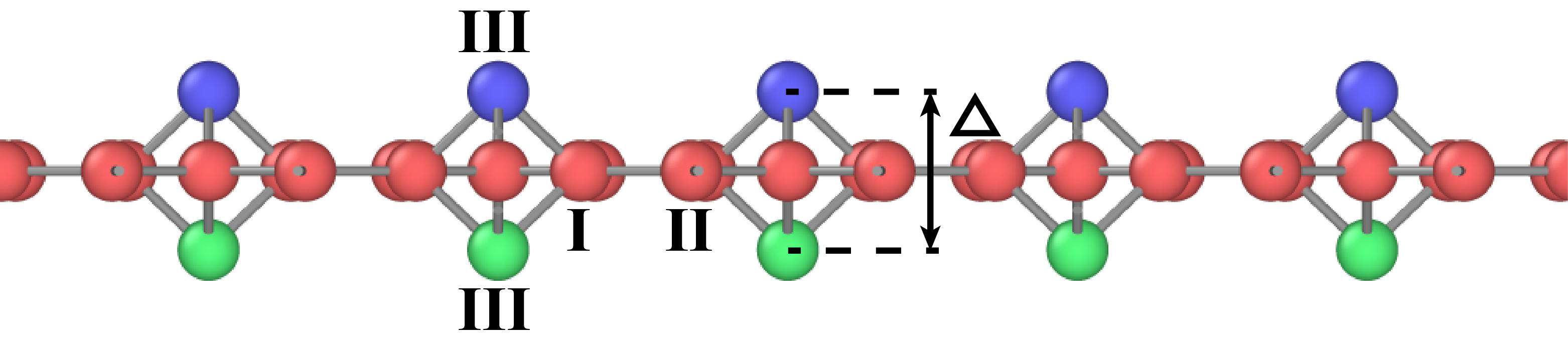} 
\par\end{centering}
\centering{}\caption{Silicene structures: (a) graphitic (planar) structure, (b,c) top and
edge views of the buckled structure, and (d,e,f) top and edge views
of the $\sqrt{3}\times\sqrt{3}$ dumbbell structure. \label{fig:Silicine-structures}}
\end{figure}

The planar (graphene-like) silicene {[}Fig.~\ref{fig:Silicine-structures}(a){]}
is mechanically unstable and spontaneously transforms to the more
stable buckled structure {[}Fig.~\ref{fig:Silicine-structures}(b,c){]}.\citep{Cahangirov:2009fu,Cahangirov:2014xw,Matusalem:2015aa}
The latter has a split width $\Delta$ of about 0.45-0.49 Å and a
first-neighbor distance $r_{1}$ slightly different from that in the
planar structure.\citep{Cahangirov:2009fu,Ni:2012aa,Sahin:2013aa,Kaltsas:2013rq,Matusalem:2015aa,Ge:2016mz}
Furthermore, adsorption of Si ad-atoms on the buckled silicene creates
a series of periodic dumbbell structures that are even more stable.\citep{Peng:2016aa,Kaltsas:2013rq,Cahangirov:2014xw}
An ad-atom pushes a nearby Si atom out of its regular position and
the two atoms form a dumbbell aligned perpendicular to the silicene
plane. The dumbbell atoms have a fourfold coordination (counting the
dumbbell bond itself) consistent with the sp$^{3}$ bonding. One of
the best studied dumbbell silicenes has the $\sqrt{3}\times\sqrt{3}$
structure shown in Fig.~\ref{fig:Silicine-structures}(d,e,f) (the
dumbbell atoms are shown in blue and green). The dumbbells distort
the hexagonal structural units and create three slightly different
nearest-neighbor distances: $r_{\textrm{I,II}}$, $r_{\textrm{II,III}}$
and $\varDelta_{\textrm{III,III}}$ {[}Fig.~\ref{fig:Silicine-structures}(f){]}.

The energies and geometric characteristics of the three silicene structures
predicted by the four potentials are listed in Table \ref{tab:prop_Silicene}.
The results of DFT calculations reported in the literature are included
for comparison. The agreement with the DFT data is reasonable, especially
considering that the 2D structures were not included in the fitting
datasets of the potentials. The present potential, MOD and MEAM demonstrate
about the same agreement with the DFT calculations. The SW potential
tends to be less accurate. For the planar structure, the MOD potential
is the most accurate, followed by the present potential, MEAM and
then SW. All four potentials correctly predict that the planar structure
is mechanically unstable and transforms to the buckled structure.
The present potential, MEAM and SW correctly predict that the $\sqrt{3}\times\sqrt{3}$
dumbbell structure has a lower energy than the buckled structure.
By contrast, the MOD potential predicts that the $\sqrt{3}\times\sqrt{3}$
dumbbell structure has a higher energy, which is contrary to the DFT
calculations. All four potentials overestimate the split width $\Delta$
in the buckled structure and the distance $\varDelta_{\textrm{III,III}}$
between the dumbbell atoms in the $\sqrt{3}\times\sqrt{3}$ structure,
the present potential being closest to the DFT data.

\begin{figure}
\textbf{(a)}\enskip{}\includegraphics[width=0.435\textwidth]{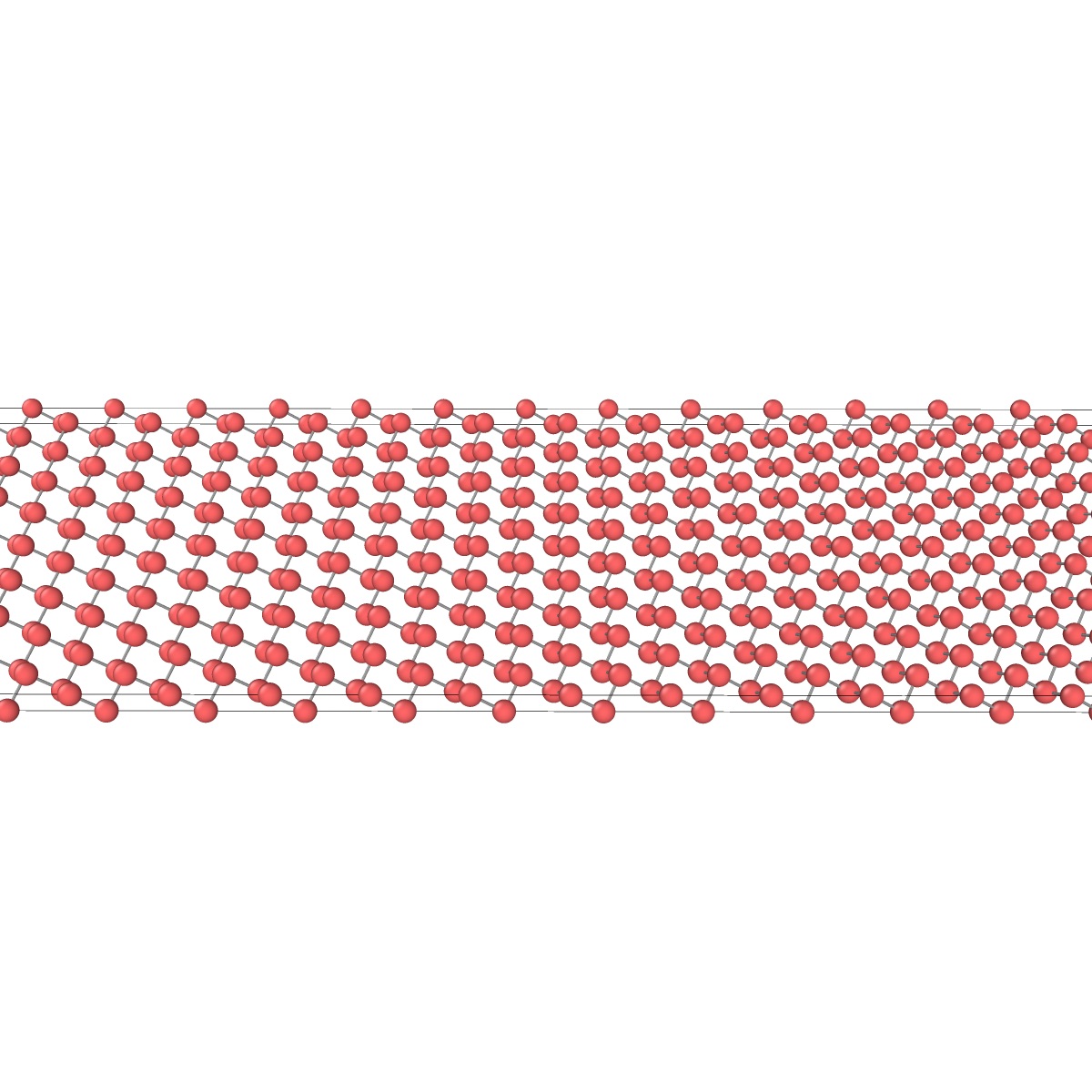}\enskip{}\enskip{}\enskip{}\textbf{(b)}\enskip{}\includegraphics[width=0.435\textwidth]{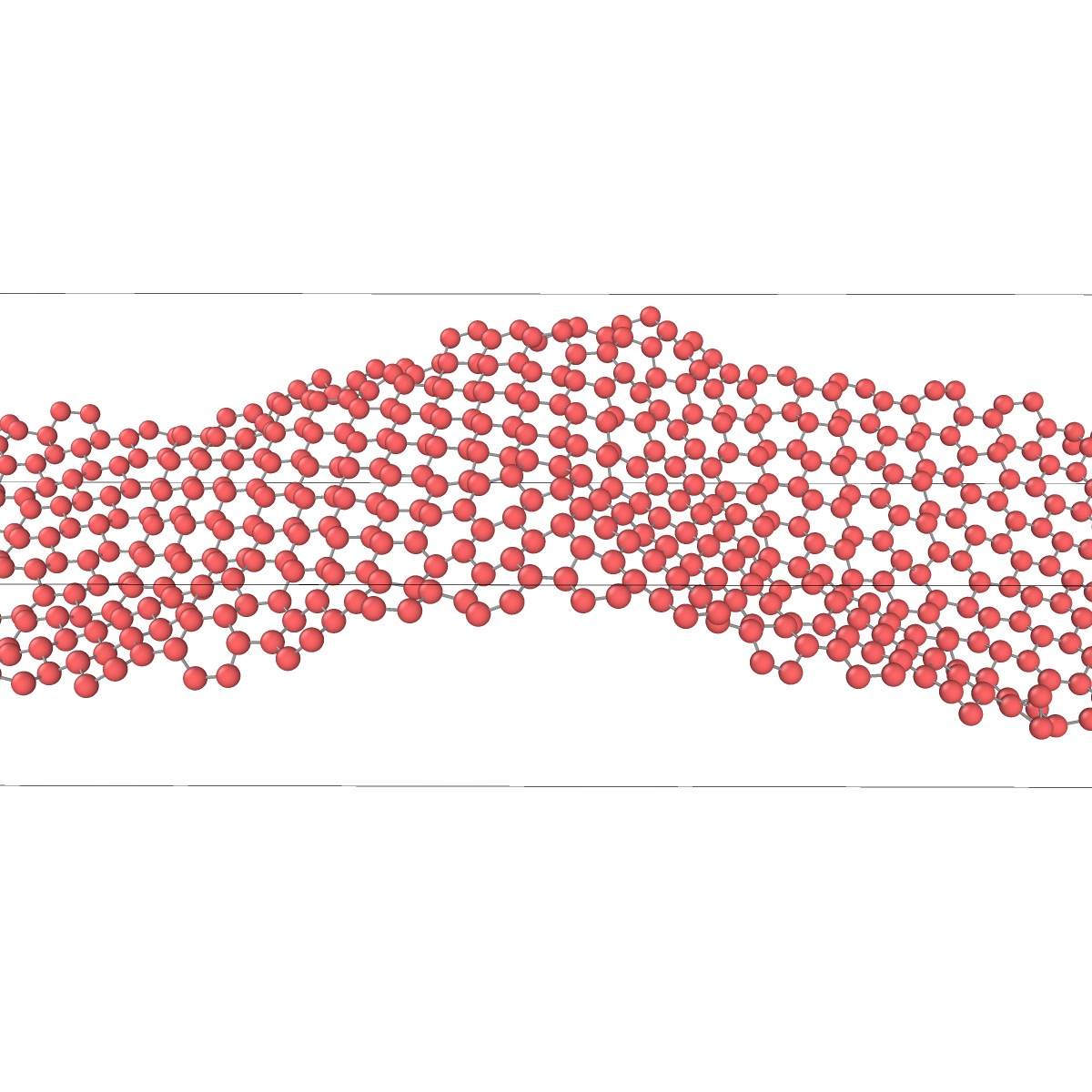} 

\textbf{(c)}\enskip{}\includegraphics[width=0.435\textwidth]{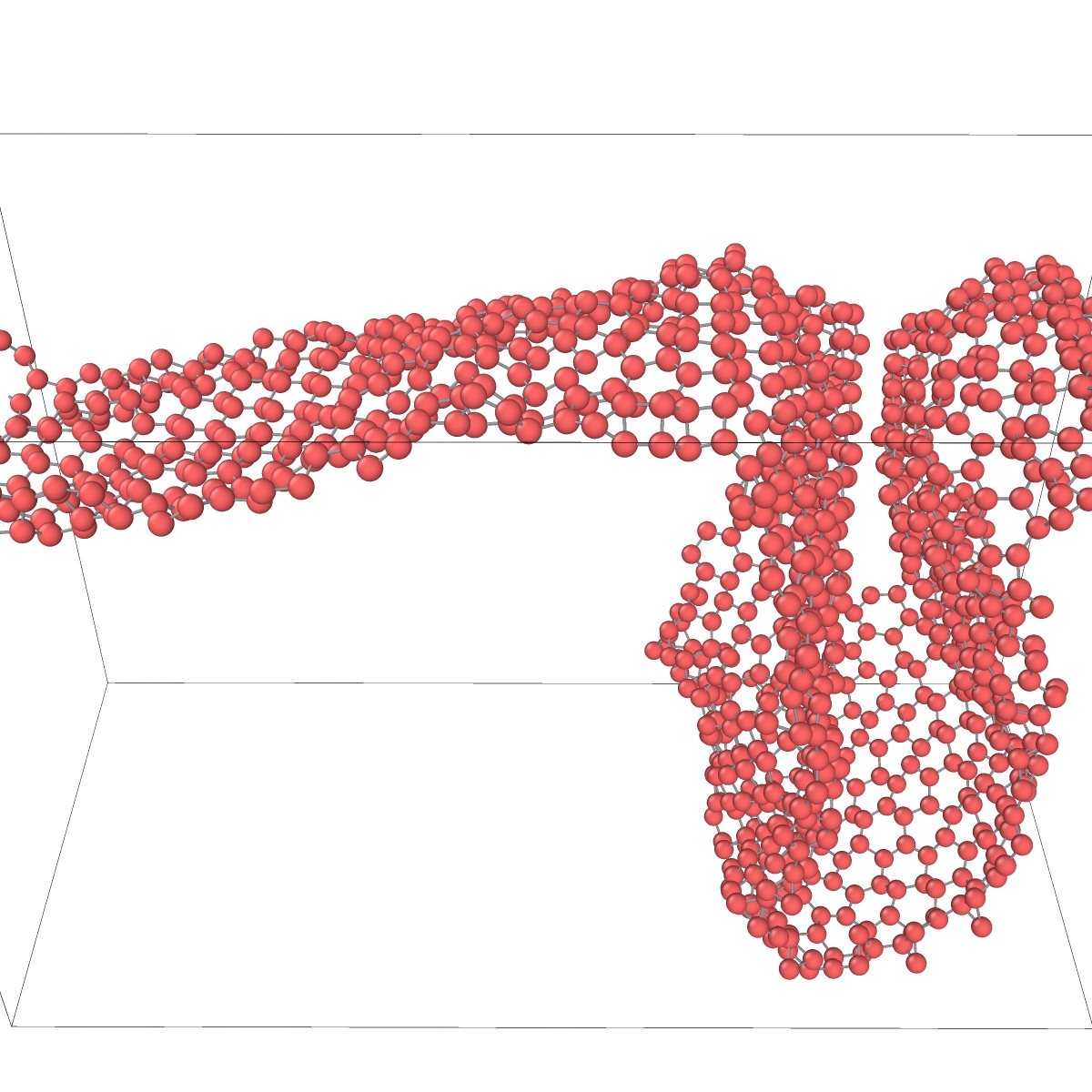}\enskip{}\enskip{}\enskip{}\textbf{(d)}\enskip{}\includegraphics[width=0.435\textwidth]{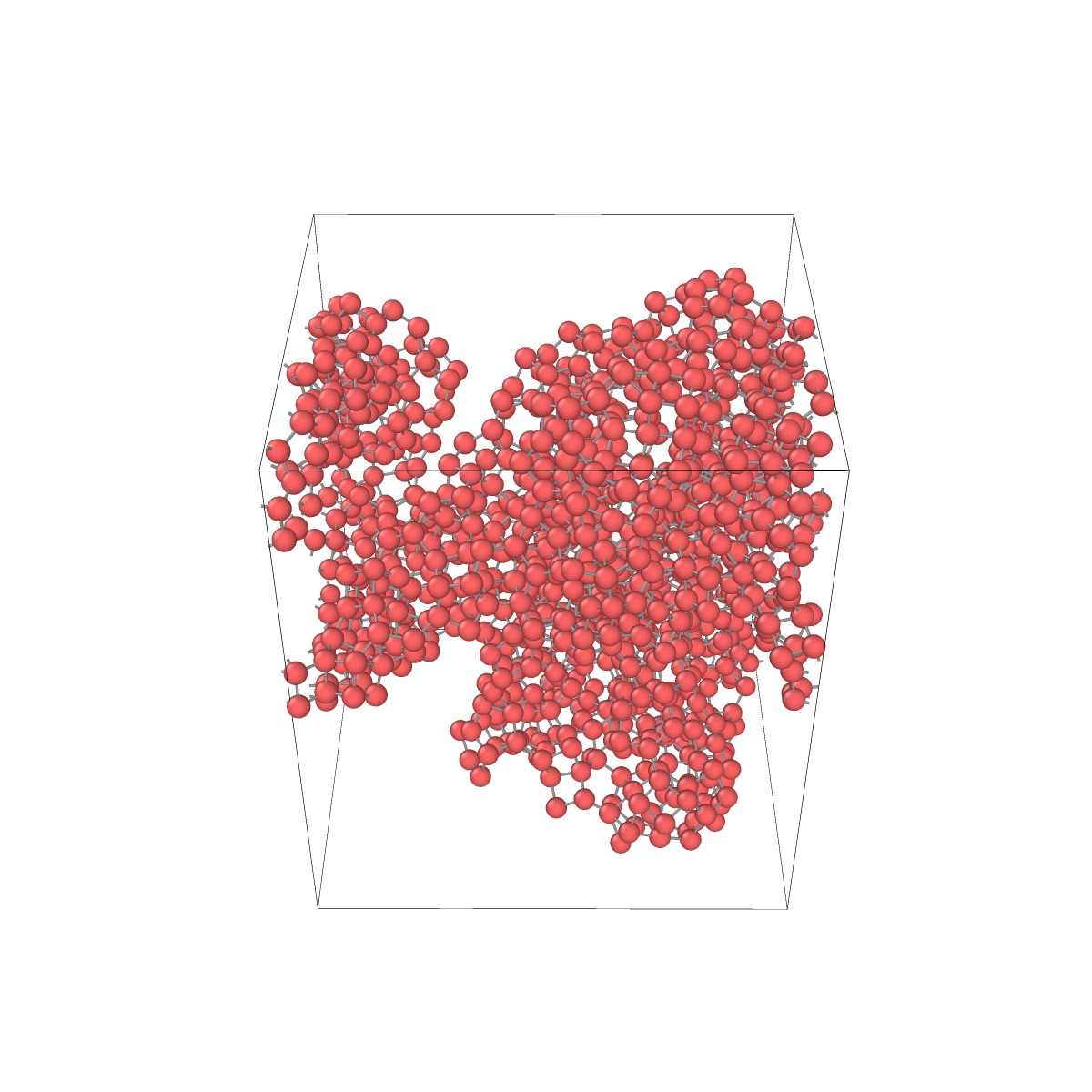}

\caption{Snapshots of MD simulations of a non-ribbon of buckled silicene modeled
with the present interatomic potential. The temperature increases
with a constant rate from 0 K to 300 K over a 1 ns time period. The
images show one repeat unit of the ribbon containing 1080 atoms. The
time increases from (a) (initial state) to (d) (final state). \label{fig:buckled_ribbon}}
\end{figure}

\begin{figure}
\textbf{(a)}\enskip{}\includegraphics[width=0.435\textwidth]{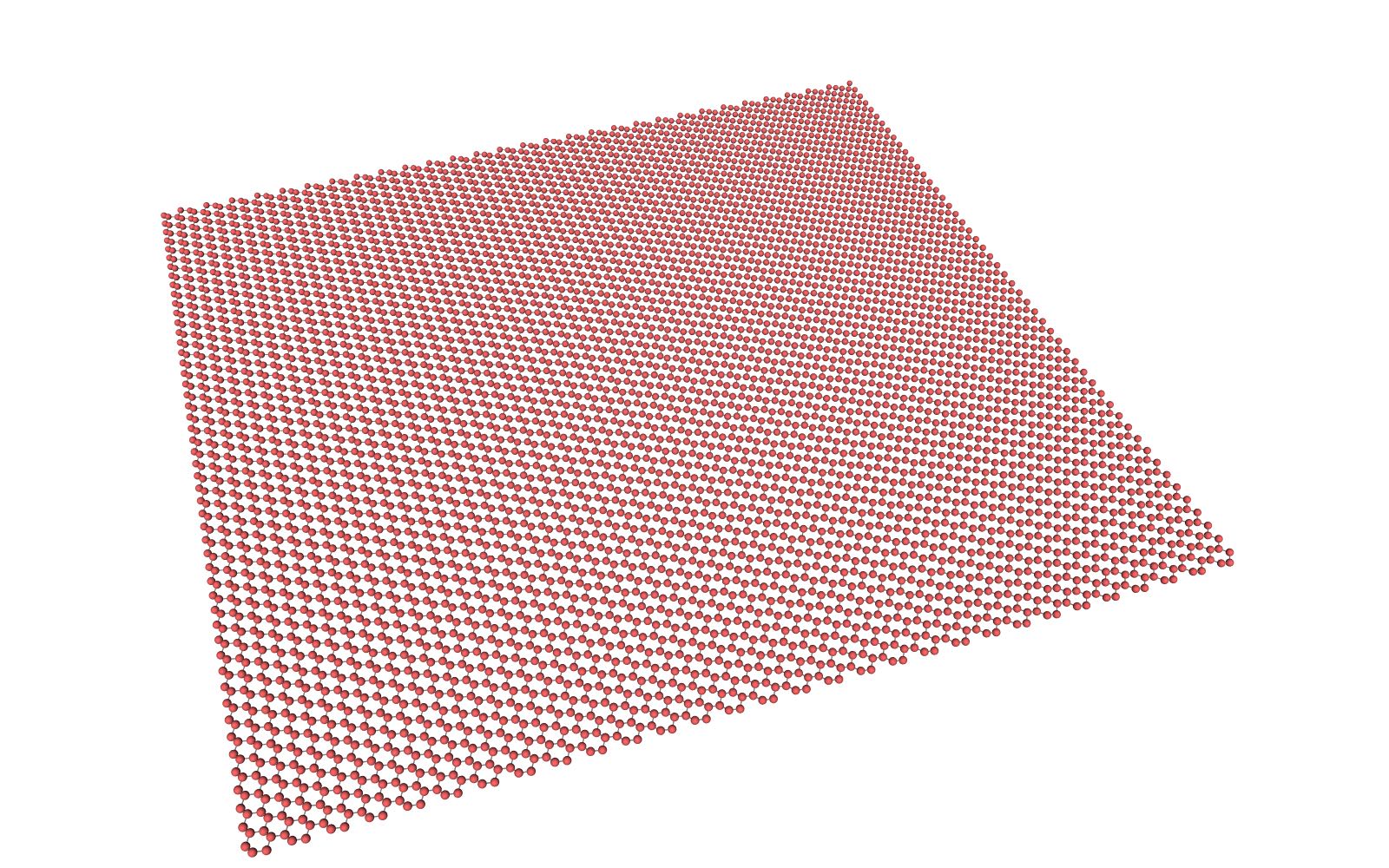}\enskip{}\enskip{}\enskip{}\textbf{(b)}\enskip{}\includegraphics[width=0.435\textwidth]{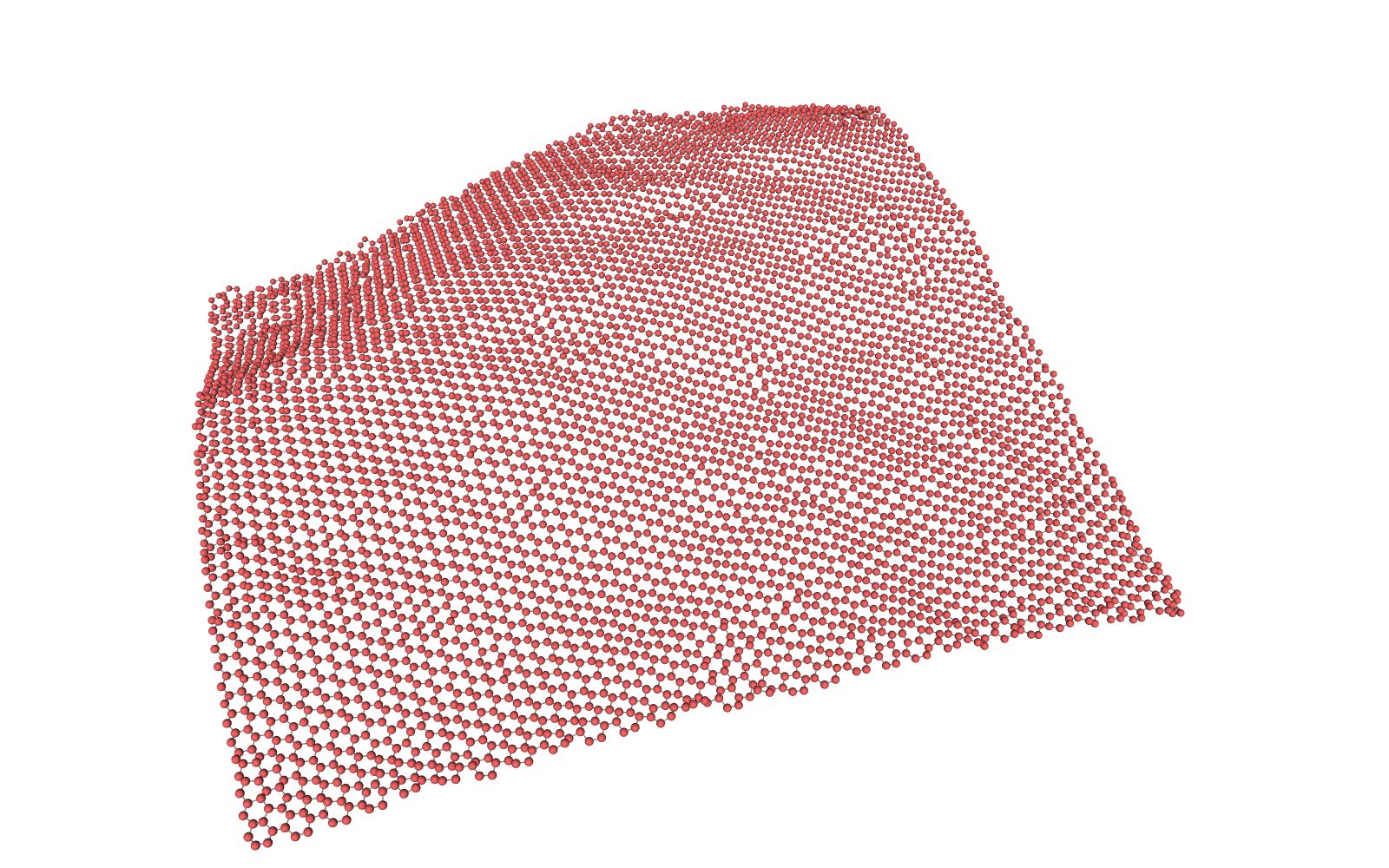} 

\bigskip{}

\textbf{(c)}\enskip{}\includegraphics[width=0.435\textwidth]{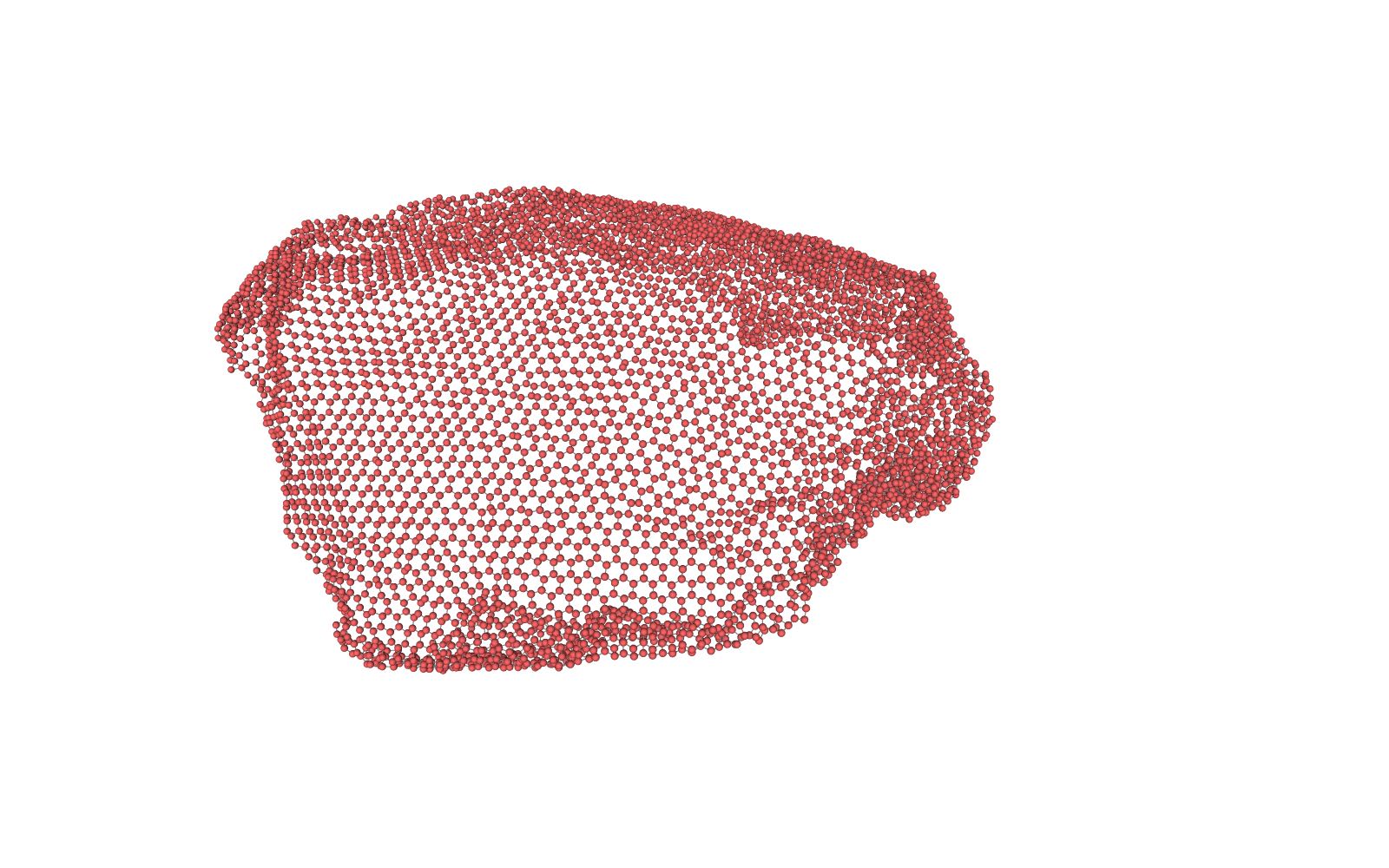}\enskip{}\enskip{}\enskip{}\textbf{(d)}\enskip{}\includegraphics[width=0.435\textwidth]{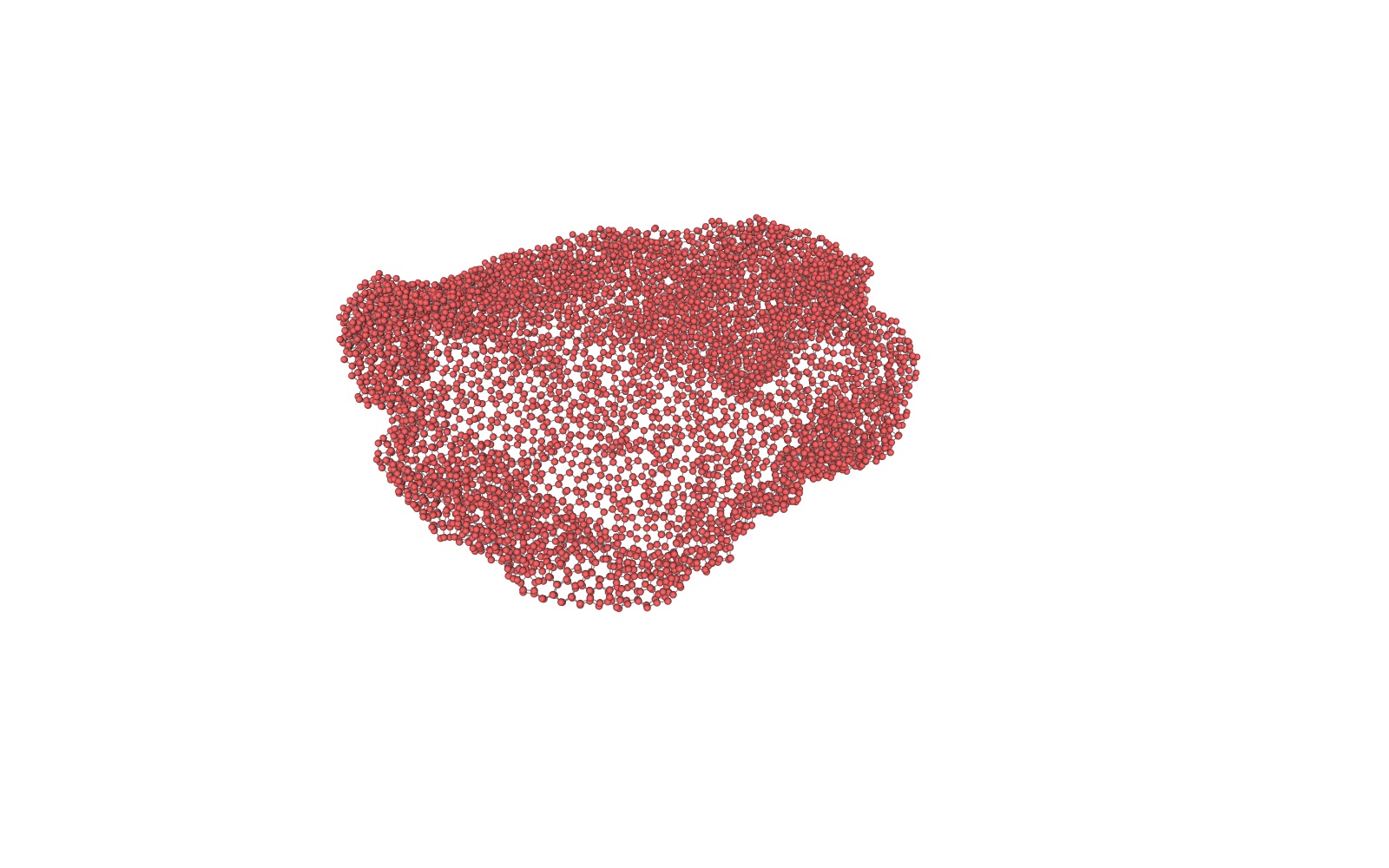}

\caption{Snapshots of MD simulations of a 6120-atom free-standing nano-sheet
(flake) of buckled silicene modeled with the present interatomic potential.
The temperature increases with a constant rate from 0 K to 300 K over
a 0.6 ns time period {[}snapshots (a), (b) and (c){]} followed by
an isothermal anneal at 300 K {[}snapshot (d) taken 0.2 ns into the
anneal{]}. \label{fig:buckled_flake}}
\end{figure}

Thermal stability of single-layer silicenes has been evaluated by
MD simulations. The simulated systems were subject to periodic boundary
conditions at zero pressure. Fig.~\ref{fig:buckled_ribbon} demonstrates
that a nano-ribbon of buckled silicene is unstable at finite temperatures
and quickly collapses to a cluster before temperature reaches 300
K. Likewise, a free-standing sheet (flake) of buckled silicene (Fig.~\ref{fig:buckled_flake})
collapses into a cluster with the shape of a bowl when temperature
reaches 300 K. The nano-ribbon and nano-flake made of the $\sqrt{3}\times\sqrt{3}$
dimerized silicene collapse as well. 

\begin{figure}
\textbf{(a)}\enskip{}\includegraphics[width=0.435\textwidth]{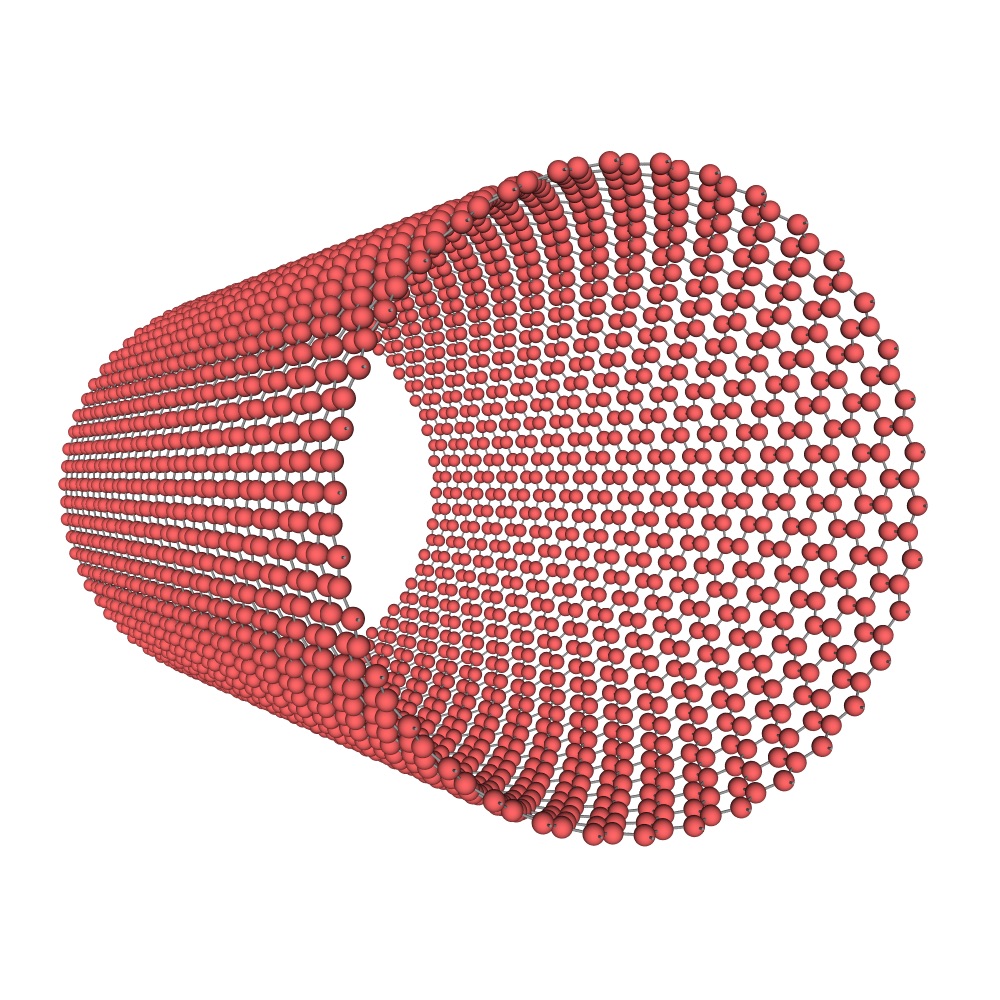}\enskip{}\enskip{}\enskip{}\textbf{(b)}\enskip{}\includegraphics[width=0.435\textwidth]{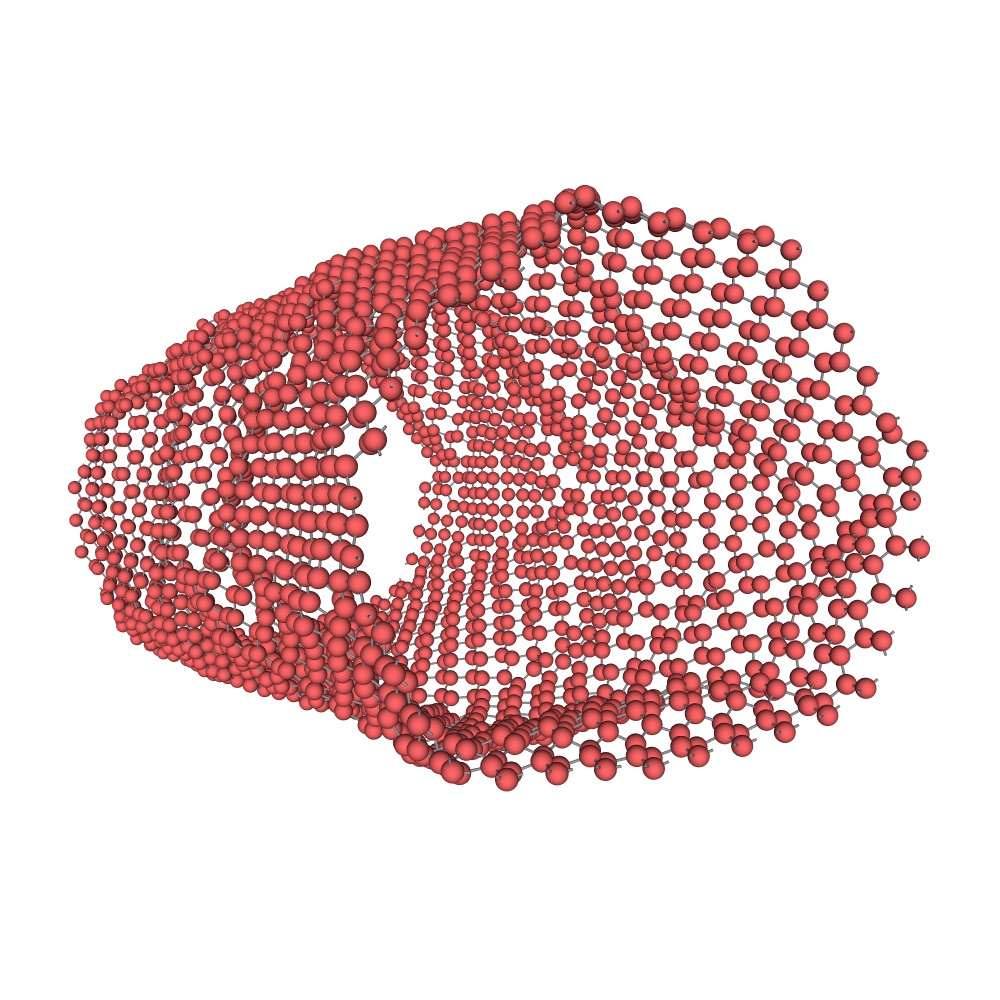} 

\textbf{(c)}\enskip{}\includegraphics[width=0.435\textwidth]{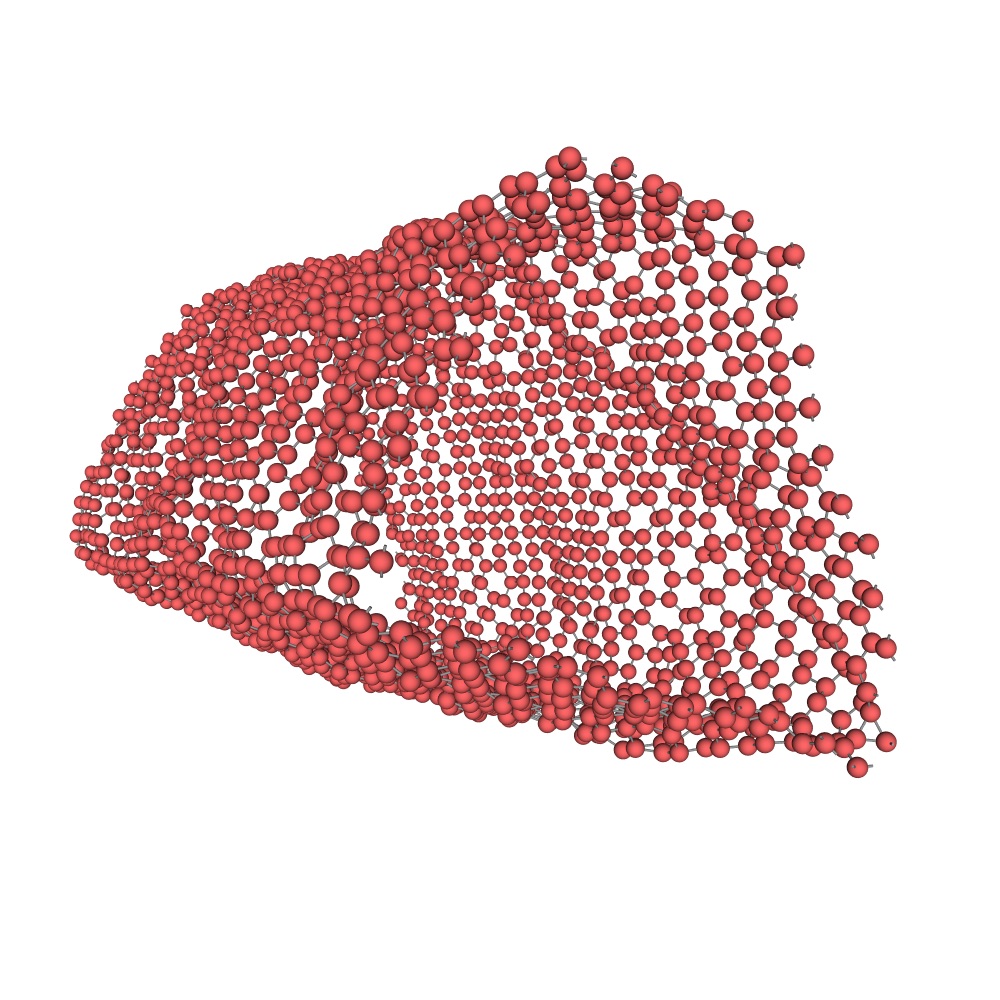}\enskip{}\enskip{}\enskip{}\textbf{(d)}\enskip{}\includegraphics[width=0.435\textwidth]{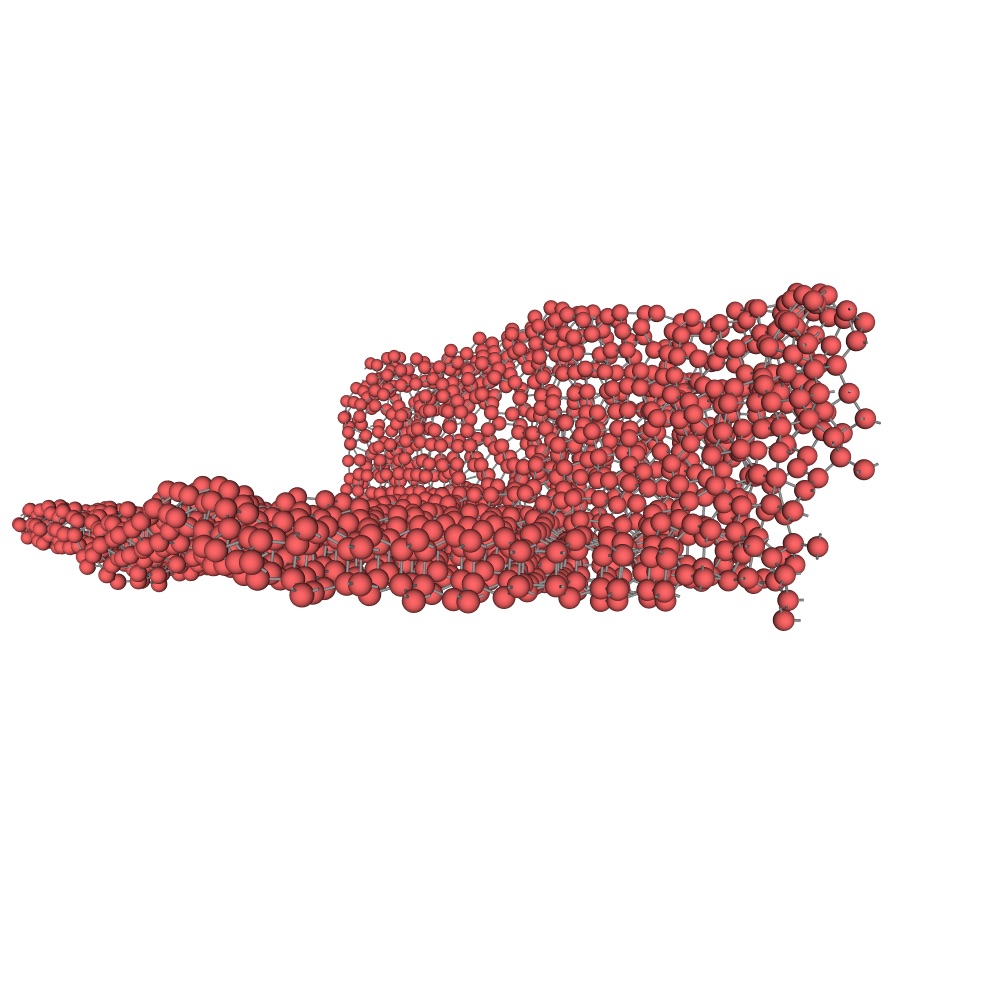}

\caption{Snapshots of MD simulations of a single-wall nano-tube of planar silicene
modeled with the present interatomic potential. The temperature increases
with a constant rate from 0 K to 300 K over a 2 ns time period. The
images show one period of the tube (diameter 49 Å, length 122 Å, 2160
atoms). The time increases from (a) (initial state) to (d) (final
state). \label{fig:buckled_tube}}
\end{figure}

A single-wall nano-tube was also tested for thermal stability. The
latter was obtained by wrapping a layer planar silicene into a tube
49 Å in diameter (Fig.~\ref{fig:buckled_tube}). The period along
the tube axis was 122 Å. As soon as temperature began to increase
starting from 0 K, the wall of the tube transformed to the buckled
structure and then collapsed before the temperature reached 300 K.
Qualitatively the same behavior of the single-layer silicene structures
was found with all four potentials. In all cases, the single layer
silicene easily developed waves due to thermal fluctuations until
neighboring surface regions came close enough to each other to form
covalent bonds. Once this happened, the bond-forming process quickly
spread over the entire surface and the structure collapsed. This chemical
reactivity and the lack of bending rigidity are the main factors that
cause the instability of free-standing single-layer silicenes at room
temperature. 

\subsection{Bilayer silicenes}

Another interesting 2D form of silicon is the bilayer silicene.\citep{Resta:2013aa,Arafune:2013cj,Liu:2013aa,Fu:2014ai,Pflugradt:2014aa,Padilha:2015ng,Yaokawa:2016hq}
Like the single-layer silicene discussed above, the bilayer silicene
was found experimentally on top of metallic surfaces such as Ag(111).\citep{Resta:2013aa,Arafune:2013cj,Pflugradt:2014aa,Yaokawa:2016hq}
By contrast to bilayer graphene, the interlayer bonds in bilayer silicene
are covalent sp$^{3}$ type. As a result, the formation of a bilayer
is accompanied by a significant energy release. It can be expected,
therefore, that bilayer silicene should be more stable than two single
layers. 

\begin{figure}
\textbf{(a)}\enskip{}\includegraphics[width=0.32\textwidth]{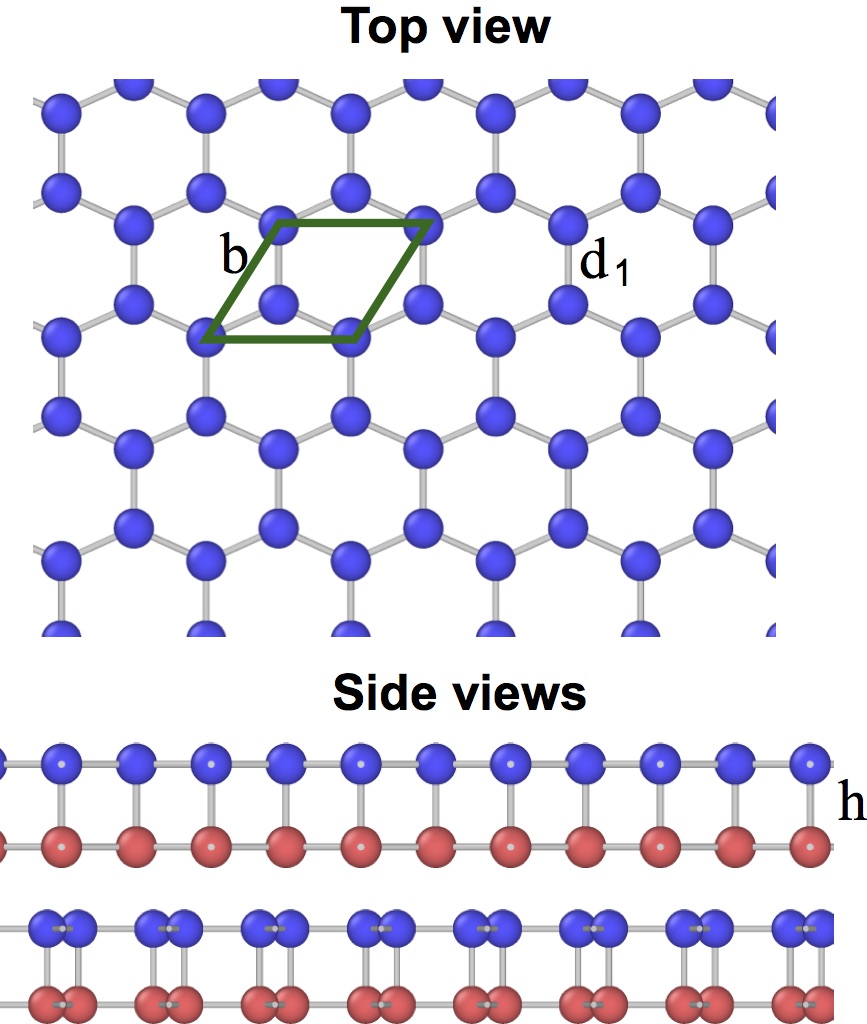}

\bigskip{}

\textbf{(b)}\enskip{}\includegraphics[width=0.33\textwidth]{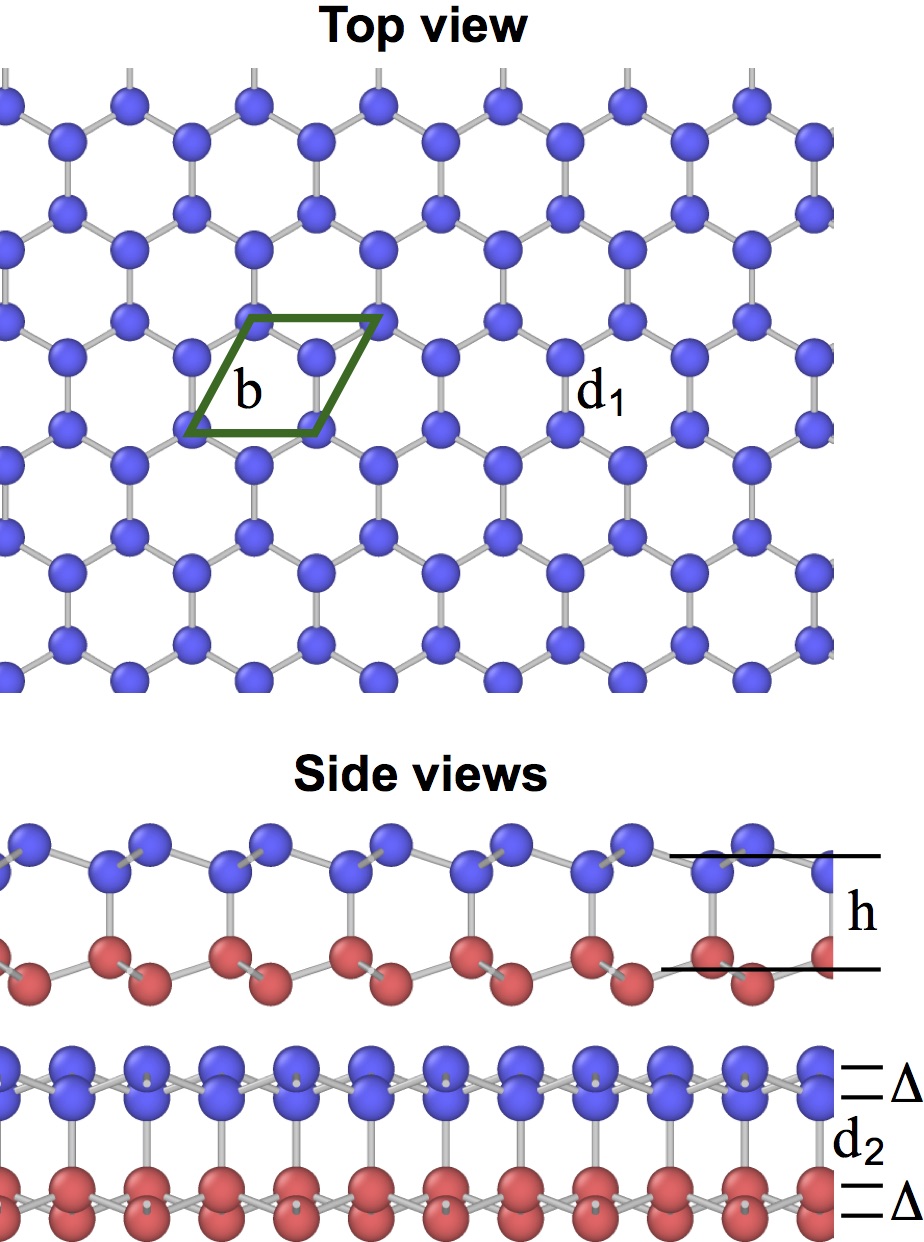}

\bigskip{}

\textbf{(c)}\enskip{}\includegraphics[width=0.33\textwidth]{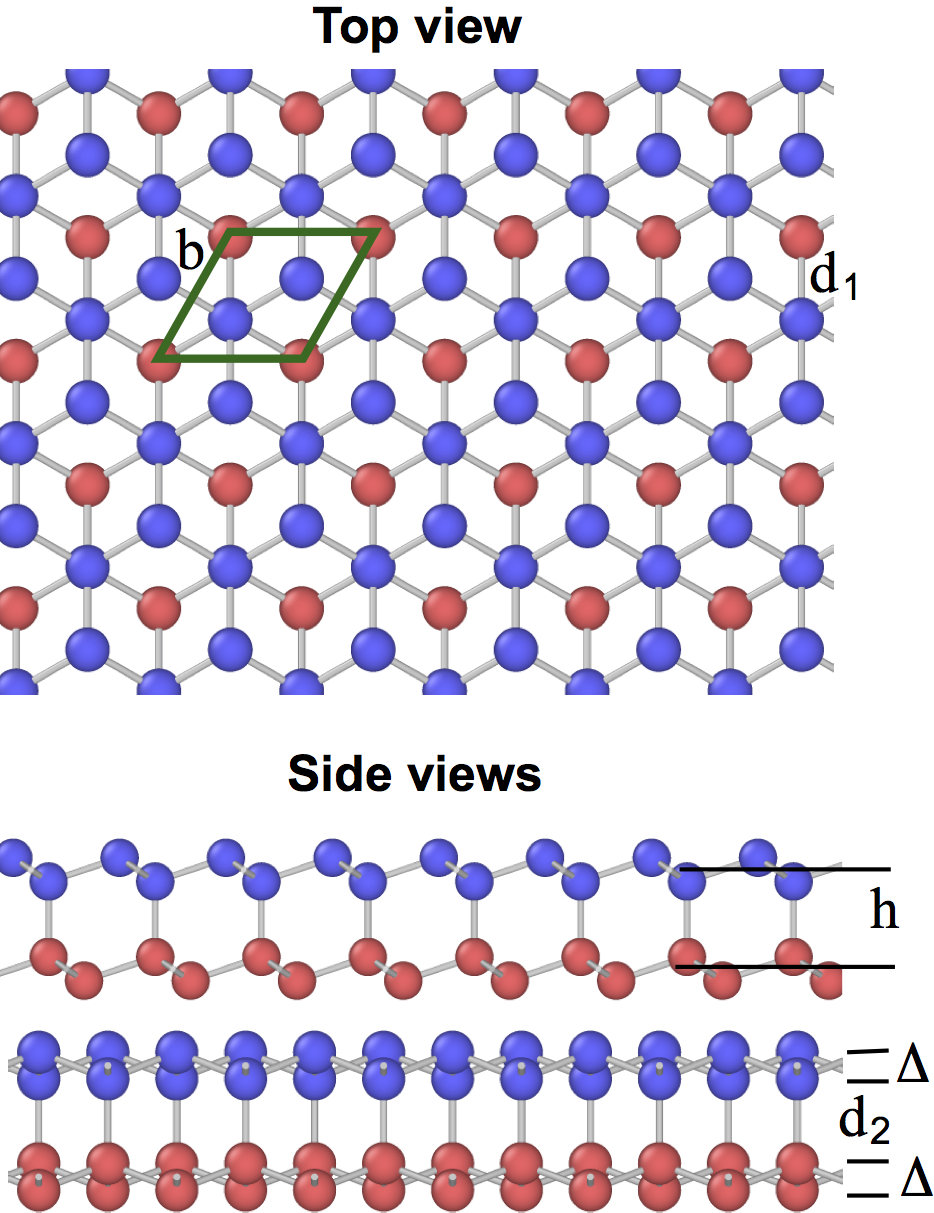}

\caption{Structures of bilayer silicenes: (a) AA$_{p}$, (b) AA$^{\prime}$,
(c) AB.\label{fig:Structures-of-bilayers}}
\end{figure}

Several structural forms of the bilayer silicene have been found in
experiments and studied by DFT calculations, depending on the type
of stacking of the two layers and whether they are planar or buckled.\citep{Resta:2013aa,Arafune:2013cj,Liu:2013aa,Fu:2014ai,Pflugradt:2014aa,Padilha:2015ng,Yaokawa:2016hq}
Three of the structures, referred to as AA$_{p}$, AA$^{\prime}$
and AB, are shown in Fig.~\ref{fig:Structures-of-bilayers}. The
AA$_{p}$ structure is obtained by stacking two planar silicene layers
(A) on top of each other and connecting them by vertical covalent
bonds {[}Fig.~\ref{fig:Structures-of-bilayers}(a){]}. This structure
is characterized by the geometric parameters $b$ (side of the rhombic
structural unit) and the interlayer spacing $h$. The bond length
between Si atoms is $d_{1}=b/\sqrt{3}$ within each layer and $h$
between the layers. In the AA$^{\prime}$ structure, both layers are
buckled, and the buckling of one layer (A$^{\prime}$) is inverted
with respect to the buckling of the other layer (A) {[}Fig.~\ref{fig:Structures-of-bilayers}(b){]}.
As a result, half of the interlayer distances are short, leading to
the formation of covalent bonds, and the other half of the distances
are longer and covalent bonds do not form. The geometric parameters
of the structure are $b$ (defined above), the in-layer bond length
$d_{1}$, the interlayer bond length $d_{2}$, and the split width
of each layer $\Delta$. The distance between the layers is $h=d_{2}+\Delta$.
Finally, in the AB structure, two buckled silicene layers A and B
are stacked together so that half of the atoms of one layer project
into the centers of the hexagonal units of the other layer {[}Fig.~\ref{fig:Structures-of-bilayers}(c){]}.
The remaining half of the atoms project onto each other and form vertical
covalent bonds. As with the single-layer silicenes, it has not been
possible so far to isolate free-standing bilayer silicene experimentally.

The cohesive energies $E_{c}$ and geometric parameters of three bilayer
silicenes computed with four interatomic potentials are compared with
DFT data in Table \ref{tab:bilayer_Silicene}. The Table also shows
the energies $\Delta E$ of the buckled bilayers AA$^{\prime}$ and
AB relative to the planar bilayer AA$_{p}$. None of the potentials
matches the DFT calculations accurately. However, the present potential
displays the closest agreement. The MOD potential incorrectly predicts
that the buckled structures AA$^{\prime}$ and AB are more stable
than AA$_{p}$ (negative $\Delta E$ values), which is contrary to
the DFT calculations. It should be noted that all four potentials
predict virtually identical properties of the AA$^{\prime}$ and AB
silicenes. This is not surprising: considering only nearest neighbor
bonds, the local atomic environments in the two structures are identical.
Their DFT lattice parameters $b$ are indeed the same (3.84 Å),\citep{Padilha:2015ng}
but the DFT energies are different (0.33 and 0.17 eV/atom, respectively;\citep{Padilha:2015ng}
our potential gives $\Delta E=0.12$ eV/atom for both). This discrepancy
apparently reflects a common feature of all short-range Si potentials. 

\begin{figure}
\textbf{(a)}\enskip{}\includegraphics[width=0.58\textwidth]{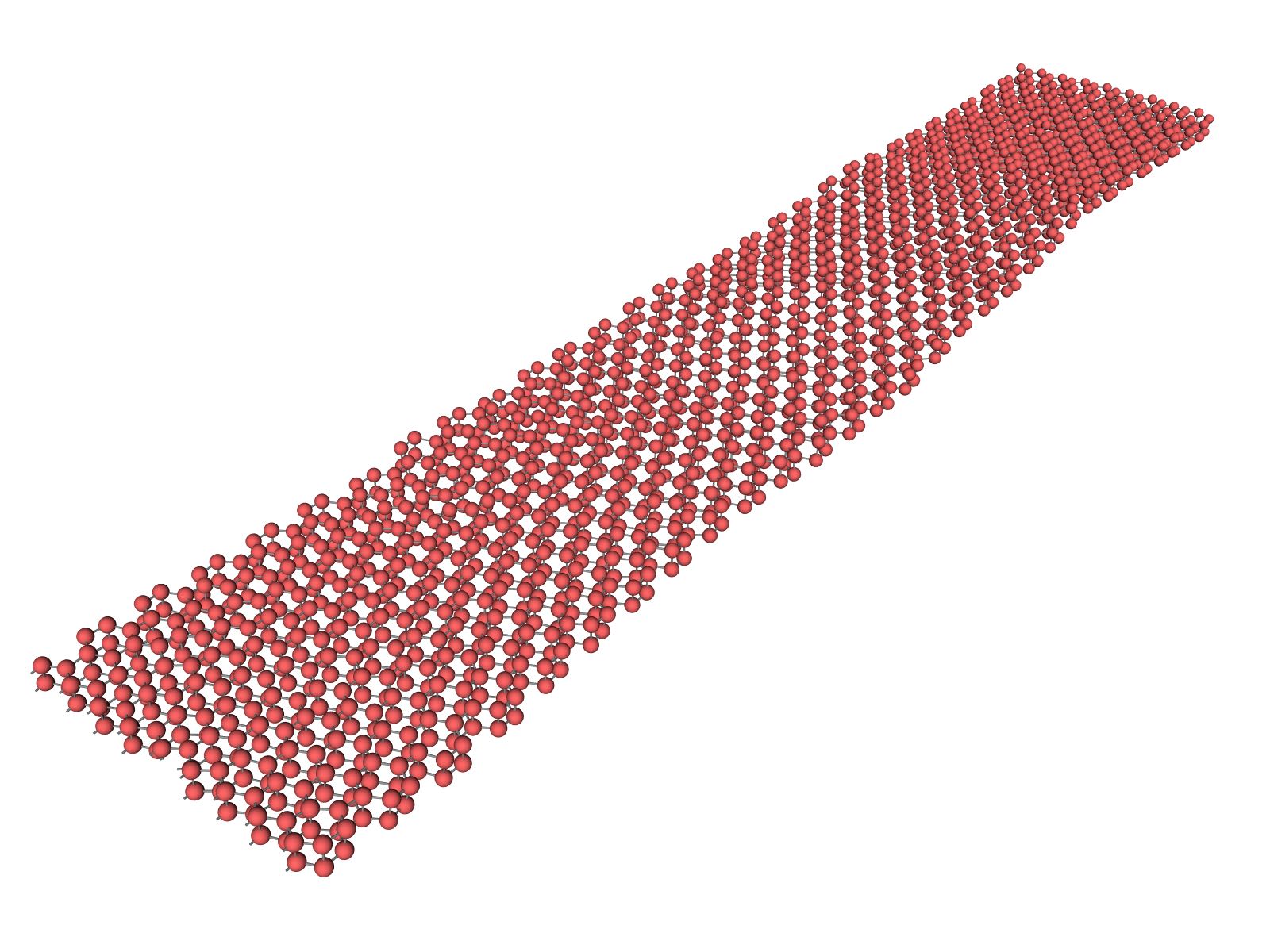}

\bigskip{}

\textbf{(b)}\enskip{}\includegraphics[width=0.62\textwidth]{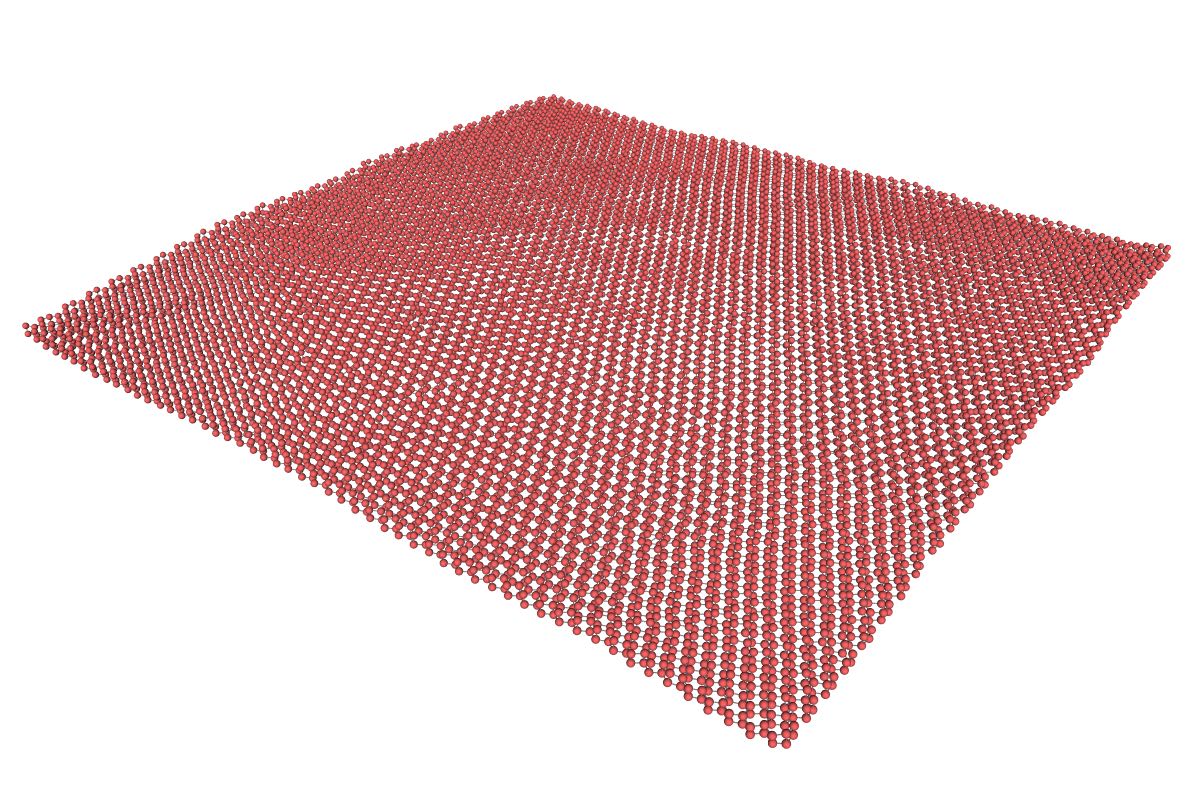}

\bigskip{}

\textbf{(c)}\enskip{}\includegraphics[width=0.43\textwidth]{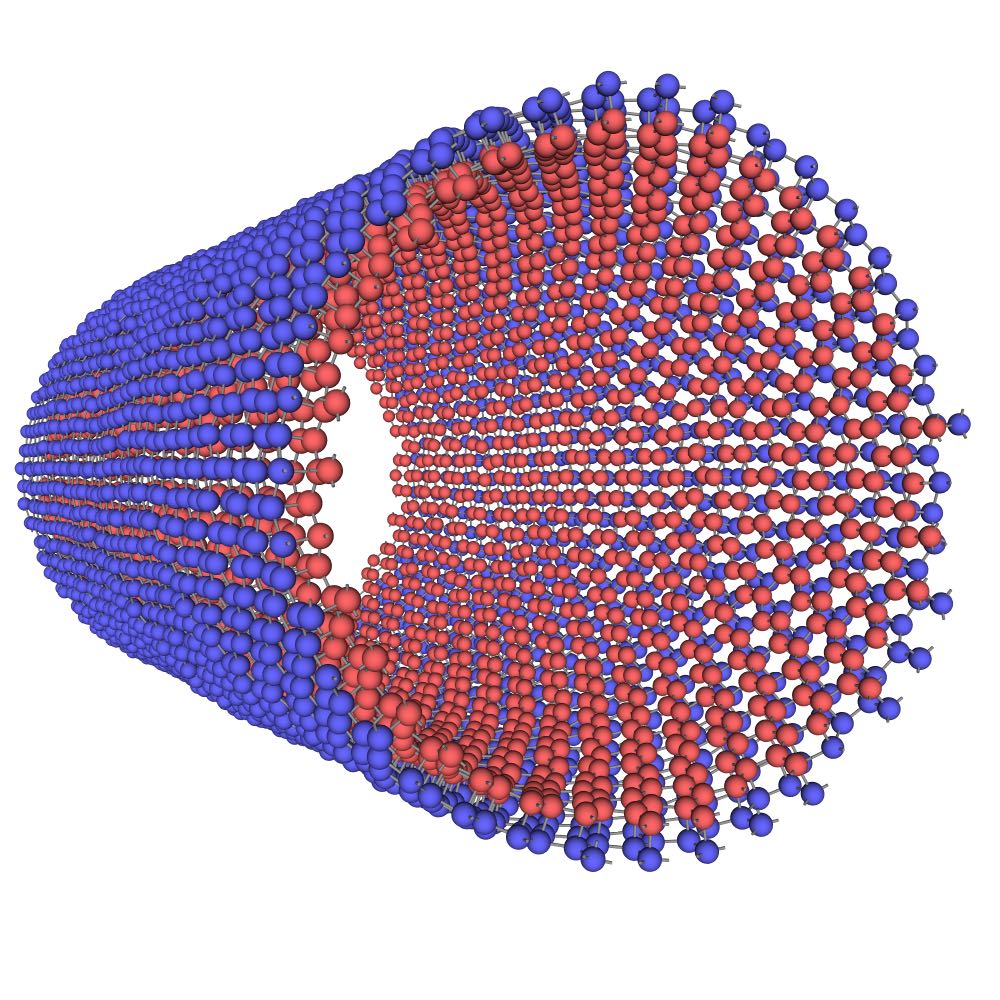}

\caption{Snapshots of MD simulations of the AA$^{\prime}$ bilayer silicene
after a 10 ns anneal at 300 K. (a) Nano-ribbon, (b) Free-standing
nano-sheet (flake), (c) Nano-tube (the two layers are shown in different
colors for clarity).\label{fig:bilayer_MD}}
\end{figure}

To assess thermal stability of bilayer silicenes, MD simulations were
conducted for the same nano-ribbon, nano-flake and nano-tube configurations
as discussed above. The most stable AA$_{p}$ silicene was chosen
for the tests. The samples were heated up to 300 K and annealed at
this temperature for 10 ns. The systems developed significant capillary
waves, especially the nano-ribbon, but none of them collapsed (Fig.~\ref{fig:bilayer_MD}).
Although 10 ns is a short time in comparison with experimental times,
these tests confirm that the bilayer silicene has a much greater bending
rigidity and smaller reactivity in comparison with its single-layer
counterpart. As such, it has a much better chance of survival in a
free-standing form at room temperature. 

In additional tests, the nano-flake was heated from 300 K to 1000
K in 6 ns followed by an isothermal anneal for 2 ns at 1000 K. The
surface of the flake developed a set of thermally activated point
defects, such as ad-atoms and locally buckled configurations, but
the flake itself did not collapse. This again confirms the significant
thermal stability of the bilayer silicene, possibly even at high temperatures.
The same tests were conducted with all four potentials and the results
were qualitatively similar. With the MOD potential, the initial AA$_{p}$
silicene quickly transformed to the more stable buckled structure,
but the system still did not collapse.

\section{Discussion and conclusions}

Silicon is one of the most challenging elements for semi-empirical
interatomic potentials. It has over a dozen polymorphs that are stable
at different temperatures and pressures and exhibit different coordination
numbers and types of bonding ranging from strongly covalent to metallic.
The diamond cubic phase displays a rather complex behavior with several
possible structures of point defects, a number of surface reconstructions,
and an increase in density upon melting. It is not surprising that
the existing Si potentials are not nearly as successful in describing
this material as some of the embedded-atom potentials for metals.\citep{Daw83,Daw84,Mishin.HMM}
In this work, we developed a new Si potential with the goal of improving
some of the properties that were not captured accurately by other
potentials. For comparison, we selected three potentials from the
literature that we consider most reliable\citep{Kumagai:2007ly,Ryu:2009dn}
or most popular.\citep{Stillinger85} 

Extensive tests have shown that the present potential does achieve
the desired improvements, in particular with regard to the vacancy
formation energies, surface formation energies and reconstructions,
thermal expansion factors and a few other properties. The potential
is more accurate, in comparison with other potentials, in reproducing
the DFT data for the novel Si polymorphs h-Si$_{6}$ and Si$_{24}$
without including them in the fitting database. But the tests have
also shown that \emph{each} of the four potentials has its successes
and failures. The present potential makes inaccurate predictions for
the energies of high-lying Si polymorphs (although their atomic volumes
are quite accurate), for the latent heat of melting, and for the short-range
order in the liquid phase. The MOD potential\citep{Kumagai:2007ly}
has its own drawbacks mentioned in Section \ref{sec:Introduction}.
The MEAM potential\citep{Ryu:2009dn} grossly overestimates the phonon
frequencies and thermal expansion factors, in addition to the incorrect
\{100\} surface reconstruction. The SW potential successfully reproduces
the surface energies and thermal expansion factors but predicts a
positive Cauchy pressure and systematically overestimates the atomic
volumes of Si polymorphs (as does the MEAM potential).

The potentials were put through a very stringent test by computing
the binding energies of small Si$_{n}$ clusters. Such clusters were
not included in the potential fitting procedure and are traditionally
considered to be out of reach of potentials unless specifically included
in fitting database. Surprisingly, the present potential, the MOD
potential,\citep{Kumagai:2007ly} and especially the MEAM potential\citep{Ryu:2009dn}
reproduce the general trends of the cluster energies reasonably well
(Fig.~\ref{fig:cluster_energies}). In many cases, the ranking of
the energies of different geometries for the same cluster size $n$
agrees with first-principles calculations. The SW potential is less
accurate: it systematically under-binds the clusters and makes more
mistakes in the energy ordering.

Encouraged by the reasonable performance for the clusters, we applied
the potentials to model single-layer and bilayer silicenes, which
were not included in the potential fitting either. While none of the
potentials reproduces all DFT calculations accurately, they generally
perform reasonably well. One notable exception is the MOD potential,
which under-binds the $\sqrt{3}\times\sqrt{3}$ dumbbell structure
of the single-layer silicene and fails to reproduce the correct ground-state
of the bilayer silicene. Furthermore, all four potentials predict
identical energies of the AA$^{\prime}$ and AB bilayer silicenes,
whereas the DFT energies are different. Other than this, the trends
are captured quite well. The present potential demonstrates the best
performance for the bilayer silicenes. 

Experimentally, silicenes have only been found on metallic substrates.
Whether they can exist in a free-standing form at room temperature
remains an open question. Evaluation of their thermal stability requires
MD simulations of relatively large systems for relatively long times
that are not currently accessible by DFT methods. Although interatomic
potentials are less reliable, they can be suitable for a preliminary
assessment. The MD simulations performed in this work indicate that
single-layer silicenes are unlikely to exist in a free-standing form.
Their large bending compliance and chemical reactivity lead to the
development of large shape fluctuations and eventually the formation
of covalent bonds between neighboring surface regions at or below
room temperature. By contrast, bilayer silicenes exhibit much greater
bending rigidity and lower surface reactivity. Nano-structures such
as nano-ribbons, nano-flakes and nano-tubes remain intact at and above
room temperature, at least on a 10 ns timescale. The fact that this
behavior was observed with all four potentials points to the generality
of these observations and suggests that free-standing bilayer silicenes
might be stable at room temperature. Of course, this tentative conclusion
requires validation by more detailed and more accurate studies in
the future.

The four potentials discussed in this work are likely to represent
the limit of what can be achieved with short-range semi-empirical
potentials. Further improvements can only be made by developing more
sophisticated, longer-range, and thus significantly slower potentials.
Analytical bond-order potentials offer one option.\citep{Oloriegbe_PhD:2008aa,Gillespie:2007vl,Drautz07a}
Recent years have seen a rising interest in machine-learning potentials.\citep{Mueller:2016aa,Behler:2016aa,Bartok:2010aa,Behler07,Behler:2008aa,Botu:2015bb}
While even slower, they allow one to achieve an impressive accuracy
of interpolation between DFT energies, in some cases up to a few meV/atom.
However, the lack of transferability to configurations outside the
training dataset remains an issue. 

\section*{Acknowledgements}

This work was supported by the U.S. Department of Energy, Office of
Basic Energy Sciences, Division of Materials Sciences and Engineering,
the Physical Behavior of Materials Program, through Grant No.~DE-FG02-01ER45871.

%\bibliographystyle{/Users/ymishin/YURI/Bibliography/ActaMatnew}
%\bibliography{/Users/ymishin/YURI/Bibliography/literat}

\newpage{}\clearpage{}

\begin{table}
\caption{Optimized parameters of the new Si potential. Parameters of the MOD
potential\citep{Kumagai:2007ly} are listed for comparison.\label{tab:parameters} }
\begin{tabular}{lcc}
\hline 
Parameter  & Present  & MOD$^{a}$ \tabularnewline
\hline 
$A$ (eV)  & 3198.51383  & 3281.5905 \tabularnewline
$B$ (eV)  & 117.780724  & 121.00047 \tabularnewline
$\lambda_{1}$ (Å$^{-1}$)  & 3.18011795  & 3.2300135 \tabularnewline
$\lambda_{2}$ (Å$^{-1}$)  & 1.39343356  & 1.3457970 \tabularnewline
$\eta$  & 2.16152496  & 1.0000000 \tabularnewline
$\eta\times\delta$  & 0.544097766  & 0.53298909 \tabularnewline
$\alpha$  & 1.80536502  & 2.3890327 \tabularnewline
$\beta$  & 3  & 1 \tabularnewline
$c_{0}$ (eV)  & -0.0059204  & 0.0 \tabularnewline
$c_{1}$  & 0.201232428  & 0.20173476 \tabularnewline
$c_{2}$  & 614230.043  & 730418.72 \tabularnewline
$c_{3}$  & 996439.097  & 1000000.0 \tabularnewline
$c_{4}$  & 3.33560562  & 1.0000000 \tabularnewline
$c_{5}$  & 25.2096377  & 26.000000 \tabularnewline
$h$  & -0.381360867  & \textendash 0.36500000 \tabularnewline
$R_{1}$ (Å)  & 2.54388270  & 2.7 \tabularnewline
$R_{2}$ (Å)  & 3.20569403  & 3.3 \tabularnewline
\hline 
$^{a}$Ref.\,\onlinecite{Kumagai:2007ly}  &  & \tabularnewline
\end{tabular}
\end{table}

\begin{table}[h]
\caption{Properties of diamond cubic Si computed with four interatomic potentials
in comparison with experimental data and DFT calculations.\label{tab:fundamental_prop} }

\begin{tabular}{lcccccc}
\hline 
Property  & Experiment  & DFT  & Present  & MOD$^{d}$  & MEAM$^{w}$  & SW$^{u}$ \tabularnewline
\hline 
$E_{c}$ (eV/atom)  & 4.63$^{c}$  & 4.84$^{r}$  & 4.630  & 4.630  & 4.630  & 4.337 \tabularnewline
$a$ (Å)  & 5.430$^{a}$  & 5.451$^{r}$  & 5.434  & 5.429  & 5.431  & 5.431 \tabularnewline
$c_{11}$(GPa)  & 165$^{a}$; 167.40$^{b}$  &  & 172.6  & 166.4  & 163.8  & 151.4 \tabularnewline
$c_{12}$ (GPa)  & 64$^{a}$; 65.23$^{b}$  &  & 64.6  & 65.3  & 64.5  & 76.4 \tabularnewline
$c_{44}$ (GPa)  & 79.2$^{a}$; 79.57$^{b}$  &  & 81.3  & 77.1  & 76.5  & 56.4 \tabularnewline
$\nu_{\textrm{max}}$ (THz)  & 15.7$^{o}$  &  & 17.6  & 17.5  & 25.6  & 17.8\tabularnewline
 &  &  &  &  &  & \tabularnewline
\multicolumn{7}{l}{Vacancy: }\tabularnewline
$E_{v}^{f}$ ($T_{d}$) (eV)  & 3.6$^{j}$  & 3.17$^{m}$; 3.69$^{t}$  & 3.54  & 2.82  & 3.57  & 2.64 \tabularnewline
 &  & 3.29\textendash 4.3$^{h}$; 3.70-3.84$^{s}$  &  &  &  & \tabularnewline
$E_{v}^{f}$ ($D_{3d}$) (eV)  &  & 3.97$^{t}$; 4.29$^{v}$; 4.37$^{n}$  & 3.61  & \textendash{}  & 3.77  & \textendash{} \tabularnewline
 &  & 3.67-3.70$^{s}$; 5.023$^{i}$  &  &  &  & \tabularnewline
 &  &  &  &  &  & \tabularnewline
\multicolumn{7}{l}{Interstitials: }\tabularnewline
$E_{i}^{f}$ (hex) (eV)  &  & 3.31\textendash 5$^{h}$; 2.87-3.80$^{s}$  & 3.51  & 4.13$^{d}$  & \textendash{}  & \textendash{} \tabularnewline
$E_{i}^{f}$ ($T_{d}$) (eV)  &  & 3.43\textendash 6$^{h}$; 3.43-5.10$^{s}$  & 3.01  & 3.27$^{d}$  & 4.12  & 4.93 \tabularnewline
$E_{i}^{f}$ (B) (eV)  &  & 4\textendash 5$^{h}$  & 4.34  & 5.03$^{d}$  & 6.78  & 5.61 \tabularnewline
$E_{i}^{f}\langle110\rangle$(eV)  &  & 3.31\textendash 3.84$^{h}$; 2.87-3.84$^{s}$  & 3.26  & 3.57$^{d}$  & 3.91  & 4.41 \tabularnewline
 &  &  &  &  &  & \tabularnewline
\multicolumn{7}{l}{Surface energy $\gamma_{\textrm{s}}$ (J/m$^{2}$): }\tabularnewline
\{111\}  & 1.24$^{q}$; 1.23$^{p}$  & 1.57$^{l}$; 1.74$^{f}$  & 1.11  & 0.89  & 1.2  & 1.36 \tabularnewline
\{100\}  &  & 2.14$^{l}$; 2.39$^{f}$; 2.36$^{k}$  & 2.19  & 1.77  & 1.74$^{e}$  & 2.36 \tabularnewline
\{100\}$_{2\times1}$  & 1.36$^{p}$  & 1.71$^{g}$; 1.45$^{f}$; 1.51$^{k}$  & 1.21  & 1.07  & 1.24  & 1.45 \tabularnewline
\{110\}  & 1.43$^{p}$  & 1.75$^{k}$  & 1.36  & 1.08  & 1.41  & 1.67 \tabularnewline
 &  &  &  &  &  & \tabularnewline
\multicolumn{7}{l}{Melting: }\tabularnewline
T$_{m}$(K)  & 1687  &  & 1687  & 1681$^{d}$; 1682  & 1687$^{w}$  & 1691$^{v}$; 1677 \tabularnewline
$\Delta V_{m}/V_{\textrm{solid}}$ (\%)  & -5.1$^{a}$  &  & -3.8  & -12.5  & -2.7  & -7.2 \tabularnewline
$L$ (kJ/mol)  & 50.6$^{a}$  &  & 24.0  & 34.7  & 43.2  & 31.1 \tabularnewline
\hline 
\multicolumn{7}{l}{$^{a}$Ref.\,\onlinecite{Smithells-metals2004}; $^{b}$Ref.\,\onlinecite{Skimin1951};
$^{c}$Ref.\,\onlinecite{Kittel}; $^{d}$Ref.\,\onlinecite{Kumagai:2007ly};
$^{e}$Constrained relaxation; $^{f}$Ref.\,\onlinecite{Stekolnikov:2002zr}}\tabularnewline
\multicolumn{7}{l}{$^{g}$Ref.\,\onlinecite{Timonova:2007fr}; $^{h}$References in
Ref.\,\onlinecite{Kumagai:2007ly}; $^{i}$Ref.\,\onlinecite{Wright:2006sf};
$^{j}$Ref.\,\onlinecite{Dannefaer:1986qd}}\tabularnewline
\multicolumn{7}{l}{$^{k}$Ref.\,\onlinecite{Lu:2005jw}; $^{l}$Ref.\,\onlinecite{Jelinek:2012ij};
$^{m}$Ref.\,\onlinecite{Goedecker:2002yf}; $^{n}$Ref.\,\onlinecite{Dabrowski:2015ss};
$^{o}$Ref.\,\onlinecite{Dolling:1963}; $^{p}$Ref.\,\onlinecite{Eaglesham:1993qa} }\tabularnewline
\multicolumn{7}{l}{$^{q}$Ref.\,\onlinecite{Gilman:1960kl}; $^{r}$Ref.\,\onlinecite{Yin:1982ef};
$^{s}$Ref.\,\onlinecite{Ganchenkova:2015aa}; $^{t}$Ref.\,\onlinecite{Centoni:2005sd};
$^{u}$Ref.\,\onlinecite{Stillinger85}; $^{v}$Ref.\,\onlinecite{Puska:1998kx};
$^{w}$Ref.\,\onlinecite{Ryu:2009dn} }\tabularnewline
\end{tabular}
\end{table}

\begin{table}[htb]
\caption{Energies (eV/atom) of alternate crystal structures of Si relative
to the cubic diamond phase in comparison with first-principles calculations.\label{tab:alternate_strucs_iso_vol} }

\centering{}%
\begin{tabular}{lccccc}
\hline 
Structure  & \emph{Ab initio}  & Present  & MOD$^{i}$  & MEAM$^{l}$  & SW$^{a}$ \tabularnewline
\hline 
FCC  & 0.449$^{c}$; 0.57$^{f}$; 0.537$^{h,m}$  & 1.113  & 0.4473  & 0.8975  & 0.3963 \tabularnewline
 & 0.6494$^{n}$; 0.5536$^{p}$  &  &  &  & \tabularnewline
HCP  & 0.55$^{f}$; 0.508$^{m}$; 0.5946$^{n}$; 0.5301$^{p}$  & 1.1019  & 0.4426  & 0.8909  & 0.3963 \tabularnewline
BCC  & 0.43$^{q}$; 0.435$^{c}$; 0.46$^{j}$;  & 0.6945  & 0.4377  & 0.5354  & 0.2810 \tabularnewline
 & 0.53$^{f}$; 0.523$^{m}$; 0.6142$^{p}$  &  &  &  & \tabularnewline
HEX  & 0.293$^{e}$  & 0.7322  & 0.3901  & 0.5591  & 0.3876 \tabularnewline
SC  & 0.276$^{c}$; 0.35$^{f}$; 0.38$^{b}$  & 0.2849  & 0.3076  & 0.4688  & 0.2745 \tabularnewline
$\beta\textrm{-Sn}$  & 0.19$^{d}$; 0.33$^{d}$; 0.414$^{d}$; 0.454$^{d}$  & 0.3725  & 0.3343  & 0.3671  & 0.2012 \tabularnewline
 & 0.3264$^{n}$; 0.27$^{f}$; 0.32$^{b}$; 0.290$^{h}$  &  &  &  & \tabularnewline
 & 0.2718$^{p}$; 0.380$^{r}$; 0.291$^{m}$  &  &  &  & \tabularnewline
BC8  & 0.13$^{s}$; 0.159$^{h}$; 0.126$^{j}$  & 0.2008  & 0.2127  & 0.2502  & 0.1880 \tabularnewline
 & 0.110$^{k}$; 0.166$^{n}$  &  &  &  & \tabularnewline
Wurtzite  & 0.011$^{h,m}$; 0.016$^{f}$  & 0.0000  & 0.0000  & 0.00001  & 0.0000 \tabularnewline
ST12  & 0.136$^{j}$; 0.1181$^{k}$  & 0.3900  & 0.4470  & 0.6031  & 0.4857 \tabularnewline
cP46  & 0.063$^{h}$; 0.0637$^{n}$  & 0.0703  & 0.0581  & 0.0625  & 0.0502 \tabularnewline
h-Si$_{6}$  & 0.35$^{g}$  & 0.5021  & 0.5863  & 0.6464  & 0.8417\tabularnewline
Si$_{24}$ & 0.09$^{t}$ & 0.1816 & 0.1864 & 0.2340 & 0.1949\tabularnewline
\hline 
RMS error  &  & 0.2883  & 0.1124 & 0.2138 & 0.1745\tabularnewline
\hline 
\multicolumn{6}{l}{$^{a}$Ref.\,\onlinecite{Stillinger85}; $^{b}$Ref.\,\onlinecite{Timonova:2007fr}
and references therein}\tabularnewline
\multicolumn{6}{l}{$^{c}$Ref.\,\onlinecite{Kumagai:2007ly}; $^{d}$Ref.\,\onlinecite{Sorella:2011xd}
and references therein}\tabularnewline
\multicolumn{6}{l}{$^{e}$Ref.\,\onlinecite{Balamane:1992fp} and references therein;
$^{f}$Ref.\,\onlinecite{Yin:1982ef} }\tabularnewline
\multicolumn{6}{l}{$^{g}$Ref.\,\onlinecite{Guo:2015aa}; $^{h}$Ref.\,\onlinecite{Materials_Project};
$^{i}$Ref.\,\onlinecite{Kumagai:2007ly}; $^{j}$Ref.\,\onlinecite{Needs:1995uk};
$^{k}$Ref.\,\onlinecite{Crain:1994lo}}\tabularnewline
\multicolumn{6}{l}{$^{l}$Ref.\,\onlinecite{Ryu:2009dn}; $^{m}$Ref.\,\onlinecite{Wolverton2013};
$^{n}$Ref.\,\onlinecite{database-cmu.edu}; $^{p}$Refs.\,\onlinecite{Curtarolo:2012kq,Curtarolo2012}}\tabularnewline
\multicolumn{6}{l}{$^{q}$Ref.\,\onlinecite{Methfessel1989}; $^{r}$Ref.\,\onlinecite{Kaltak:2014ee};
$^{s}$Ref.\,\onlinecite{Yin:1984gm}; $^{t}$Ref.\,\onlinecite{Kim:2015aa}}\tabularnewline
\end{tabular}
\end{table}

\begin{table}[htb]
\caption{Equilibrium volume per atom (Å$^{3}$) of alternate crystal structures
of Si in comparison with experiment and first-principles calculations. }
\label{tab:vol_strucs} \centering{}%
\begin{tabular}{lcccccc}
\hline 
Structure  & Experiment  & \emph{Ab initio}  & Present  & MOD$^{e}$  & MEAM$^{h}$  & SW$^{a}$ \tabularnewline
\hline 
Diamond  & 20.024$^{f}$  & 20.264$^{c}$; 20.444$^{d}$; 20.439$^{i}$; 20.33$^{b,l}$;  & 20.052  & 20.002  & 20.024  & 20.023 \tabularnewline
 &  & 19.59$^{l}$; 20.46$^{m}$; 19.03$^{g}$; 16.686$^{f}$; 20.385$^{j}$  &  &  &  & \tabularnewline
 &  & 19.77$^{b}$; 20.42$^{b}$; 20.124$^{b}$; 20.21$^{b}$; 20.08$^{b}$  &  &  &  & \tabularnewline
FCC  &  & 14.678$^{c}$; 14.484$^{d}$; 14.504$^{i}$; 14.810$^{j}$;  & 14.448  & 14.262  & 17.312  & 17.824 \tabularnewline
 &  & 14.337$^{k}$  &  &  &  & \tabularnewline
HCP  &  & 14.477$^{c}$; 14.313$^{i}$; 14.68$^{j}$  & 14.439  & 14.257  & 17.279  & 17.824 \tabularnewline
BCC  &  & 14.738$^{c}$; 14.2427$^{k}$  & 14.483  & 14.045  & 15.592  & 17.082 \tabularnewline
HEX  &  & 15.21$^{l}$; 14.56$^{l}$; 13.15$^{p}$  & 15.423  & 14.992  & 17.457  & 18.230 \tabularnewline
SC  &  & 16.179$^{c}$; 15.7653$^{k}$  & 15.639  & 15.581  & 18.194  & 17.822 \tabularnewline
$\beta\textrm{-Sn}$  & 14.0$^{f}$; 14.2$^{f}$  & 15.479$^{c}$; 15.334$^{d}$; 16.0$^{f}$; 15.292$^{i}$  & 15.016  & 15.085  & 16.560  & 17.275 \tabularnewline
 &  & 14.92$^{b}$; 15.45$^{b}$; 15.25$^{b}$; 15.34$^{b}$  &  &  &  & \tabularnewline
 &  & 15.31$^{b}$; 15.405$^{j}$; 15.35$^{m}$; 14.8859$^{k}$  &  &  &  & \tabularnewline
BC8  & 18.13$^{f}$; 18.26$^{f}$  & 17.724$^{f}$; 17.48$^{g}$; 18.44$^{j}$; 18.427$^{d}$  & 18.112  & 18.079  & 19.374  & 17.902 \tabularnewline
 &  & 18.2619$^{k}$; 18.082$^{n}$  &  &  &  & \tabularnewline
Wurtzite  &  & 20.324$^{c}$; 20.440$^{d}$; 20.380$^{i}$; 19.7575$^{k}$  & 20.052  & 20.002  & 20.024  & 20.023 \tabularnewline
ST12  &  & 17.65$^{g}$; 17.57$^{g}$  & 18.083  & 18.123  & 20.931  & 18.325 \tabularnewline
cP46  &  & 23.256$^{d}$; 23.214$^{i}$; 23.128$^{j}$  & 22.746  & 22.663  & 23.042  & 22.663 \tabularnewline
h-Si$_{6}$  &  & 27.188$^{q}$  & 28.575  & 28.725  & 33.460  & 31.667\tabularnewline
Si$_{24}$ & 21.52$^{r}$ & 21.934$^{r}$ & 21.861  & 21.809 & 23.189 & 22.083\tabularnewline
\hline 
RMS error  &  &  & 0.6758  & 0.6609 & 1.9147 & 2.0452\tabularnewline
\hline 
\multicolumn{6}{l}{$^{a}$Ref.\,\onlinecite{Stillinger85}; $^{b}$Ref.\,\onlinecite{Sorella:2011xd}
and references therein; $^{c}$Ref.\,\onlinecite{Yin:1982ef};
$^{d}$Ref.\,\onlinecite{Materials_Project} } & \tabularnewline
\multicolumn{6}{l}{$^{e}$Ref.\,\onlinecite{Kumagai:2007ly}; $^{f}$Ref.\,\onlinecite{Needs:1995uk}
and references therein; $^{g}$Ref.\,\onlinecite{Crain:1994lo};
$^{i}$Ref.\,\onlinecite{Wolverton2013}} & \tabularnewline
\multicolumn{6}{l}{$^{j}$Ref.\,\onlinecite{database-cmu.edu}; $^{k}$Ref.\,\onlinecite{Oloriegbe_PhD:2008aa};
$^{l}$Ref.\,\onlinecite{Gaal-Nagy:2004nx}; $^{m}$Ref.\,\onlinecite{Kaltak:2014ee};
$^{n}$Ref.\,\onlinecite{Biswas:1984mg}; $^{p}$Ref.\,\onlinecite{Balamane:1992fp};
$^{q}$Ref.\,\onlinecite{Guo:2015aa}; $^{r}$Ref.\,\onlinecite{Kim:2015aa}} & \tabularnewline
\multicolumn{6}{l}{} & \tabularnewline
\multicolumn{6}{l}{} & \tabularnewline
\end{tabular}
\end{table}

\begin{table}[htb]
\caption{Energies $\gamma_{us}$ (in Jm$^{-2}$) of unstable stacking faults
computed with the present interatomic potential in comparison with
other potentials and first-principles calculations.\label{tab:GSF}}
\begin{tabular}{lccccc}
\hline 
Property  & \textit{Ab initio}  & Present  & MOD$^{a}$  & MEAM$^{b}$  & SW$^{c}$ \tabularnewline
\hline 
(111)$\langle110\rangle$ shuffle  & 1.81$^{d,e}$  & 1.09  & 1.04  & 1.40  & 0.87 \tabularnewline
(111)$\langle110\rangle$ glide  & 4.97$^{f}$  & 5.25  & 5.00  & 4.58  & 6.37 \tabularnewline
(111)$\langle211\rangle$ glide  & 2.02$^{d,e}$  & 2.39  & 2.05  & 2.86  & 3.09 \tabularnewline
(100)$\langle110\rangle$  & 2.15$^{e}$  & 2.44  & 1.77  & 2.19  & 1.61 \tabularnewline
\hline 
\hline 
$^{a}$Ref.\,\onlinecite{Kumagai:2007ly}; $^{b}$Ref.\,\onlinecite{Ryu:2009dn};
$^{c}$Ref.\,\onlinecite{Stillinger85}  &  &  &  &  & \tabularnewline
$^{d}$Ref.\,\onlinecite{Kaxiras:1993fk}; $^{e}$Ref.\,\onlinecite{Juan:1996uo};
$^{f}$Digitized from Ref.\,\onlinecite{Juan:1996uo}  &  &  &  &  & \tabularnewline
\end{tabular}
\end{table}

\begin{table}[htbp]
\caption{Cohesive energies (eV/atom) of Si clusters relative to isolated atoms
computed with four interatomic potentials in comparison with first-principles
calculations. The asterisk marks mechanically unstable structures
whose energies were obtained by anisotropic volume relaxation without
local atomic displacements.\label{tab:cluster_coh_eng} }

\centering{}%
\begin{tabular}{lcccccc}
\hline 
Cluster  & Experiment  & \emph{Ab initio}  & Present  & MOD$^{a}$  & MEAM$^{b}$  & SW$^{c}$ \tabularnewline
\hline 
Si$_{2}$  & 1.62$^{d}$  & 1.53$^{d}$; 1.81$^{e}$  & 1.327  & 1.788  & 2.473  & 1.084\tabularnewline
Si$_{3}$.1  &  & 2.03$^{d}$; 2.41$^{e}$  & 1.710  & 2.003  & 2.519  & 1.267\tabularnewline
Si$_{3}$.2  & 2.6$^{d}$  & 2.39$^{d}$; 2.58$^{e}$  & 1.757  & 2.197  & 2.672  & 1.446\tabularnewline
Si$_{3}$.3  &  & 2.61$^{e}$  & 2.259  & 2.147  & 2.815  & 1.480\tabularnewline
Si$_{4}$.1  &  & 1.82$^{d}$; 2.48$^{e}$  & 1.901  & 2.121  & 2.593  & 1.372 \tabularnewline
Si$_{4}$.2  &  & 2.02$^{d}$; 2.49$^{e}$  & 2.457  & 2.325{*}  & 2.984  & 1.669{*} \tabularnewline
Si$_{4}$.3  &  & 2.21$^{d}$; 2.73$^{e}$  & 2.571  & 2.810  & 3.021  & 2.035 \tabularnewline
Si$_{4}$.4  &  & 2.22$^{d}$  & 2.219  & 2.232  & 2.759  & 1.525 \tabularnewline
Si$_{4}$.5  &  & 2.68$^{d}$; 3.09$^{e}$  & 2.579  & 2.441{*}  & 2.995  & 1.746{*}\tabularnewline
Si$_{5}$.1  &  & 2.02$^{d}$; 2.62$^{e}$  & 2.613  & 3.013  & 3.075  & 2.168 \tabularnewline
Si$_{5}$.2  &  & 2.69$^{d}$; 3.04$^{e}$  & 2.800  & 2.731  & 3.159  & 2.062 \tabularnewline
Si$_{5}$.3  &  & 3.09$^{e}$  & 2.678  & 2.549{*}  & 3.037  & 1.845{*}\tabularnewline
Si$_{5}$.4  &  & 2.78$^{d}$; 3.30$^{e}$  & 2.836  & 2.821  & 3.124  & 2.146 \tabularnewline
Si$_{5}$.5  &  &  & 2.017  & 2.192  & 2.626  & 1.433 \tabularnewline
Si$_{6}$.1  &  & 2.22$^{d}$  & 2.618  & 3.023  & 3.075  & 2.168\tabularnewline
Si$_{6}$.2  &  & 3.33$^{e}$  & 2.862  & 2.793  & 3.269  & 2.142 \tabularnewline
Si$_{6}$.3  &  & 3.04$^{d}$; 3.448$^{e}$  & 2.664{*}  & 2.658{*}  & 3.225  & 1.970{*}\tabularnewline
Si$_{6}$.4  &  & 3.453$^{e}$  & 2.706{*}  & 2.771{*}  & 3.260  & 2.139{*}\tabularnewline
Si$_{6}$.5  &  &  & 2.606  & 2.975  & 3.045  & 2.132\tabularnewline
Si$_{6}$.6  &  &  & 2.093  & 2.239  & 2.651  & 1.475 \tabularnewline
Si$_{7}$.1  &  & 3.56$^{e}$  & 2.938  & 2.960  & 3.344  & 2.321{*}\tabularnewline
Si$_{8}$.1  &  & 3.22$^{e}$  & 2.919  & 3.006{*}  & 3.267  & 2.379{*}\tabularnewline
Chain  &  & 2.260$^{a}$  & 2.477  & 2.475  & 2.771  & 1.680\tabularnewline
\hline 
\multicolumn{5}{l}{$^{a}$Ref.\,\onlinecite{Kumagai:2007ly}; $^{b}$Ref.\,\onlinecite{Ryu:2009dn};
$^{c}$Ref.\,\onlinecite{Stillinger85}} &  & \tabularnewline
\multicolumn{3}{l}{$^{d}$Ref.\,\onlinecite{Raghavachari:1986yg} and references therein;
$^{e}$Ref.\,\onlinecite{Fournier:1992tw}} & \multicolumn{2}{c}{} &  & \tabularnewline
\end{tabular}
\end{table}

\begin{table}[htb]
\caption{Properties of single-layer silicenes computed with four interatomic
potentials in comparison with DFT calculations.\label{tab:2D_single}}
\label{tab:prop_Silicene} \centering{}%
\begin{tabular}{lccccc}
\hline 
Property  & \textsl{\emph{Ab initio}}\emph{ } & Present  & MOD$^{a}$  & MEAM$^{b}$  & SW$^{c}$ \tabularnewline
\hline 
\multicolumn{6}{l}{Honeycomb planar:}\tabularnewline
$E_{c}$ (eV/atom)  & 3.96$^{f}$  & 3.6955  & 3.8280  & 3.6234  & 3.1450 \tabularnewline
$b$ (Å)  & 3.895$^{f}$  & 4.042  & 4.019  & 4.306  & 4.104\tabularnewline
$r_{1}$ (Å)  & 2.249$^{f}$  & 2.332  & 2.321  & 2.486  & 2.369 \tabularnewline
 &  &  &  &  & \tabularnewline
\multicolumn{6}{l}{Honeycomb buckled:}\tabularnewline
$\Delta E_{c}^{\textrm{buckled}-{\normalcolor \textrm{diamond}}}$
(eV/atom)  & 0.76$^{e}$  & 0.88  & 0.69  & 0.89  & 1.09 \tabularnewline
$\Delta E_{c}^{\textrm{buckled}-\sqrt{3}\times\sqrt{3}}$ (eV/atom)  & 0.048$^{d}$  & 0.14  & -0.08  & 0.08  & 0.07 \tabularnewline
$b$ (Å)  & 3.88$^{k}$; 3.87$^{d,g}$; 3.83$^{e}$  & 3.870  & 3.820  & 3.944  & 3.840 \tabularnewline
$r_{1}$ (Å)  & 2.28$^{d}$; 2.25$^{e,l}$  & 2.328  & 2.312  & 2.449  & 2.352 \tabularnewline
$\Delta$ (Å)  & 0.44$^{d,e}$; 0.45$^{g,j}$  & 0.655  & 0.694  & 0.901  & 0.784\tabularnewline
 & 0.46$^{i}$; 0.49$^{l}$  &  &  &  & \tabularnewline
 &  &  &  &  & \tabularnewline
\multicolumn{6}{l}{$\sqrt{3}\times\sqrt{3}$ Dumbbell:}\tabularnewline
$b$ (Å)  & 6.52$^{d,h}$  & 6.475  & 6.471  & 6.312  & 6.604 \tabularnewline
$r_{\textrm{II,III}}$ (Å)  & 2.40$^{d,h}$  & 2.393  & 2.425  & 2.526  & 2.513 \tabularnewline
$r_{\textrm{I,II}}$ (Å)  & 2.28$^{d}$  & 2.333  & 2.425  & 2.456  & 2.359 \tabularnewline
$\Delta_{\textrm{III,III}}$ (Å)  & 2.76$^{h}$  & 3.0564  & 3.111  & 3.160  & 3.261 \tabularnewline
\hline 
\hline 
\multicolumn{6}{l}{$^{a}$Ref.\,\onlinecite{Kumagai:2007ly} ; $^{b}$Ref.\,\onlinecite{Ryu:2009dn};
$^{c}$Ref.\,\onlinecite{Stillinger85}; $^{d}$Ref.\,\onlinecite{Kaltsas:2013rq};
$^{e}$Ref.\,\onlinecite{Cahangirov:2009fu};$^{f}$Ref.\,\onlinecite{Balamane:1992fp}; }\tabularnewline
\multicolumn{6}{l}{$^{g}$Ref.\,\onlinecite{Ge:2016mz}; $^{h}$Ref.\,\onlinecite{Cahangirov:2014xw};
$^{i}$Ref.\,\onlinecite{Ni:2012aa}; $^{j}$Ref.\,\onlinecite{Matusalem:2015aa};
$^{k}$Ref.\,\onlinecite{Roome:2014aa}; $^{l}$Ref.\,\onlinecite{Sahin:2013aa}}\tabularnewline
\end{tabular}
\end{table}

\begin{table}[htb]
\caption{Properties of three structures of bilayer silicenes computed with
interatomic potentials and DFT calculations. }
\label{tab:bilayer_Silicene} 
\centering{}%
\begin{tabular}{lccrcc}
\hline 
Property  & \textsl{Ab initio}  & Present  & MOD$^{a}$  & MEAM$^{b}$  & SW$^{c}$ \tabularnewline
\hline 
\multicolumn{6}{l}{Bilayer planar silicene AA$_{p}$:}\tabularnewline
$E_{c}$ (eV/atom)  & 4.16$^{d}$; 4.27$^{d}$  & 4.3067  & 4.2183  & 4.1739  & 3.8542 \tabularnewline
$b$ (Å)  & 4.12$^{e}$; 4.13$^{d}$ & 4.3264; 3.9804  & 4.0913  & 4.2685  & 4.1497\tabularnewline
 &  4.14$^{d}$  &  &  &  & \tabularnewline
$d_{1}$ (Å)  & 2.38$^{d,e}$; 2.39$^{d}$  & 2.3641,2.3737  & 2.3621  & 2.4644  & 2.3958 \tabularnewline
$h$ (Å)  & 2.41$^{d,e}$  & 2.3916  & 2.4393  & 2.4869  & 2.4428 \tabularnewline
 &  &  &  &  & \tabularnewline
\multicolumn{6}{l}{Bilayer buckled silicene AA$^{\prime}$:}\tabularnewline
$E_{c}$ (eV/atom)  &  & 4.1866  & 4.2776  & 4.1626  & 3.7945 \tabularnewline
$\Delta E{}^{\textrm{buckled-planar}}$ (eV/atom)  & 0.33$^{e}$  & 0.1201  & -0.0593  & 0.0113  & 0.0597\tabularnewline
$b$ (Å)  & 3.84$^{e}$  & 3.8430  & 3.8245  & 3.9155  & 3.8402\tabularnewline
$d_{1}$ (Å)  &  & 2.3405  & 2.3311  & 2.4081  & 2.3517\tabularnewline
$d_{2}$ (Å)  &  & 2.3543  & 2.3515  & 2.3801  & 2.3517\tabularnewline
$h$ (Å)  &  & 3.0994  & 3.0990  & 3.2101  & 3.1356\tabularnewline
$\Delta$ (Å)  &  & 0.7451  & 0.7475  & 0.8300  & 0.7839\tabularnewline
 &  &  &  &  & \tabularnewline
\multicolumn{6}{l}{Bilayer buckled silicene AB:}\tabularnewline
$E_{c}$ (eV/atom)  & 4.10$^{d}$; 4.25$^{d}$  & 4.1866  & 4.2776  & 4.1626  & 3.7945 \tabularnewline
$\Delta E{}^{\textrm{buckled-planar}}$ (eV/atom)  & 0.17$^{e}$  & 0.1201  & -0.0593  & 0.0113  & 0.0597 \tabularnewline
$b$ (Å)  & 3.84$^{d,e}$; 3.86$^{d}$  & 3.8429  & 3.8245  & 3.9155  & 3.8402 \tabularnewline
$d_{1}$ (Å)  & 2.32$^{d}$  & 2.3405  & 2.3311  & 2.4082  & 2.3517 \tabularnewline
$d_{2}$ (Å)  & 2.51$^{d}$; 2.54$^{d}$  & 2.3543  & 2.3515  & 2.3801  & 2.3517 \tabularnewline
$h$ (Å)  & 3.19$^{d}$; 3.20$^{d}$  & 3.0994  & 3.0990  & 3.2101  & 3.1359 \tabularnewline
$\Delta$ (Å)  & 0.66$^{d}$; 0.68$^{d}$  & 0.7451  & 0.7475  & 0.8300  & 0.7839 \tabularnewline
\hline 
\hline 
\multicolumn{6}{l}{$^{a}$Ref.~\onlinecite{Kumagai:2007ly}; $^{b}$Ref.\,\onlinecite{Ryu:2009dn};
$^{c}$Ref.\,\onlinecite{Stillinger85}; $^{d}$Ref.\,\onlinecite{Fu:2014ai};
$^{e}$Ref.\,\onlinecite{Padilha:2015ng}}\tabularnewline
\end{tabular}
\end{table}

\newpage{}

\clearpage{}

\end{document}